\documentclass[prb,twocolumn,showpacs,preprintnumbers,amsmath,amssymb,floatfix,nofootinbib]{revtex4-2}

\usepackage[resetlabels,labeled]{multibib}

\usepackage{graphicx}
\usepackage{dcolumn}
\usepackage{bm}
\usepackage{color}
\usepackage{ulem}
\usepackage{multirow}
\usepackage{gensymb}
\usepackage{epstopdf}
\usepackage[colorlinks=true]{hyperref}
\usepackage[dvipsnames]{xcolor}
\usepackage{amsmath}
\usepackage{amssymb}
\usepackage{textcomp}
\usepackage{placeins}
\usepackage{braket}
\usepackage[dvipsnames]{xcolor}
\usepackage{amsmath}
\usepackage{amssymb}
\usepackage{bm}
\usepackage{epsfig}
\usepackage{graphicx}
\usepackage{color}
\usepackage{hyperref}
\usepackage[toc,page,titletoc]{appendix}
\usepackage[capitalise,nameinlink]{cleveref}

\def\be{\begin{equation}}
\def\ee{\end{equation}}
\def\bea{\begin{eqnarray}}
\def\eea{\end{eqnarray}}

\newcommand{\specialcell}[2][c]{%
  \begin{tabular}[#1]{@{}c@{}}#2\end{tabular}}

\begin{document}
\newcount\timehh  \newcount\timemm
\timehh=\time \divide\timehh by 60
\timemm=\time
\count255=\timehh\multiply\count255 by -60 \advance\timemm by \count255
\title{Land\'e $g$-factors and spin dynamics of charge carriers \\in CuCl nanocrystals in a glass matrix}

\author{Dennis~Kudlacik$^{1}$,  Evgeny~A.~Zhukov$^{1,2}$, Dmitri~R.~Yakovlev$^{1,2}$,  Gang~Qiang$^{1}$,  \\ Marina~A.~Semina$^{2}$, Aleksandr A.~Golovatenko$^{2}$, Anna~V.~Rodina$^{2}$, \\ Alexander L. Efros$^{3}$,  Alexey~I. Ekimov$^{4}$,   and Manfred~Bayer$^{1}$}

\affiliation{$^{1}$Experimentelle Physik 2, Technische Universit\"at Dortmund, 44221 Dortmund, Germany}
\affiliation{$^{2}$Ioffe Institute, Russian Academy of Sciences, 194021 St. Petersburg, Russia}
\affiliation{$^{3}$Naval Research Laboratory, Washington, DC 20375, USA}
\affiliation{$^{4}$Nanocrystals Technology Inc., New York, NY, USA }

\date{\today}

\begin{abstract}
The spin properties of charge carriers confined in CuCl semiconductor nanocrystals (NCs) of different sizes (radius from 1.8~nm up to 28~nm) crystallized in a glass matrix are studied experimentally and theoretically. By means of photoluminescence, spin-flip Raman scattering,  time-resolved Faraday ellipticity, and time-resolved differential transmission performed at temperatures in the range of $1.6 - 120$~K at magnetic fields up to 8~T, comprehensive information on the Land\'e $g$-factors as well as the population and spin dynamics is received. The spin signals are contributed by confined electrons with a $g$-factor close to 2, which shows a weak increase with decreasing NC size, i.e. increasing electron confinement energy. 
We revisit the theory of exciton confinement as a whole in spherical NCs within the six-band valence band model in order to describe the size dependence of the $Z_3$ and $Z_{1,2}$ exciton energies in CuCl NCs. We demonstrate theoretically that the stronger increase of the $Z_{1,2}$ energy transitions with decreasing radius can be explained by the strong absorption from the excited exciton state caused by strong heavy hole-light hole mixing in the exciton. The parameters of the six-band Hamiltonian describing both the exciton and hole kinetic energies are estimated from the comparison of the calculated and experimental size dependences of the exciton transitions. A theoretical model of the size-dependent Land\'e $g$-factors for electron and hole confined in spherical NCs of semiconductors with negative spin-orbit splitting of the valence band is developed.    

\end{abstract}


\maketitle

\section{Introduction}

CuCl nanocrystals (NCs) most propbably were the first material system for which the quantum confinement effect was observed by Gross and Kaplyanskii in 1957 \cite{KG1957}. Puzzling at the time of their discovery, these observations are now well understood in retrospect. The authors were studying the absorption spectra of NaCl crystals strongly doped with Cu and observed two lines corresponding to the spectra of bulk CuCl. The absorption spectra of samples with smaller Cu concentration also showed two CuCl-like lines, however, shifted to higher energies compared to the lines in highly doped crystals. The CuCl lines disappeared from the spectra after heating and subsequent rapid quenching of the sample. One year later, they remeasured the absorption of that sample and again found the two additional lines. Their splitting was practically the same as in bulk CuCl, but the lines were shifted by 100 meV to higher energies from their bulk values. 

All these data were reported in great detail, but the authors did not recognize the quantum confinement effect as the origin of their observations. 24 years later in 1981, after the first publication of Ekimov and Onuschenko~\cite{ekimov1981quantum} on the size dependent optical properties of CuCl NCs in a glass matrix, it became clear that the energy shift observed by Gross and Kaplyanskii  is due to the spatial confinement of excitons in CuCl nanocrystals, which formed in the NaCl:Cu crystals. The reappearance of the CuCl absorption lines without sample heat treatment is connected with the slow growth of CuCl NCs at room temperature due to the phase decomposition of a supersaturated NaCl:Cu solid solution.    

The exciton radius in bulk CuCl is $a_{\rm ex}=0.7$\,nm~\cite{Nikitine1967}. Consequently, the exciton has to be considered in the weak confinement regime~\cite{EE1982,Ekimov1985} in all investigated NCs because $a_{\rm ex}$ is much smaller than the NC radius, $a$. The size dependence of the exciton optical transition energy, $\hbar\omega(a) \equiv  E_{\rm X}(a)$, 
in this regime can be described by the model of a particle confined in a spherical box:
\be \label{eq:1}
 E_{\rm X}(a)=E_g- E_{\rm b}+ {\hbar^2\pi^2\over 2 Ma^2} \,.
\ee 
Here $E_g=3.399$~eV is the band gap energy of CuCl at liquid helium temperature (3.25 eV at room temperature), $E_{\rm b}= 197$~meV is the exciton binding energy~\cite{Saito1995}, and $M$ is the exciton translational mass.

In addition to the size dependence of the optical transition energy, the initial practical interest in CuCl NCs was provided by their strong optical nonlinearity. Indeed, in the weak confinement regime, the exciton oscillator strength is proportional to the ratio $(a/a_{\rm ex})^3$ \cite{EE1982}. The enhancement factor of the exciton oscillator strength due to localization -- resulting in the so-called "the giant oscillator strength" of localized excitons -- was predicted by Rashba and Gurgenishvili in 1962~\cite{Rashba}. In CuCl nanocrystals, this enhancement results in fast exciton radiative recombination within a few tens of picoseconds. The size dependence of the exciton decay time $\propto 1/a^3$ was measured by Itoh et al. \cite{Itoh1990} and Kataoka et al.~\cite{Kataoka1989,Kataoka1993}. Optical excitation of the biexciton in CuCl NCs also should have a giant oscillator transition strength, leading to a strong nonlinear optical effect in absorption~\cite{Kataoka1993,Yano1997,Sato2016biexciton}. After significant efforts to understand inhomogeneous broadening of the exciton line in CuCl NCs~\cite{Itoh1988,Itoh1991}, the first observation of superradiance was reported for an ensemble of CuCl NCs~\cite{Miyajima2009}. 

Surprisingly, however, there are very few reports o fresearch work on spin related phenomena and the exciton fine structure in CuCl NCs. We have found only several papers that investigate the optical properties of CuCl NCs in a magnetic field. Nomura et al.  measured the exciton $g$-factor for CuCl NCs embedded in a polymer matrix using magnetic circular dichroism~\cite{NomuraPRB1993}. In NCs with radius of $a=4.6$~nm they found a $g$ factor equal to  0.447  for  the lowest exciton line and $-0.18$ for the upper exciton line.  Miyajima et al. studied CuCl NCs embedded in a NaCl matrix in magnetic field and measured the magnetic-field-induced mixing of the bright and dark excitons, identified through the acceleration of the emission decay~\cite{ItohPSSC2008}. 
 
\begin{figure*}[hbt]
\includegraphics[width=15 cm]{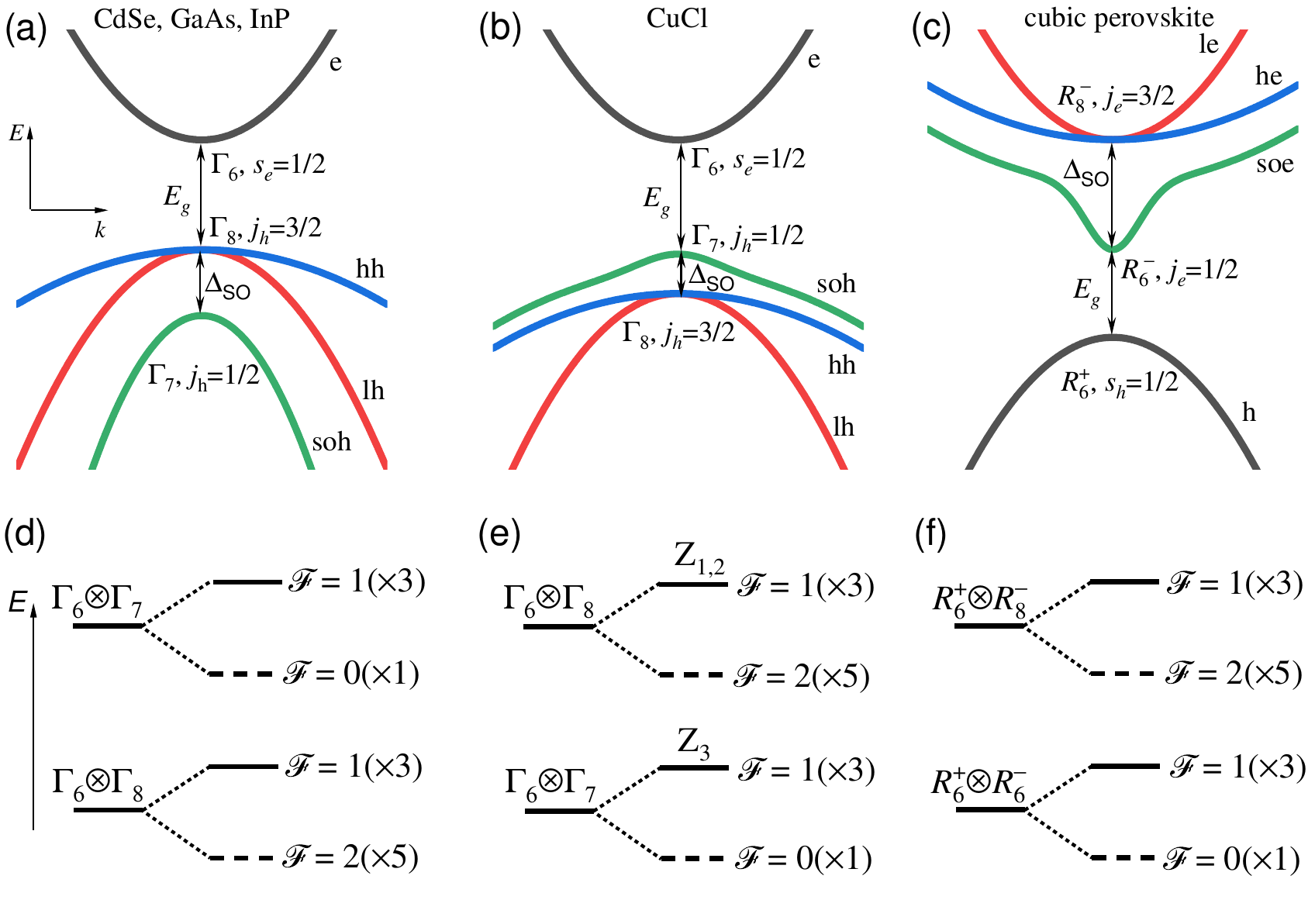}
\caption{Valence and conduction bands in bulk semiconductors: (a) typical II-VI and III-V semiconductors with zinc-blend crystal structure, like  CdSe, GaAs, and InP ($\Delta_{\rm SO}>0$), (b) CuCl ($\Delta_{\rm SO}<0$), and  (c) lead halide perovskites in the cubic phase. Black lines show the dispersion (isotropic energy dependence on the wave vector $k$) for the $s$-type energy bands. Blue, red and green lines show the dispersion of the $p$-type energy bands corresponding to heavy-, light-, and spin-orbit-split hole (electron) states.  Panels (d-f) show the corresponding lowest exciton states at rest with total angular momentum ${\cal F}=0,1,2$   (\bm{${\cal F}$}  =${\bm s}_e+{\bm s}_h + {\bm I}$), comprising the electron and hole spins and the internal orbital angular momentum.  The degenerate states are split by the short range exchange interaction $\propto ({\bm \sigma}_e{\bm \sigma}_h)$. Solid and dashed lines are for the bright and dark exciton states, respectively. The notation details are given in the text. } 
\label{Fig:ebands}
\end{figure*}

CuCl is a direct-band-gap semiconductor with zinc-blend crystal structure without inversion symmetry (point group T$_d$). Typically, in zinc-blend semiconductors, for example, the cubic-phase II-VI and III-V semiconductors, the conduction and valence bands are formed by $s$-type and $p$-type atomic orbitals, respectively. They have extrema at the $\Gamma$-point of the Brillouin zone, as shown in  Figs.~\ref{Fig:ebands}(a) and \ref{Fig:ebands}(b). The respective basis functions are the invariant orbital Bloch function $S$ for the conduction band and the $X, Y, Z$ Bloch functions transforming as the coordinates $x,y,z$ by the operations of the point group T$_d$ for the valence band. The $\Gamma_{6}$ conduction band states are two-fold degenerate with respect to the electron spin projection on the direction of motion: ${s}_{ez} =\pm 1/2$ corresponding to the eigenstates $\uparrow$ and $\downarrow$ of the spin operator ${\bm s}_e = {\bm \sigma}^e/2$. Here, the $\sigma^e_{x,y,z}$ are the Pauli matrices describing the electron spin. The electrons have an isotropic parabolic dispersion $E(k)$ at small wave vectors $k$, as shown by the black curves marked by 'e' in Figs.~\ref{Fig:ebands}(a) and \ref{Fig:ebands}(b).   

The valence band is characterized by the angular momentum operator ${\bm j}_h = {\bm I} +{\bm s}_h$, where $\bm I$ ($I=1$) is the internal orbital angular momentum operator with eigen functions being linear combinations of the $X, Y, Z$,  and the hole spin operator ${\bm s}_h = {\bm \sigma}_h/2$, where $s_h=1/2$ ($\sigma^h_{x,y,z}$ are the Pauli matrices describing the hole spin  states $\uparrow$ and $\downarrow$).  The spin-orbit interaction splits the top of the valence band into the four-fold degenerate states with $j_h=3/2$ ($j_{hz}=\pm3/2, \pm1/2$) transforming according to the $\Gamma_8$ irreducible representation and the two-fold degenerate $\Gamma_7$ states  with $j_h=1/2$ ($j_{hz}= \pm1/2$ ). The respective dispersion of the heavy hole (hh) with projections $\pm 3/2$, the light hole (lh) with projections $\pm 1/2$, and the spin-orbit-split holes (soh) are shown in Figs.~\ref{Fig:ebands}(a) and \ref{Fig:ebands}(b) by the blue, red, and green lines, respectively. Importantly, in contrast to GaAs, CdSe, or InP, in CuCl the $\Gamma_7$ valence band with the total angular momentum ${j}_h=1/2$ has a higher energy than the $\Gamma_8$ valence band with ${j}_h=3/2$, so that the spin-orbit energy splitting $\Delta_{\rm SO}<0$, see Fig.~\ref{Fig:ebands}(b),  while it is positive for conventional semiconductors, see Fig.~\ref{Fig:ebands}(a). 

The energy band structure of CuCl has many features in common with the lead halide perovskites in the cubic phase, shown in Fig.~\ref{Fig:ebands}(c). In the perosvkites, the band structure is inverted in comparison to CuCl. Namely, the conduction and valence bands are formed by $p$-type and $s$-type atomic orbitals, respectively. They have extrema at the ${\cal R}$-point  of the Brillouin zone  with the wave vector group O$_h$ having the same character tables of irreducible representations as the point group O$_h$ at the $\Gamma$ point \cite{Even2015}.  Thus, the respective Bloch functions $X$, $Y$, $Z$ and $S$  of the conduction and valence bands are transforming according to the point group O$_h$. The hole  ${\cal R}_6^+$ states at the top of the valence band have spin $s_h=1/2$ as well as a parabolic dispersion, see the black curve (h) in Fig.~\ref{Fig:ebands}(c). The conduction band is split by the spin-orbit interaction into the lowest ${\cal R}_6^-$ electron states ('soe', green) with total angular momentum $j_e=1/2$ and  projections $j_{ez}= \pm 1/2$ on the direction of motion, and the ${\cal R}_8^-$ electron states with $j_e=3/2$ and projections $j_{ez}=\pm3/2$ for the heavy electrons ('he', blue) and $j_{ez}=\pm1/2$  for the light electrons ('le', red). Here ${\bm j}_e= {\bm I}+ {\bm s}_e$ is the total electron angular momentum.
 
The band-edge exciton in CuCl is formed by the electron with spin $s_e=1/2$ and the hole with internal angular momentum $j_h=1/2$. The resulting ground state exciton, similar to perovskites, has a $2\times 2 = 4$-fold degeneracy without accounting for the electron-hole exchange interaction or other perturbations, see Figs.~\ref{Fig:ebands}(e,f).  On the other hand, in semiconductors with a $\Gamma_8$-top valence band, the exciton ground state is formed with the hole $j_h=3/2$ and has a $2\times 4 = 8$-fold degeneracy, as shown in Fig.~\ref{Fig:ebands}(d). In these diagrams we label a degeneracy $\nu$ of an exciton state with $(\times \nu)$, where $\nu=1,3,5$. The symmetry analysis shows that in all considered semiconductors both the ground state and the excited exciton manifolds have only three  optically active states. Indeed, the electromagnetic field vector components transform as vector coordinates $x,y,z$ according to the three-dimensional irreducible representation $\Gamma_5$ of the $T_d$ point group. According to the multiplication table \cite{Koster}, we obtain for the  exciton with $\Gamma_7$ holes $\Gamma_6\times \Gamma_7=\Gamma_2+\Gamma_5$ and for the exciton with $\Gamma_8$ holes   $\Gamma_6\times \Gamma_8= \Gamma_3+\Gamma_4+\Gamma_5$. Only $\Gamma_5$ excitons are optically active.  They are three-fold degenerate and described by the  total exciton angular momentum $\bm{{\cal F}}=1$, where $\bm{{\cal F}}  ={\bm s}_e+{\bm s}_h + {\bm I}$. The $\Gamma_2$ state with $\bm{{\cal F}}=0$ and the $\Gamma_3+\Gamma_4$ states with  $\bm{{\cal F}}=2$  are optically forbidden dark excitons.  In Figures~\ref{Fig:ebands}(d,e,f) the bright and dark excitons split by the isotropic electron-hole exchange interaction $\propto ({\bm \sigma}^e{\bm \sigma}^h)$ are shown by solid and dashed lines, respectively. Exactly the same ordering of the exciton states with the lowest four-fold exciton manifold and the excited eight-fold exciton manifold occurs in cubic-phase lead halide perovskites \cite{Sercel2019,Rodina2024}. However, in perovskites the energy splitting between them, corresponding to the absolute value of the spin-orbit energy splitting $|\Delta_{\rm SO}|$, is comparatively large (about 1.5~eV \cite{Sercel2019}), so that the excitons can be considered separately, without mixing between them. In CuCl, this energy is about 70 meV \cite{Cardona1963}, and the interaction between holes and thus excitons formed with the holes from different subbands may become important, especially due to their confinement in nanocrystals.
In this paper, we consider only the confinement in spherical-symmetry nanocrystals, so that the symmetry of the hole and exciton states is not lowered in comparison with the bulk states. The quasi-cubic model for the confinement of carriers and excitons with lower symmetry due to both the crystal phase and cubic shape of perovskite nanocrystals was developed in Ref.~\onlinecite{Sercel2019}. 

Two exciton absorption peaks in CuCl crystals are related to the $\Gamma_7$ and $\Gamma_8$ valence bands and are commonly labeled  as $Z_3$ and $Z_{1,2}$ excitons, respectively \cite{Cardona1963}. They were also observed in the absorption spectra of CuCl NCs embedded in glass \cite{ekimov1981quantum,Ekimov1982,Ekimov_jtp,Ekimov1985} and in a NaCl matrix \cite{Itoh1988}. Both peaks shift to higher energy with decreasing NC size due to exciton confinement. 

In this paper, we study experimentally and theoretically the spin properties of charge carriers confined in CuCl semiconductor nanocrystals of different sizes (radius from 1.8~nm up to 28~nm) embedded in a glass matrix. Spin-flip Raman scattering and time-resolved Faraday ellipticity (or rotation) are used to obtain information on the carrier Land\'e $g$-factors and their coherent spin dynamics. Time-resolved differential transmission is used to measure the carrier population dynamics in NCs. Theoretically, we reexamine the level structure of exciton transitions involving two valence band states, and calculate its modification in NCs of various sizes. A theory of the electron and hole $g$-factors in CuCl NCs is developed and model calculations with various band parameters are made. These results are compared with the experimental data.    

The paper is organized as follows. The sample description and description of experimental techniques are given in Sec.~\ref{sec:experimentals}. In Section~\ref{sec:ER}, the experimental results on the studied CuCl NCs are given, starting with the optical properties (absorption) in zero magnetic field, followed by the spin properties measured by spin-flip Raman scattering (SFRS) and by the opulation and spin dynamics assessed by time-resolved techniques. In Section~\ref{sec:theory}, the theoretical description of the optical spectra of confined excitons and the developed model of the $g$-factors of electrons, holes and excitons in CuCl NCs are given. The theoretical considerations are accompanied by calculations using various band parameters, whose values are not well established for CuCl NCs. 
The comparison of theoretical and experimental results is given in Sec.~\ref{sec:discussion}.  

\section{Experimentals}
\label{sec:experimentals}

We studied CuCl NCs of various sizes embedded in a multicomponent silicate glass matrix, with an initial composition including a compound of copper and chlorine at a concentration on the order of 1\%. Other components of the glass were selected to increase the diffusion coefficients of Cu and Cl in the matrix. Five samples with average NC radii of 1.8~nm, 2.9~nm, 4.0~nm, 8.1~nm, 15.3~nm, and 28~nm are measured. The samples were prepared during the Oswald-ripening stage in the CuCl NCs growth from a supersaturated solution of the glass matrix, when large NCs grow at the expense of the small ones \cite{LS1959}. The technique developed in Ref.~\cite{Ekimov1982} allows one to control the NC size and change the average radius of the CuCl NCs from 1.2 nm to 30 nm. The radius was measured by small angle X-ray scattering. The size distribution of the NCs grown by this technique is about 15\%. Although the investigated samples were prepared in 1982  \cite{Ekimov1982}, they do not show any significant aging effect. For example, the exciton resonances are at the same energies as in the original experiments.  

Absorption measurements are performed with the use of a halogen lamp which light is transmitted through the sample and analyzed by an 0.5 meter spectrometer with a Silicon based charge-coupled-device (CCD) camera, cooled by liquid nitrogen, attached. The samples are placed in a helium bath cryostat, where they are in direct contact with pumped superfluid helium at the temperature of $T=1.7$~K.

The spin-flip Raman scattering (SFRS) experiments are performed in strong magnetic fields up to $8$~T applied parallel to the laser propagation direction. the samples are mounted in a cryostat with a superconducting split-coil solenoid and kept in direct contact with superfluid helium at $T=1.9$~K. Continuous-wave laser radiation with the photon energy tunable in the spectral range of $3.1-3.3$~eV is used for resonant excitation of the excitons in the CuCl NCs. The laser light is provided by a single-frequency Ti:Sa laser (Matisse DS) equipped with a second harmonic generation unit, which is based on a low-loss, ultra compact enhancement cavity using a patented triangle-shaped ring resonator. The Raman signals are detected in backscattering geometry and analyzed with a double monochromator (U1000) equipped with a Peltier-cooled GaAs photomultiplier. Raman signals are measured in close vicinity to the laser line in a range of $0.3-1.6$~meV, which corresponds to typical Zeeman splittings of excitons and charge carriers in semiconductors in not too small fields in the available range of field strengths. The Stokes and anti-Stokes components are analyzed, which are shifted to lower and higher energies from the laser line, respectively. The laser is circularly polarized ($\sigma_{exc}^+$ or $\sigma_{exc}^-$) and the Raman signal is analyzed for the circular polarizations $\sigma_{det}^+$ or $\sigma_{det}^-$. The light polarization is set by a combination of a $\lambda/4$ retardation plate and a Glan-Thompson prism.

For time-resolved pump-probe measurements, the emission of a Ti:Sapphire mode-locked laser (pulse duration 1.5~ps, spectral width about 1~meV, repetition frequency of 75.8~MHz corresponding to a repetition period of 13.2~ns) is frequency doubled with a nonlinear BBO crystal to be tunable in spectral range of $3.1-3.3$~eV. The time delay between the pump and probe pulses is provided by mechanical delay line. The population dynamics of the photogenerated carriers are measured by time-resolved differential transmission (TRDT). For that, linearly polarized pump and probe beams are used. The probe beam transmitted through the sample is sent to a balanced photodetector, where it is compared with the reference beam. The laser intensity is modulated at 50~kHz, which is used for lock-in detection.

The coherent spin dynamics of carriers are measured by time-resolved Faraday ellipticity (TRFE) in the transmission geometry. It represents a variation of the commonly used time-resolved Faraday rotation technique~\cite{Awschalom2002,Yakovlev_ch6_2017} providing, however, similar information. Spin oriented carriers are generated by the circularly polarized pump pulses. The pump helicity is modulated between $\sigma^+$ and $\sigma^-$ polarizations at 50~kHz frequency, using a photoelastic modulator. The pump-excited area of the sample (with a diameter of about 300~$\mu$m) is tested by the linearly polarized probe pulses with the beam size slightly smaller than the pump beam size. The induced ellipticity after transmission of the linearly polarized probe beam is measured using a balanced photodetector connected to a lock-in amplifier, for whcih a double modulation scheme is used. For that the probe beam is modulated in intensity at a frequency of 84~kHz and the signal is recorded at the difference frequency. The pump power is in the range of $P_{pump}=0.3-10$~mW and probe power is $P_{probe}=0.3$~mW.

For time-resolved measurements, the samples are placed in an optical cryostat with a split pair coils for magnetic field application. The temperature in the sample insert is variable from 1.7~K up to 300~K. For $T=1.7$~K, the samples are immersed in superfluid helium, while for higher temperatures they are neld in cold helium gas. The measurements are performed in magnetic fields up to 2.5~T, applied perpendicular to the direction of pump beam propagation (Voigt geometry $B_{\text{V}}$).

\section{Experimental results}
 \label{sec:ER}

\subsection{Absorption spectra of CuCl NCs}
 \label{sec:abs}
 
\begin{figure*}[hbt]
\begin{center}
\includegraphics[width=9cm]{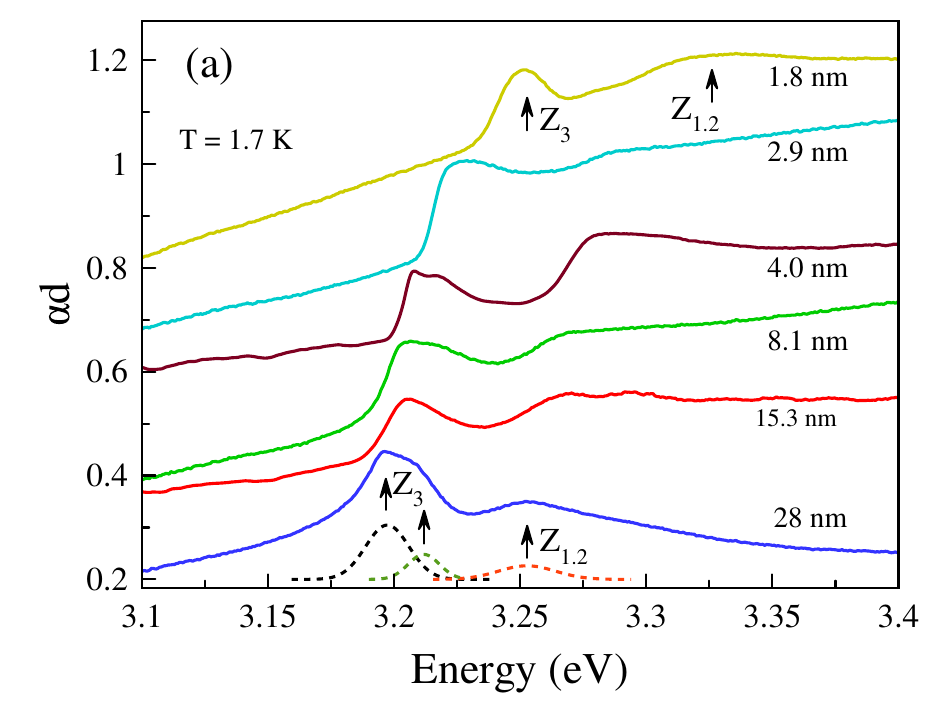}
\includegraphics[width=8.5cm]{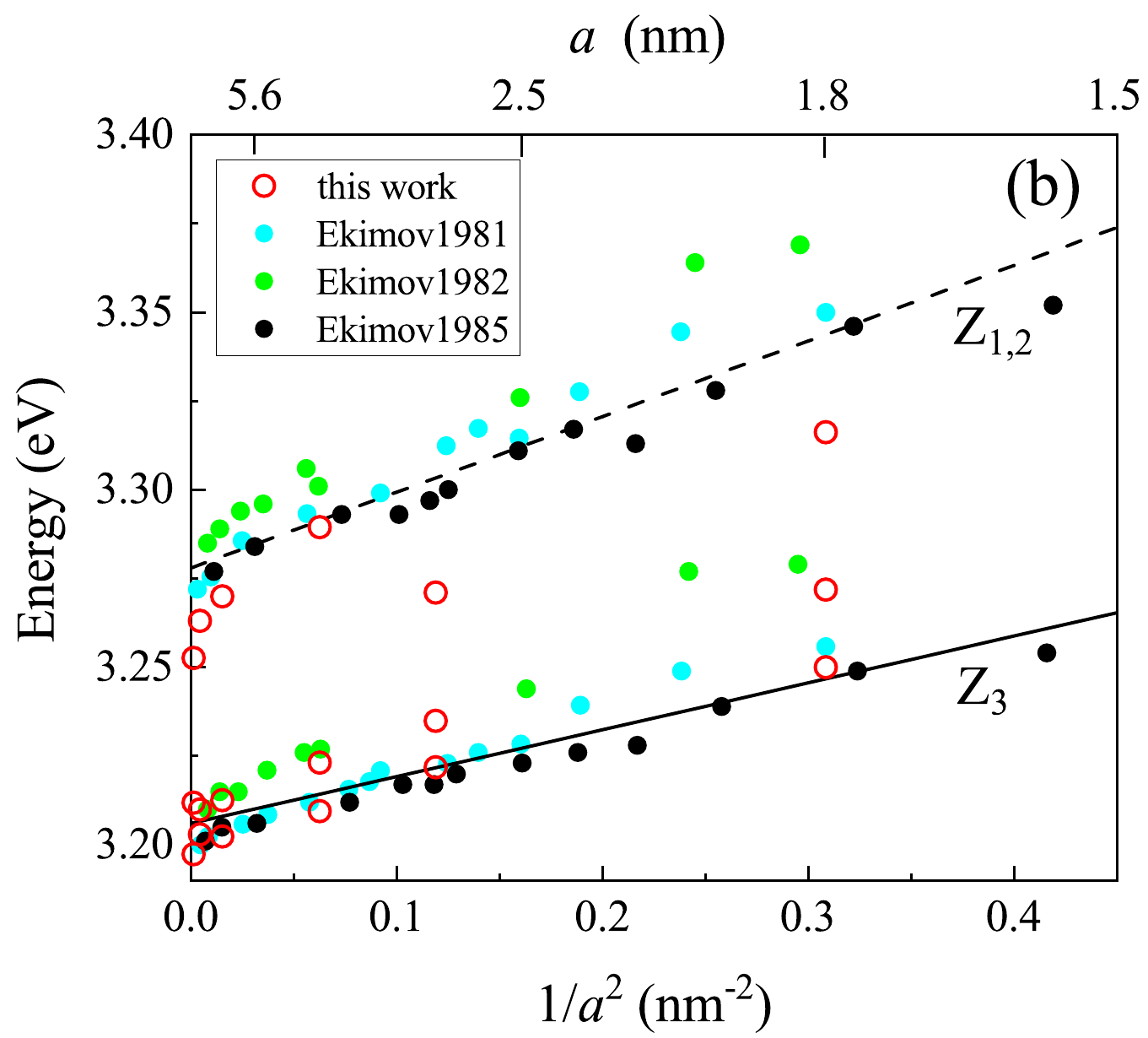}
\caption{(a)  Absorption spectra of CuCl NCs in glass, measured at  $T=1.7$~K The spectra for NCs with different radii $a$ are vertically shifted for clarity. The dashed lines in the lower part of the panel show the components obtained from fitting the absorption spectrum of the sample with NC radius $a=28$~nm, the arrows show the position of the exciton peaks.
(b) Energies of the exciton absorption lines, measured in this work and in Refs. \cite{ekimov1981quantum} (Ekimov1981), \cite{Ekimov1982} (Ekimov1982), and \cite{Ekimov_jtp,Ekimov1985} (Ekimov1985) plotted as function of $1/a^2$. Lines are linear fits to the data for $Z_3$ (solid line) and $Z_{1,2}$ (dashed line) with the parameters given in the text. 
}
\label{Fig:Abs}
\end{center}
\end{figure*}

The absorption spectra of the six studied samples, measured at $T=1.7$~K, are shown in  Fig.~\ref{Fig:Abs}(a). They show pronounced exciton transitions related to the lowest in energy $Z_3$ exciton, which involves the $\Gamma_7$ hole, and the $Z_{1,2}$ exciton contributed by the $\Gamma_8$ hole. Note that the $Z_3$ exciton spits into two lines.  
In order to the evaluate exciton energies, we fit the absorption spectra with three Gaussian lines. An example of such a fit is shown for the 28 nm NCs by the dashed lines. The exciton energies for all studied samples are given in Table~\ref{Abs:Z}. The effect of exciton confinement is evidences as high energy shift of the exciton energies with decreasing NC radius $a$. We plot this shift in Fig.~\ref{Fig:Abs}(b) as a function of  $1/a^2$. Here the measured data are shown by open red circles and the closed symbols give the results for CuCl NCs at $T=4.2$~K, taken from Refs.~\cite{ekimov1981quantum,Ekimov1982,Ekimov_jtp,Ekimov1985}.  

\begin{table}[hbt]
\begin{tabular}{|c|>{\centering\arraybackslash} m{0.12\textwidth}|>{\centering\arraybackslash} m{0.12\textwidth}|}
\hline
NC radius & $Z_3$ & $Z_{1,2}$ \\
(nm)     & (eV)& (eV) \\ \hline
$1.8$     &$3.252; \,\,  3.276$ &    $3.316$  \\ \hline
$2.9$     &$3.222; \,\,  3.235$ &    $3.283$  \\ \hline
$4.0$     &$3.207; \,\,  3.216$ &    $3.280$  \\ \hline
$8.1$     &$3.202; \,\,  3.217$ &    $3.279$  \\ \hline
$15.3$    &$3.204; \,\,  3.213$ &    $3.263$ \\ \hline
$28.0$    &$3.197; \,\,  3.212$  &    $3.253$ \\ \hline
\end{tabular}
\caption{Energies of the exciton transitions in the absorption spectra of the CuCl NCs shown in Fig.~\ref{Fig:Abs}(a). $T=1.7$~K. }
\label{Abs:Z}
\end{table}

Taken together all experimental data for the excitons $X=Z_3$ and $X=Z_{1,2}$ in Fig.~\ref{Fig:Abs}(b), their energies are well described by linear dependences according to 
\begin{equation}\label{eq:fit}
    E_{X=Z_i}(a)=E_0(Z_i)+C(Z_i)\frac{\hbar^2\pi^2}{2 m_0 a^2}.
\end{equation}
Here $m_0$ is the free electron mass, $i=1,2,3$ numbers the exciton lines $Z_{1,2}$ and $Z_3$, $E_0(Z_i)$ are the exciton energies in bulk CuCl, and the $C(Z_i)$ are the coefficients, which characteriz the slopes of the dependences. For the $Z_3$ line we obtain $E_0 (Z_3)=3.206$~eV and $C(Z_3)=0.35$. For the $Z_{1,2}$ line, $E_0 (Z_{1,2})=3.278$~eV  and $C(Z_{1,2})=0.56$. 

For an exciton formed with two parabolic bands, resulting in Eq.~\eqref{eq:1}, one would expect to obtain the $\Gamma_7$ exciton translational mass $M_s=m_0/C(Z_3)$, where $M_s=m_e+m_h$ is the sum of the electron and hole effective masses, correspondingly. One has to be careful here, as it was shown in Refs.~\cite{EE1982,Ekimov_jtp,Ekimov1985} that in the case of a NC ensemble with size dispersion described by the Lifshitz--Slesov distribution ~\cite{LS1959}, an additional coefficient 0.67 needs to be taken into account for excitons with the oscillator transition strength $\propto a^3$. In this case, $M_s$ was determined to be $0.67 m_0/C(Z_3)=1.9 m_0$ in Refs.~\cite{Ekimov_jtp,Ekimov1985}.  

A similar fit of the higher $Z_{1,2}$ exciton line is more complicated because of the complex structure of the $\Gamma_8$ valence band. The considerations, conducted in Refs. \cite{Ekimov_jtp,Ekimov1985} within the four-band model, describing this line as originating from the lowest $\Gamma_8$ exciton transition, however, is not correct. As we show in the present work in Sec.~\ref{sec:param}, several $\Gamma_8$ exciton transitions contribute to the $Z_{1,2}$ line. This phenomenon explains the stronger shift of the $Z_{1,2}$ line with cedresing size $a$, so that $C(Z_{1,2})>C(Z_{3})$.

Photoluminescence spectra of the studied samples are measured for continuous-wave and pulsed excitation at cryogenic temperatures of 1.9~K and 4.2~K, respectively. The associated details can be found in the Supplementary Information. They show exciton lines, which are in good agreement with the published data for CuCl NCs, so that we do not have to describe details in the main text.

\subsection{Spin-flip Raman scattering}

The spin-flip Raman scattering technique measures Raman spectra in close vicinity of the laser line, typically in range of $0.3-1.6$~meV. The spectral shift of the scattered photons from the laser photon energy equals to the Zeeman splitting of charge carriers or excitons that is described by 
\begin{equation}
\Delta E_{\rm Z} =|g| \mu_B B \,.
\label{Zeeman}
\end{equation}
Here $\mu_B$ is the Bohr magneton, and $g$ is the Land\'e $g$-factor whose value thus can be evaluated from experiment, but not the sign. This technique requires strong magnetic fields up to $7-10$~T and was successfully applied for epitaxially-grown (In,Ga)As/GaAs quantum dots~\cite{Debus2014}, colloidal CdS NCs~\cite{Sirenko1998}, and colloidal CdSe nanoplatelets~\cite{Shornikova2018,Kudlacik2020,Yakovlev2023_chapter}. The theoretical analysis of the underlying mechanisms and the polarization selection rules was done for CdSe nanoplatelets~\cite{Kudlacik2020,Rodina2020} and perovskite NCs~\cite{Rodina2022,Rodina2024}, whose spin level structure of the electronic states in vicinity of the band gap is the same as in CuCl NCs, see Fig.~\ref{Fig:ebands}.

\begin{figure*}[hbt]
\begin{center}
\includegraphics[width=16cm]{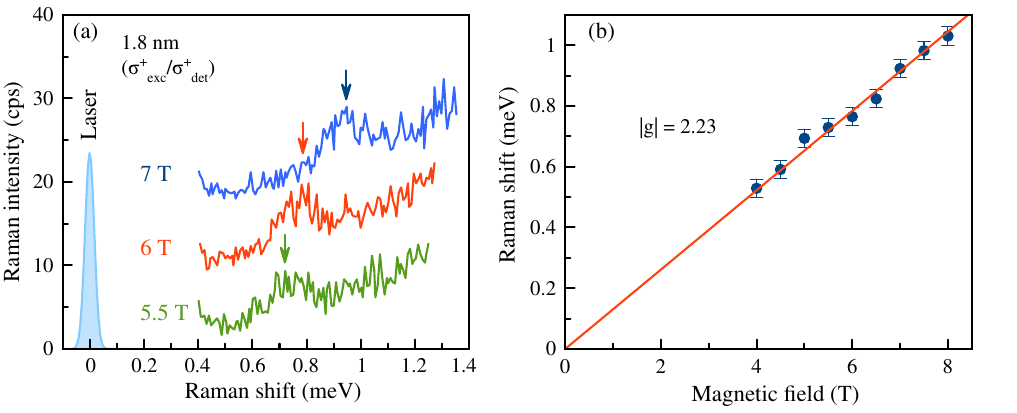}
\caption{(a) SFRS spectra of the 1.8~nm NCs measured at various magnetic fields in co-circular polarization configuration ($\sigma_{exc}^+$, $\sigma_{det}^+$) at $T = 1.9~$K. The spectra are shifted vertically for better visibility. The maxima of the spin-flip lines are marked by arrows. Blue line at zero Raman shift is the laser, which in this experiment has the photon energy of $3.252$~eV with a power of $1$~mW.  (b) Magnetic field dependence of the Raman shift. The red line is a fit with Eq.~\eqref{Zeeman} and $|g| =2.23$.  }
\label{fig2}
\end{center}
\end{figure*}

We measure all studied samples using SFRS. The signals associated with spin-flip lines, shifting with increasing magnetic field away from the laser, are relatively weak and only detectable in the 1.8~nm, 2.9~nm and 28~nm NCs. SFRS spectra from the 1.8~nm NCs are shown in Fig.~\ref{fig2}(a) for different magnetic fields. They are measured for resonant excitation with the exciton PL maximum at 3.252~eV. These traces are measured in the co-polarized configuration with the laser circular polarization set to $\sigma_{exc}^+$ and also the signal detection in $\sigma_{det}^+$ polarization. The spin-flip line, whose maximum is marked with an arrow, appears on a background from resonant photoluminescence. The signals are shown for the Stokes spectral range when the spin-flip line is shifted to lower energies relative to the laser line, which is taken as a positive Raman shift in Fig.~\ref{fig2}(a).

\begin{figure}[t!]
\begin{center}
\includegraphics[width=8cm]{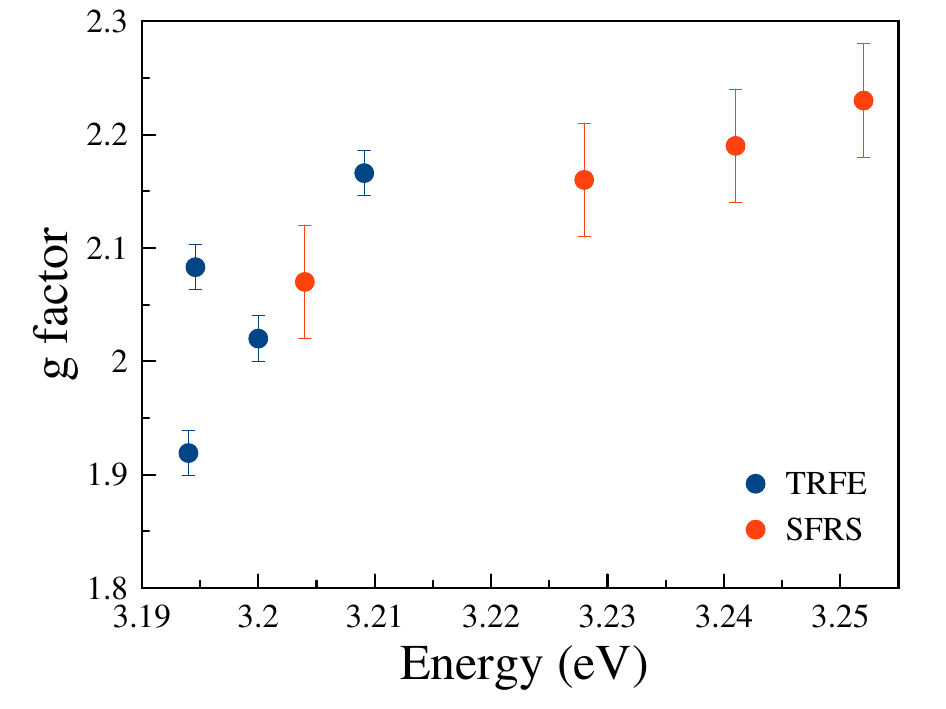}
\caption{Excitation energy dependence of the $g$-factor values $|g|$ in CuCl NCs with sizes ranging from 1.8 to 28~nm, evaluated from: (a) spin-flip Raman scattering (red circles) and (b) time-resolved Faraday ellipticity (blue circles). $T=1.9~$K.  }
\label{fig gfactor}
\end{center}
\end{figure}

Figure~\ref{fig2}(b) illustrates the dependence of the Raman shift on the magnetic field. A linear fit using Eq.~\eqref{Zeeman} yields the $g$-factor value of $|g| = 2.23\pm 0.05$. Note that the data set exhibits zero offset when extrapolated to zero magnetic field.  The experimental data for NCs with radii of 2.9~nm and 28~nm are comparable to those obtained for NCs with radius of 1.8~nm. The data are provided in the Supplementary Information, Figs.~S3, S4, S5, and S6. The evaluated $g$-factors are collected in Table~\ref{Tab_g-factors} and presented in Fig.~\ref{fig gfactor}. One can see that the $g$-factor depends on the laser photon energy, exhibiting an increase from 2.07 to 2.23 as the photon energy grows from 3.204 to 3.252~eV. It should be noted that larger energies correspond to smaller NCs.  

\begin{table}[hbt]
\begin{tabular}{|c|>{\centering\arraybackslash} m{0.15\textwidth}|>{\centering\arraybackslash} m{0.15\textwidth}|}
\hline
NC radius & Laser energy & $g$-factor value  \\
(nm)     & (eV) & $|g|$ \\ \hline
1.8      &3.252 &    $2.23\pm 0.05$  \\ \hline
2.9      &3.241 &    $2.19\pm 0.05$ \\ \hline
2.9      &3.228 &    $2.16\pm 0.05$ \\ \hline
28     &3.204 &    $2.07\pm 0.05$ \\ \hline
\end{tabular}
\caption{$g$-factors in CuCl NCs of various sizes measured by SFRS. The $g$-factor is evaluated from the shift of the Stokes-shifted Raman line with increasing magnetic field. $T=1.9$~K.}
\label{Tab_g-factors}
\end{table}

In order to clarify whether the $g$-factor of the electron, the hole, or the exciton is measured in SFRS, it is first necessary to gain an understanding of the process involved. In a neutral NC, a spin-flip of the bright exciton may take place, which corresponds to a $g$-factor of $g_X=g_e-g_h$, where $g_{e(h)}$ are electron (hole) $g$-factors. Note that we use here the definition of the $g_h$ sign often used for colloidal NCs~\cite{Efros_book}, see also Sec. \ref{sec:gfactors}. Note that in literature, the opposite sign of the hole $g$-factor is often used for bulk CuCl~\cite{Certier1969,Koda1970,khan1970g} and CuCl NCs~\cite{NomuraPRB1993} with $g_X=g_e+g_h$ for the bright and $g_X=g_e-g_h$ for the dark $Z_3$ excitons, respectively. Therefore, one has to carefully clarify the used definitions for the comparison of results by different authors. Alternatively, the spin-flip may be attributed to the flips of either the electron or the hole within an exciton, each with the corresponding carrier $g$-factor. In charged NCs, either with a resident electron or a resident hole, the spin-flip of the resident carrier interacting with the bright exciton provides Raman shifts waccording to $g_e$ or $g_h$, respectively. In NCs and quantum dots this process is the most efficient one, as was shown experimentally~\cite{Debus2014,Shornikova2018,Kudlacik2020}.  

The experimental information on exciton and carrier $g$-factors in CuCl bulk and NCs is rather limited. For the bright $Z_3$ exciton in bulk one can find values of 0.9~\cite{Certier1969}, 0.61~\cite{Koda1970,Suga1971}, and about 0.3~\cite{Yasui1996}. And for the exciton in CuCl NCs with $a=4.6$~nm the reported $g_X=0.447$~\cite{NomuraPRB1993}. They are considerably smaller than the values we measured in SFRS. Additionally, the spin-flip of the bright exciton necessitates a change in the exciton magnetic moment by 2, which significantly reduces the process efficiency. This allows us to the exclude the exciton spin-flip from consideration. Consequently, the measured $g$-factors should be attributed either to the electron or the hole. We propose that the signals originate from charged NCs, as the carrier-exciton mechanism is more efficient than the spin-flip of a carrier within the exciton.       

In principle, the polarization selection rules may allow for the distinction of various spin-flip mechanisms through the measurement of signals in different polarization configurations. However, in the studied CuCl NCs the polarization properties are weakly pronounced, see Supplementary Information Figs.~S3(a,b) and S6(a). This lack of pronounced polarization selectivity precludes us from drawing unique conclusions from the data.

\subsection{Time-resolved measurements}
\label{sec:TR}

Time-resolved techniques give access to the population and spin dynamics of excitons and carriers in semiconductor NCs~\cite{Yakovlev2023_chapter,Gupta1999,Gupta2002,Gang2022}. They were used for investigating the dynamics of exciton-polaritons in bulk CuCl~\cite{Rahimpour2004,Rahimpour2005,Cronenberger2006} and the population dynamics of excitons and biexcitons in CuCl NCs~\cite{Yano1997}. We use pump-probe methods with resonant excitation of the Z$_3$ excitons in CuCl NCs. In time-resolved differential transmission ($\Delta T/T$) we use linearly polarized laser pulses for both pump and probe, which allows us to obtain information on the population dynamics. In turn, time-resolved Faraday ellipticity exploys a circularly polarized pump, which induces spin polarization. The decay of the polarization contains information on the spin lifetime, which is determined by either recombination or spin relaxation. Application of a perpendicular magnetic field (Voigt geometry) extends the potentiality of TRFE towards evaluation of $g$-factors and spin dephasing times in spin ensembles.  

We perform TRDT and TRFE measurements on all studied CuCl NCs and find different phenomenology in small (1.8, 2.9, and 4.0~nm) and large (8.1, 15.3, and 28~nm) NCs. In the small NCs, the population and spin relaxation dynamics are short-lasting (in the range of $10-130$~ps) and the spin dynamics do not demonstrate any influence of magnetic field (no appearance of Larmor precession or change of decay time). Contrary to that, the large NCs show spin dynamics, which significantly outlast the population dynamics and show pronounced spin oscillations in magnetic field. Below we present characteristic results for the 1.8~nm and 15.3~nm NCs, detailed information for the other samples can be found in the Supplementary Information, Sec.~S4.      

The results on the 1.8~nm NCs are collected in Fig.~\ref{Fig:1,8 nm}. The traces are detected at $E_{pump}=3.248$~eV, which corresponds to the maximum of the exciton absorption. The population dynamics measured via $\Delta T/T$ is fast, Fig.~\ref{Fig:1,8 nm}(a). The decay has two characteristic times of $\tau_1=13$~ps and $\tau_2=58$~ps, which we evaluate by fitting the experimental data with   
\begin{equation}
A(t)=\sum_{i}{A_{0,i}\exp\left(-\frac{t}{\tau_{i}}\right)} \, .
 \label{n1}
\end{equation}
Here, the $A_{0,i}$ are the amplitudes of the signal components and the $\tau_{i}$ are the decay times. These times are in a range typical for exciton recombination in CuCl NCs~\cite{Kataoka1989,Itoh1990, Kataoka1993}. The spin dynamics measured by TRFE behaves very similarly to the population dynamics with decay times of $\tau_1= 11$~ps and $\tau_2=64$~ps, see Fig.~\ref{Fig:1,8 nm}(b). No modifications are found in the magnetic field of $B_{\rm V} = 0.5$~T, Fig.~\ref{Fig:1,8 nm}(c), as well as in stronger fields up to 2.5~T.   

The dynamics in the 2.9~nm and 4.0~nm NCs closely resemble those in the 1.8~nm NCs, see Supplementary Information, Fig.~S7. The only differences for the in 2.9~nm NCs are the emergence of a very long-lived component with $\tau_2=6$~ns in TRDT and 430~ps in TRFE. It is not affected by an external magnetic fields of 1~T applied either in Voigt or Faraday configuration, see Fig.~S7(e). It can be tentatively assigned to the recombination of dark excitons in neutral NCs or to photocharging of the NCs. 

\begin{figure*}[hbt]
\includegraphics[width=14cm]{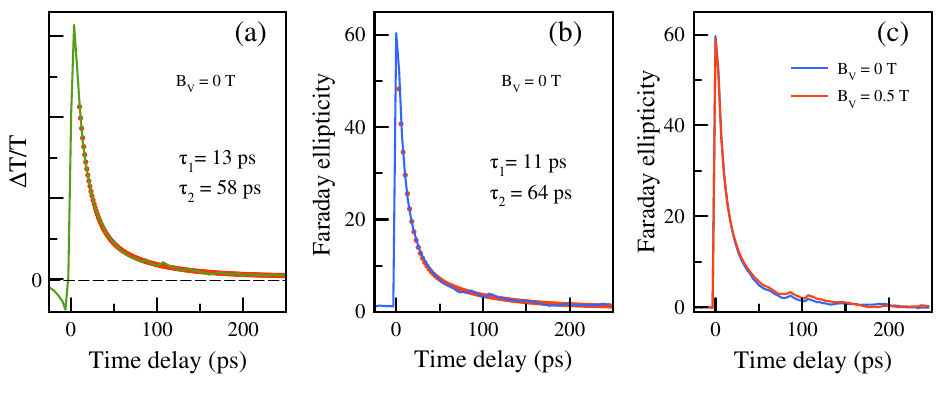}
\caption{Population and spin relaxation dynamics in 1.8~nm NCs measured at $T=1.7$~K for $E_{pump}=3.248$~eV. (a) Dynamics of differential transmission at zero magnetic field (green line) and fit with Eq.~\eqref{n1} (red circles). (b) Faraday ellipticity dynamics at $B_{\rm V} = 0$~T (blue line) and fit with Eq.~\eqref{n1} (red circles).
(c) Faraday ellipticity dynamics in various magnetic fields $B_{\rm V}$.    
}
\label{Fig:1,8 nm}
\end{figure*}

Let us turn to the dynamics properties in large NCs, which are exemplified for the 15.3~nm NCs. For them, the population dynamics measured by TRDT are shown in Figs.~\ref{Fig:15,3-1 nm}(a,b). It shows a fast decay of 33~ps corresponding to exciton recombination and also weak in intensity, but also long-lived signal with a decay time of about 1.3~ns. One can see in Fig.~\ref{Fig:15,3-1 nm}(b) that it extends over 4~ns delay. This long-lasting dynamics can be related either to dark excitons or to photocharging processes, when one of the carrier escapes to the glass matrix. 

\begin{figure*}[hbt]
\begin{center}
\includegraphics[width=14cm]{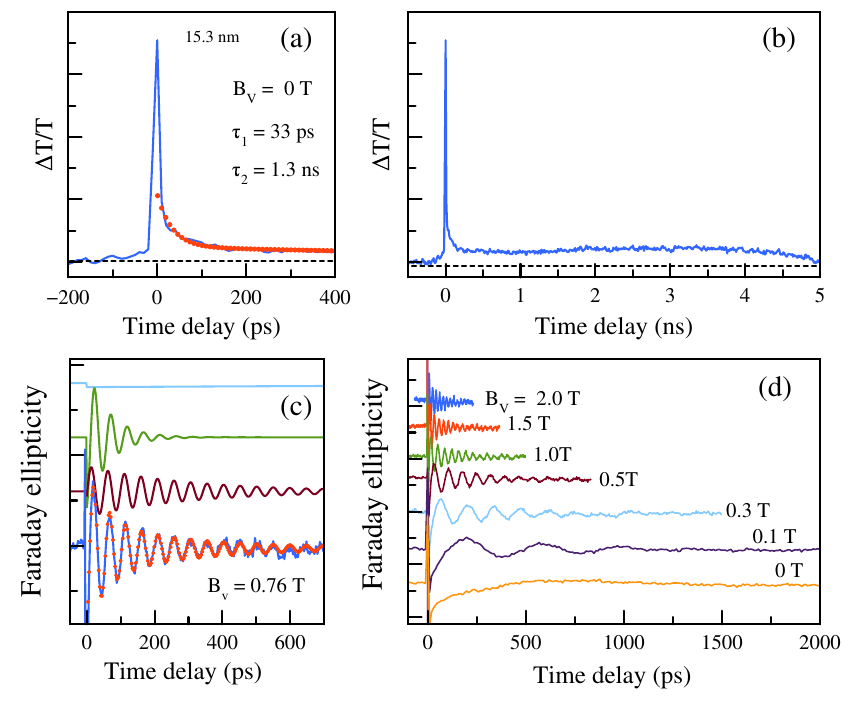}
\caption{Population and spin dynamics in 15.3~nm NCs measured at $T=1.7$~K for $E_{pump}=3.195$~eV.  
(a) TRDT dynamics in a short time interval (blue line) and its fit with Eq.~\eqref{n1} (red circles).
(b) TRDT dynamics in a long delay range.
(c) Faraday ellipticity dynamics at $B_{\rm V}=0.76$~T (blue line), fit to these data using Eq.~\eqref{n2} (red circles) and components of this fit (cyan, green, and brown lines). 
(d) Faraday ellipticity dynamics at various magnetic fields $B_{\rm V}$.
}
\label{Fig:15,3-1 nm}
\end{center}
\end{figure*}

The photocharging scenario is confirmed by the spin dynamics measured by TRFE in magnetic field, see Fig.~\ref{Fig:15,3-1 nm}(d). One can see, that at $B_{\rm V}=0.1$~T the characteristic oscillations due to Larmor spin precession can be detected for delays exceeding 1~ns, i.e. much longer that the bright exciton lifetime. Note that the dark exciton cannot contribute to such spin oscillations in TRFE signals. Coherent spin precession of a resident carrier can be excited via the charged exciton state in NC, as was shown for (In,Ga)As/GaAs quantum dots~\cite{Yakovlev2023_chapter}, CdSe NCs~\cite{Gang2022}, and CsPbBr$_3$ perovskite NCs~\cite{Grigoryev2021}.

With increasing magnetic field up to 2~T, the oscillations frequency in the FE dynamics increases as expected, see Fig.~\ref{Fig:15,3-1 nm}(d), which corresponds to an increase of the Zeeman splitting of the resident carrier. Also the decay of the oscillation amplitude accelerates, evidencing faster spin dephasing in stronger magnetic fields. In order to quantitatively analyze the coherent spin dynamics, we fit them with the following function:
\begin{eqnarray}
\label{n2}
A_{FE}(t)=\sum_{i}{A_{0,i} \cos(\omega_{L,i}t)} \exp{\left(-\frac{t}{T^*_{2,i}}\right)}+
\\ \nonumber A_{non}\exp\left(-\frac{t}{T_{non}}\right).
\end{eqnarray}
Here $A_{0,i}$ and $\omega_{L,i}$ are the amplitude and Larmor precession frequency of the oscillating components, $T_{2,i}^*$ are the spin dephasing times, $A_{non}$ and $T_{non}$ are the amplitude and spin relaxation time of the non-oscillating component. The Larmor precession frequency is related to the Zeeman spin splitting via $\hbar \omega_L=\Delta E_{Z}$ and, therefore, with the $g$-factor through 
\begin{equation}
|g|=\hbar \omega_{L}/(\mu_{B}B)\,.
\label{g-factor}
\end{equation}

An example of such a fit made for the spin dynamics at $B_{\rm V}=0.76$~T is shown in Fig.~\ref{Fig:15,3-1 nm}(c). Here, the red circles result from a fit to the experimental dynamics (blue line). Three components can be identified in the signal and are shown also in the same figure. Two of them oscillate with the same frequency corresponding to $|g|=2.08$, but have different spin dephasing times of $T_{2,1}^*=114$~ps and $T_{2,2}^*=187$~ps.
The third non-oscillating component has a small intensity with the decay time of 2.1~ns. 

The parameters evaluated from the spin dynamics and their dependences on magnetic field are shown in Fig.~\ref{Fig:15,3-2 nm}. The Larmor precession frequency shows a linear dependence on magnetic field, which fit gives $|g|=2.08\pm 0.02$, see Fig.~\ref{Fig:15,3-2 nm}(a). Note that this dependence has zero offset when extrapolated to zero magnetic field, which allows us to exclude spin precession of the carriers within the exciton, as in this case a finite frequency offset, determined by the electron-hole exchange splitting, is expected.  

\begin{figure*}[hbt]
\begin{center}
\includegraphics[width=13cm]{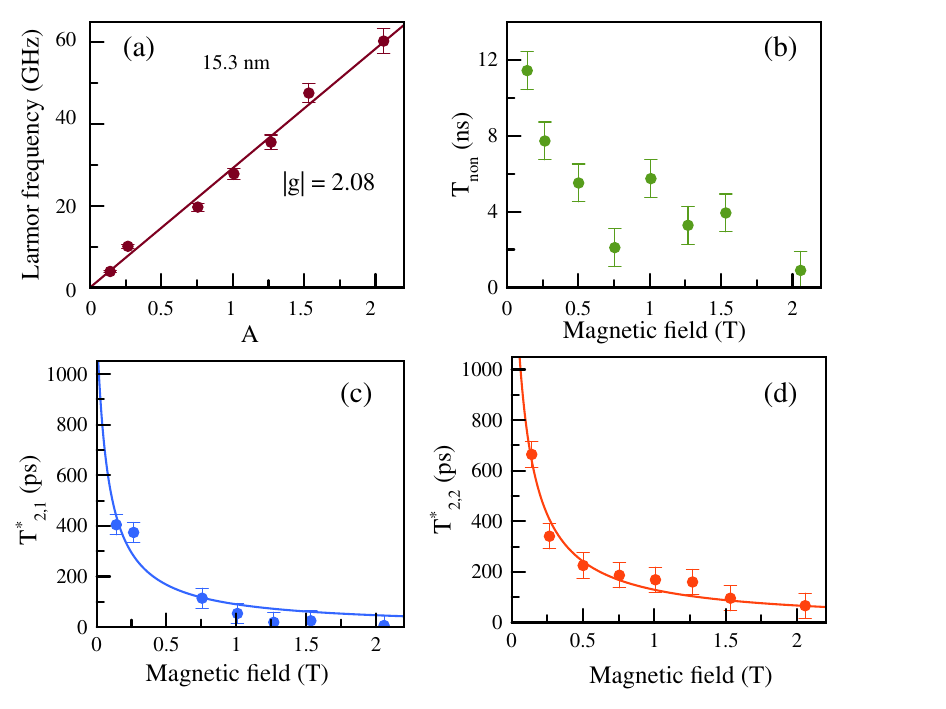}
\caption{Parameters of the spin dynamics of carriers in the 15.3~nm NCs, measured at $T=1.7$~K for $E_{pump}=3.195$~eV.  
(a) Dependence of the Larmor precession frequency on magnetic field (circles) with a linear fit giving $|g|=2.08$.
(b) Dependence of the spin relaxation time on magnetic field for the non-oscillating component. 
(c,d) Dependence of the spin dephasing time on magnetic field for the two oscillating electron components and fit of these data using Eq.~\eqref{n4} (solid lines).
}
\label{Fig:15,3-2 nm}
\end{center}
\end{figure*}

The magnetic field dependences of the spin dephasing times for the two oscillating components are shown in Figs.~\ref{Fig:15,3-2 nm}(c,d). The strong decrease of the dephasing times with growing magnetic field is provided by the dispersion of the Larmor precession frequencies of the carriers in the NC ensemble. We fit the dependences with the following equation, which allows us to evaluate the spread of the $g$-factors, $\Delta g_i$:  
\begin{equation}
T_{2,i}^*=1/[1/T_{2,i}^*(0)+(\Delta g_i\mu_{B}B)/\hbar].
\label{n4}
\end{equation}
The obtained parameters are $\Delta g_1=0.12$ ($5.6\%$ of the $g$-factor value) and $T_{2,1}^*(0)=1.2$~ns for the faster oscillating component, and $\Delta g_2=0.08$ ($3.8\%$) and $T_{2,2}^*(0)=1.8$~ns for the slower oscillating component.

As both oscillating components have the same $g$-factor, we assign them to the same carrier type, either electrons or holes. Variation of $\Delta g$ may be related to the presence of two close distributions in NC sizes, which are also evident as two components of Z$_3$ exciton resonance in absorption, see Fig.~\ref{Fig:Abs}(b). 
It should be noted, that the $T_{2,i}^*(0)$ values evaluated by this fit are typically longer that the real spin dephasing times at zero magnetic field, as there other mechanisms come in action, e.g. spin dephasing at nuclear spin fluctuations. 

It is noteworthy that, the non-oscillating component has a smaller amplitude than the oscillating components, but it consistently is present in all measured signals at low temperatures. Its decay time is considerably longer than the spin dephasing times of the oscillating components, reaching $T_{non}=7$~ns in low magnetic fields and decreasing to 1~ns at 2.1~T, see Fig.~\ref{Fig:15,3-2 nm}(b).

The spin dynamics in the 8.1~nm and 28~nm NCs show a similar phenomology as in the 15.3~nm NCs. Their details are given in the Supplementary Information Figs.~S8 and S9. Their spin parameters are collected in Table~\ref{Table 1} and the evaluated $g$-factor values are presented in Fig.~\ref{fig gfactor}. 

\begin{table*} [hbt]
\caption{Decay times $T_{2,i}^*$, $T_{non}$ and $g$-factor values obtained from the FE signals, and $\tau_i$ obtained from the FE and $\Delta T/T$ signal for NCs of various sizes. $E_{pump}$ is laser photon energy. $T=1.7$~K.} 

\label{Table 1}
\begin{tabular}{p{1.6cm} p{1.2cm} p{1.2cm} p{2.5cm} p{2.5cm} p{2.5cm} p{1.9cm}}
\hline
\hline
\\
NC radius & $E_{pump}$ &  $|g|$ & FE decay times &$\Delta T/T$ decay times & FE dephasing times & $T_{non}$  \\
(nm) & (eV) &   & $B_V=0.5$~T   & $B_V=0$~T & $B_V=0.14$~T  & $B_V=0.14$~T  \\
\\
\hline
\\
1.8 & 3.248 & - & $\tau_1=11$~ps &$\tau_1=13$~ps &  &   \\
 &  &  & $\tau_2=64$~ps & $\tau_2=58$~ps &  &  \\
\\
2.9 & 3.228 &  - & $\tau_1=21$~ps & $\tau_1=130$~ps &  \\
 &  &  & $\tau_2=434$~ps & $\tau_2=6000$~ps & & \\
\\
4.0 & 3.202 &  - & $\tau_1=21$~ps & $\tau_1=21$~ps &  &    \\
 &  &  & $\tau_2=60$~ps & $\tau_2=73$~ps  & & \\
\\
8.1 & 3.209 & 2.17&   & $\tau_1=51$~ps   & $T_{2,1}^*=564$~ps &  \\
 &  &  &  & $\tau_2=492$~ps & $T_{2,2}^*=630$~ps &\\
\\
8.1 & 3.200 & 2.02&   & $\tau_1=38$~ps  & $T_{2,1}^*=380$~ps &\\
 &  &  &  &$\tau_2=364$~ps  & $T_{2,2}^*=480$~ps &\\
 \\
15.3 & 3.195 &  2.08 &  &  &  $T_{2,1}^*=400$~ps  &7~ns \\
 &  &   &   &      & $T_{2,2}^*=660$~ps  &\\
\\
28 & 3.194 &  1.92 &  &   &  $T_{2,1}^*=340$~ps  &2.5~ns  \\
 &  &   &  &   &  $T_{2,2}^*=830$~ps  &\\
\\

\hline
\\

\end{tabular}
\end{table*}

\begin{figure*}[hbt]
\begin{center}
\includegraphics[width=13cm]{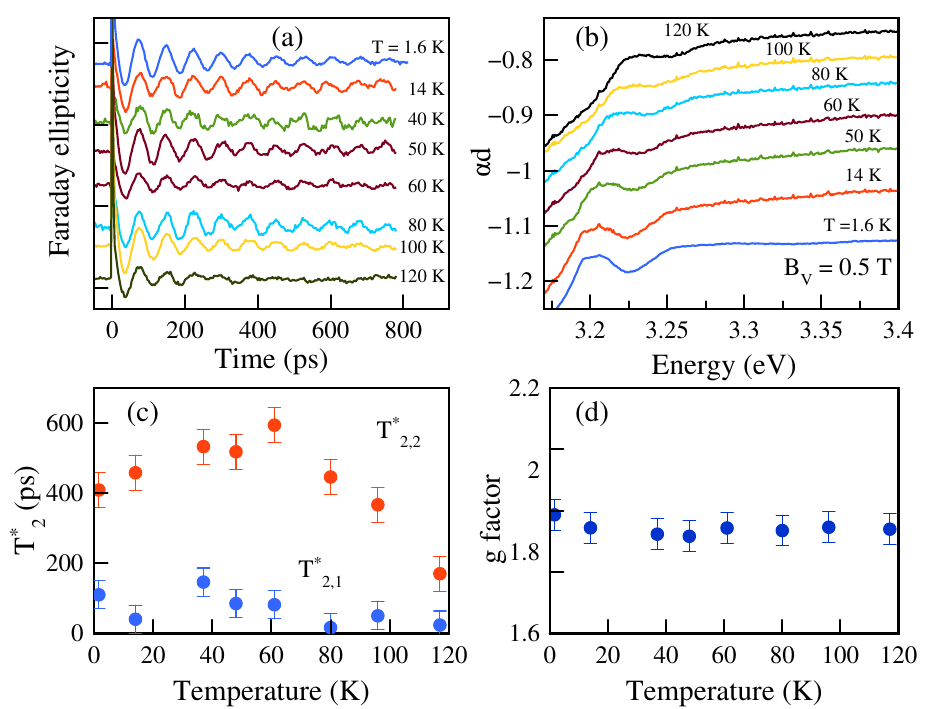}
\caption{Spin dynamics of carriers in the 28~nm NCs measured at various temperatures at $B_{\rm V}=0.5$~T. (a) TRFE dynamics at various temperatures. (b) Absorption spectra at various temperatures. 
(c) Temperature dependence of the spin dephasing times.  
(d) Temperature dependence of the $g$-factor.   
}
\label{Fig:28nm temper}
\end{center}
\end{figure*}

The strong carrier confinement in colloidal NCs is favorable for maintaining spin coherence at elevated temperatures. For example, spin precession at room temperature was demonstrated for CdSe NCs~\cite{Hu2019,Gang2022} and CsPbBr$_3$ perovskite NCs~\cite{Crane2020,Lin2022,Cheng2024}. Therefore, it is instructive to examine the coherent spin dynamics in CuCl NCs at various temperatures. We perform these measurements for the 28~nm NCs and show the results in Fig.~\ref{Fig:28nm temper}. In the magnetic field of $B_{\rm V}=0.5$~T, spin precession is detectable up to 120~K, see Fig.~\ref{Fig:28nm temper}(a). With the temperature increase from 1.6~K up to 120~K Z$_3$, the exciton resonance slightly shifts to higher energies by 27~meV, see Fig.~\ref{Fig:28nm temper}(b). Note that the same behavior was found for the CuCl NCs in Ref.~\cite{Ekimov1982}. In order to keep the resonant excitation conditions, the laser photon energy is shifted accordingly. The $g$-factor values exhibit a minimal change with increasing temperature, see Fig.~\ref{Fig:28nm temper}(d). The spin dephasing time of the fast oscillating component $T_{2,1}^*$ decreases with growing temperature from 110~ps to 24~ps. For the slow oscillating component $T_{2,2}^*$ increases from 410~ps at 11.6~K to 600~ps at 60~K above which it decreases to 170~ps at 120~K, see Fig.~\ref{Fig:28nm temper}(c).  

Let us summarize the experimental results on the spin properties of CuCl NCs. By using the two experimental techniques of spin-flip Raman scattering and time-resolved Faraday ellipticity, we have been able to address the carrier spin properties in all studied NCs with radii varied from 1.8~nm up to 28~nm. The measured $g$-factors are presented in Fig.~\ref{fig gfactor}. Their values increase with decreasing NC size (i.e. increasing confinement) from $1.92$ in the 28~nm NCs up to $2.23$ in the 1.8~nm NCs. The relatively small variations of the $g$-factor may be related to the fact that its value is close to the Land\'e $g$-factor of fthe ree electron of $g_0 \approx 2.003$ and the band mixing in CuCl NCs has a moderate effect on the $g$-factors in the studied structures in the weak confinement regime. 

The measured $g$-factor considerably exceeds the reported values for the bright state of the Z$_3$ excitons. Also the coherent spin dynamics have been measured for delay times of 1~ns, which considerably exceeds the exciton lifetime. These facts allow us to conclude that the spin signal is related to resident carriers in (photo)charged NCs. On the basis of our experimental results we cannot make a unique assignment of the carrier type, namely whether electrons or holes are involved. Theoretical consideration are needed at this point to shed light on the identification. These considerations will be presented in Section \ref{sec:gfactors}. 

\clearpage

\section{Theory and Modeling} 
\label{sec:theory}
To describe the size dependence of the absorption spectra and of the $g$-factors measured in the CuCl NCs, we will  start this section with developing a theory for the energy level spectra of  electrons, holes, and excitons confined in spherical  CuCl NCs.

\subsection{Quantum confinement effect in CuCl NCs and evaluation of valence band parameters}
\label{sec:ehex}

\begin{figure*}[hbt]
\includegraphics[width=18cm]{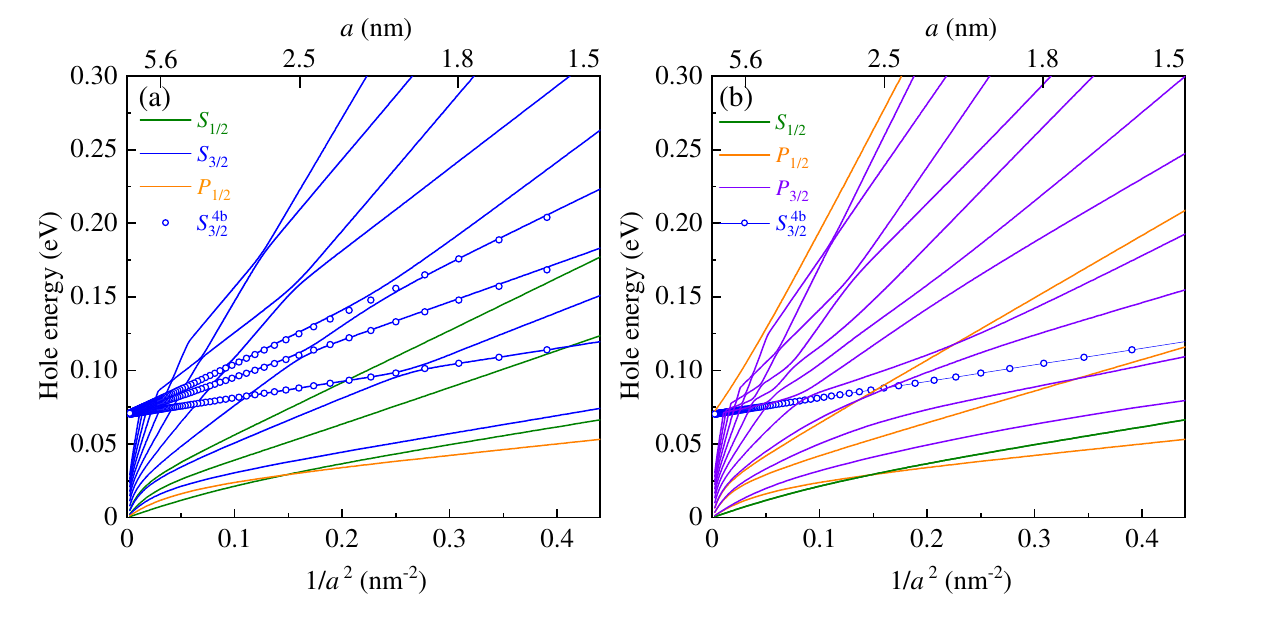}
\caption{Hole energy level energies versus $1/a^2$ in CuCl nanocrystals, calculated from the top of the valence band. (a) Green lines correspond to the first three $S_{1/2}$ levels, blue lines correspond to the $S_{3/2}$ levels, circles correspond to the first three $S_{3/2}^{4b}$ levels, calculated within the four-band model neglecting the $\Gamma_7$ and $\Gamma_8$  mixing, orange line corresponds to the first $P_{1/2}$ levels arising from the $\Gamma_8$ subband. (b) Green and blue lines show the first $S_{1/2}$ and $S_{3/2}^{4b}$ levels from panel (a), violet lines correspond to the $P_{3/2}$ levels, orange lines correspond to several lowest lying $P_{1/2}$  levels starting from the top of the $\Gamma_7$ and $\Gamma_8$ subbands.
} 
\label{ehsize}
\end{figure*}

\subsubsection{Quantum size effect for electron}

The large value of the band gap energy in CuCl of 3.4~eV allows us to neglect the contribution from the valence band states to the electron wave function as well as the energy dependence of the electron effective mass in the calculations of the ground energy level of the electron confined in spherical NCs. The respective electron Hamiltonian can be written as:
\begin{equation}
\widehat{H}_e({\bm r}_e)=-\frac{\hbar^2}{2m_e} \hat {\bm k}^2_e +V_{\rm ext}({\bm r}_e).
\end{equation} 
Here $m_e$ is the electron effective mass at the bottom of the conduction band, ${\bm k}_e = -i \nabla_{{\bm r}_e}$  is the electron wave vector (the operator $\bm \nabla$ is acting on electron coordinate ${\bm r}_e$), $V_{\rm ext}({\bm r}_e) = V_{\rm ext}(r_e)$  is the localizing potential of the NC. We will consider it as infinite box-like spherical potential $V_{\rm ext}(r_e)=0$ for $r_e<a$ and $V_{\rm ext}(r_e)=\infty$ for $r_e>a$.

The electron states in a spherical NC are characterized by the orbital angular momentum ${\bm L}_e = {\bm r}_e \times {\bm k}_e$. We neglect the spin-orbit interaction for these states which can be induced by the spherical surface \cite{Rodina2003}. Then, all electron states are two-fold degenerate with respect to the spin projection on the quantization axis defined as the $z$-axis, and $2L_e+1$ time degenerated with respect to the angular momentum projection $L_{ez}$. The respective electron wave functions can be written in the form
 \be
 \label{Psie}
 \Psi^{e}_{L_{e}}({\bm r}_e) =  R^c_{L_{e}}(r_e) Y_{l,m}(\theta,\phi) u^c_{1/2,\pm 1/2} \, . 
 \ee
Here the $u^c_{1/2,+1/2}=S\uparrow$ and $u^c_{1/2,-1/2}=S\downarrow$ are the conduction band Bloch functions, $Y_{lm}(\theta,\phi)$  are the spherical harmonics \cite{Edmonds} with $l=L_e$, $m=L_{ez}$ and the spherical electron coordinates $(\theta,\phi)$. The radial wave functions $R^c_{L_{e}}(r_e)$ are normalized through the condition $\int_0^a [R^c_{L_{e}}(r_e)]^2 r_e^2 dr_e=1 $. The energy levels with given $L_e$ are determined by the boundary conditions at $r_e=a$ and can be numbered by the main quantum number $n=1,2,...$. For the standard boundary conditions $R^c_{L_{e}}(a)=0$ that we use here, they are determined by the $n$-th root of the spherical Bessel function $j_l$ with $l=L_e$. 
 
The energy of the ground electron levels $1S_e$ ($n=1$, $L_e=0$) calculated from the bottom of the conduction band as function of the NC radius is given by
\begin{equation}\label{eq:electron}
    E_{1S_e}\equiv E_0^e=\frac{\hbar^2\pi^2}{2 m_e a^2} \, .
\end{equation}
For the calculations hereafter, we  take the value $m_e=0.4m_0$, which is close to the experimentally measured electron effective mass $m_e=0.415m_0$ from Ref.~\cite{Nikitine1967}. The other experimentally measured and theoretically estimated values of the electron effective mass are collected in Table \ref{e_masses}.

\subsubsection{Quantum size effect for hole}

To describe the hole states, we consider a six-band model taking into account both the two-fold degenerate $\Gamma_7$ topmost valence subband and the four-fold degenerate $\Gamma_8$ valence subband, see Fig.~\ref{Fig:ebands}(b). Here we neglect the effects related to the cubic symmetry of the CuCl crystal lattice, and describe the hole states by the following Hamiltonian:
\begin{equation}
\label{Hamilt}
\widehat{H}_h({\bm r}_h)=\widehat{H}_{6\times6}+V_{\rm ext}({\bm r}_h).
\end{equation}
Here $V_{\rm ext}({\bm r}_h) = V_{\rm ext}(r_h)$ is the infinite box-like spherical potential for holes and 
\begin{widetext}
\begin{eqnarray}\label{lutt6}
\widehat{H}_{6\times6}=\widehat{H}_{\rm kin}({\bm r}_h)-\frac{1}{3}\Delta_{\rm SO}\left[(\hat{\bm I}\hat{\bm s}_h)-1\right] \, , \\
\widehat{H}_{\rm kin}({\bm r}_h)=\frac{\hbar^2}{2 m_0}\left[\left(\gamma_1+4\gamma\right){\bm k}_h^2-6\gamma\left(\sum_{\alpha=x,y,z}k_{h\alpha}^2\hat{I}_{\alpha}^2+2\sum_{\alpha\neq\beta}\{k_{h\alpha} k_{h\beta}\}\{\hat{I}_\alpha \hat{I}_\beta\}\right)\right] \label{Hkin}
\end{eqnarray}
\end{widetext}
is the six-band Luttinger Hamiltonian \cite{Bir1974book,Rodina2001f,Semina2021} describing the hole kinetic energy and the spin-orbit coupling in spherical approximation in the hole representation. Here $\gamma_1$ and $\gamma=(2\gamma_2+3\gamma_3)/5$ are the Luttinger parameters, ${\bm k}_h = -i \nabla_{{\bm r}_h}$  is the hole wave vector (the operator $\bm \nabla$ is acting on the hole coordinate ${\bm r}_h$), $\hat{\bm {I}} \equiv \hat{\bm {I}}_h$ is the hole orbital angular momentum operator $I=1$, ${\bm s}_h =1/2 \hat{\bm {\sigma}}^h$ is the hole spin 1/2, and the operator $\{ab\}=(ab+ba)/2$. In CuCl, the spin-orbit splitting of the valence band, $\Delta_{\rm SO}$, has a negative sign. The hole Hamiltonian is written in the hole representation, in which the hole energy increases with increasing $k_h$ and the heavy ($m_{hh}$), light ($m_{lh}$), and spin-orbit-split ($m_{s}$) hole masses are positive. The relations between the Luttinger parameters and the hole masses, describing the valence band dispersion at small $k_h$ are given by \cite{Ekimov1993}:
\begin{eqnarray} \label{holemass}
    \frac{m_0}{m_{hh}} = \gamma_1-2\gamma, \quad  \frac{m_0}{m_{lh}} = \gamma_1+2\gamma, \quad \frac{m_0}{m_{\rm s}}=\gamma_1\,.
    \end{eqnarray}
The hole states in spherical NCs are characterized by the total angular momentum ${\bm J}_h = {\bm j_h}+{\bm L}_h =  {\bm s_h} + {\bm I} +{\bm L}_h$, where ${\bm L}_h = {\bm r}_h \times {\bm k}_h$ is the orbital angular momentum related to the hole moving in a spherical-symmetry potential. They are $(2J_h+1)$-fold degenerate with respect to the projection $J_{hz} = -J_h, ...,+J_h$.

The six-component hole envelope function, following Ref.~\onlinecite{Ekimov1993}, can be written as
\begin{multline}
\Psi^h_{J_h,J_{hz}}({\bm r}_h) = \sqrt{2J_h+1}\sum_{L_h} (-1)^{L_h-j_h+J_{hz}}\times \\ \times (i)^{L_h} R^{J_h,j_h}_{L_h}(r_h)\sum_{ J_{hz}, j_h}
\left(
\begin{array}{ccc}
L_h & j_h&J_h \\ m&\mu&-J_{hz}
\end{array}
\right)
Y_{L_h,m} u_{j_h,\mu}^v \, .
\label{PsiJJ}
\end{multline}
Here $L_h$ is the hole orbital angular momentum, the $Y_{L_hm}$ are the spherical harmonics \cite{Edmonds}, $\left(_{m~n~p}^{i~~k~~l}\right)$ are the $3j$ Wigner symbols. The Bloch functions  $u_{j_h,\mu}^v$ at the top of the $\Gamma_8$ and $\Gamma_7$  valence subbands are linear combinations of the $(X\uparrow,Y\uparrow,Z\uparrow)$ and  $(X\downarrow,Y\downarrow,Z\downarrow)$  functions, being the eigen states of the total momentum ${\bm j}_h$  and its projection $j_{hz}=\mu$. We use the same basis functions $u_{j_h,\mu}^v$ as in Ref.~\cite{Semina2021}.  Note that the basis functions and, therefore, the matrix form of the six-band Hamiltonian (see Appendix B in Ref. \cite{Semina2021}) are slightly different from those in Ref.~\cite{Ekimov1993}. However, the equations describing  the size dependence of the hole  energy levels  derived in Refs.~\cite{Grigoryan1990,Ekimov1993} are valid. For convenience, these equations that we use in the present calculations are reproduced in \ref{Ap:eqs}. Importantly, when the absence of inversion symmetry in CuCl is neglected, the hole Hamiltonian including the external spherical-symmetry potential is an invariant with respect to the spatial inversion. Therefore, the hole wave functions \eqref{PsiJJ} with given $J_h$ are formed either with  even, $L_h=0,2,...$, or odd, $L_h=1,3...$ values of the orbital momentum.    

In Fig.~\ref{ehsize}(a) we show the dependence of the lowest $nS_{1/2}$ (green lines) and $nS_{3/2}$  (blue lines) hole level energies on $1/a^2$.  The calculations are done with $\Delta_{\rm SO}=-0.07$~eV and the Luttinger  parameters  $\gamma_1=0.67$ and  $\gamma=0.29$. Their choice is explained in Sec.~\ref{sec:param}. The notation is the same as in Ref.~\cite{Ekimov1993}: $S_{1/2}$ and $S_{3/2}$ denote the states with total momentum $J_h=1/2$ and $J_h=3/2$, respectively, which wave functions contain the orbital angular harmonics with $L_h=0$ ($s$-symmetry), the numbers $n=1,2...$ numerate the levels with given symmetry. Due to the spin-orbit-driven hole state mixing the $S_{J_h}$ states are in fact mixed $s (L_h=0)$ - $d(L_h=2)$ symmetry states, and the $s$ symmetry comes always with the Bloch functions $u_{j_h=J_h,\mu=J_{hz}}$. One can see in Fig.~\ref{ehsize}(a) the anti-crossing of the $S_{3/2}$ valence subbands arising from the $\Gamma_7$ (mostly $d$-like) and $\Gamma_8$ (mostly $s$-like). The open blue circles show the $E_{nS_{3/2}}^{\rm 4b}$ energy levels calculated within the four-band model for the $\Gamma_8$ band (without accounting for the mixing with the $\Gamma_7$ band). One can see that the four-band model reproduces well the size dependence of the energies of the  $S_{3/2}$ states with the main contribution from the $\Gamma_8$ band (upper branches after anti-crossing).

For comparison, we also show by the orange line the size dependence of the lowest  $1P_{1/2}$ level with $p(L_h=1)$ symmetry. In  these states the mixing of the spin-orbit ($\Gamma_7$) and the light hole (($\Gamma_8$) states is substantial resulting in the strong deviation from a $1/a^2$ law. One can see that in small NCs, where the quantum confinemet energy is comparable to $|\Delta_{\rm SO}|$, the  $1P_{1/2}$ level becomes the lowest one. This effect is illustrated additionally in Fig.~\ref{ehsp}. Figure~\ref{ehsize}(b) shows the energies of the lowest  $nP_{1/2}$ (orange lines) and  $nP_{3/2}$ with mixed $p(L_h=1)-f(L_h=3)$ symmetry (violet lines) as fucntion of $1/a^2$. One can see in Fig.~\ref{ehsize}(b) the anti-crossing of the $P_{3/2}$ arising from the $\Gamma_7$ (mostly $f$-like) and the $\Gamma_8$ (mostly $p$-like) valence subbands. The four-band model does not allow one to describe the $P_{3/2}$ states in the studied range of NCs sizes.    

\begin{figure*}[h!]
\begin{center}
\includegraphics[width=16cm]{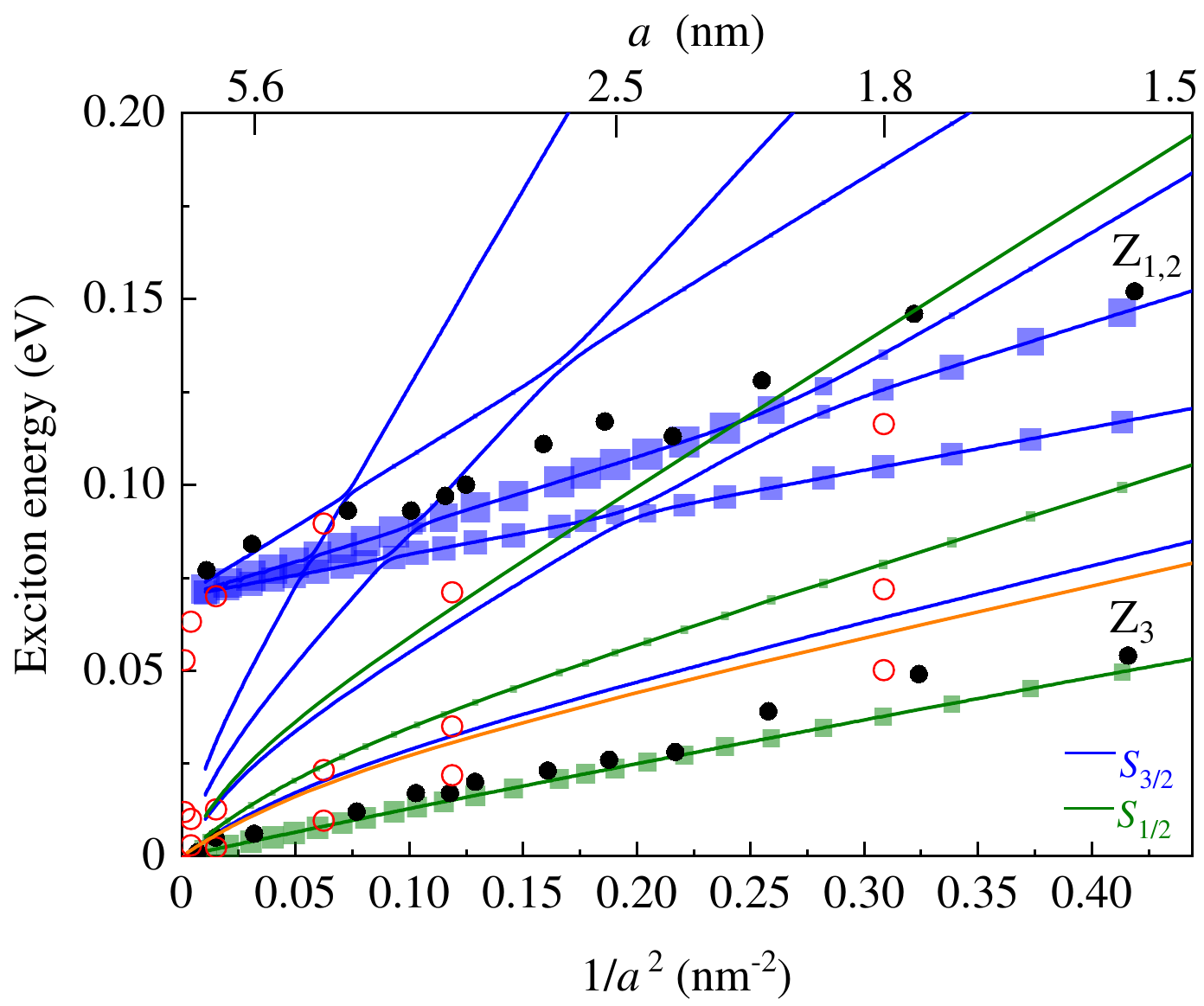}
\caption{Calculated energies of the ${\cal S}_{1/2}$ and ${\cal S}_{3/2}$ exciton states with $n=1,2,3...$ (green and blue lines, respectively) and the corresponding relative oscillator strengths ${\tilde f}_{{\cal J},n}$  (green and blue squares). Orange line corresponds to the first $P_{1/2}$ level arising from the $\Gamma_7$ exciton subband. Black circles are the data from Refs.~\cite{Ekimov_jtp,Ekimov1985}, open red circles show the experimental data from the present work. The energy $E_g-E_{\rm bind}=3.2$ eV is set as zero. The calculations are done with $\gamma_1^{\rm ex}=0.53$,  $\gamma^{\rm ex}=0.18$, and $\Delta_{\rm SO}=-0.07$~eV. 
}
\label{Fig:AbsModel}
\end{center}
\end{figure*} 

\subsubsection{Quantum confinement of exciton}

The exciton Hamiltonian can be written as
\begin{equation}
\label{ExcitonHamilt}
\widehat{H}_{\rm ex}({\bm r}_e,{\bm r}_h)=E_g+ \widehat{H}_{e}({\bm r}_e)+\widehat{H}_{h}({\bm r}_h) + V_{\rm C}({\bm r}_e-{\bm r}_h)\, ,   
\end{equation}
where $V_{\rm C}({\bm r})$ describes the Coulomb interaction between electron and hole and, in general, may include also the effects of the dielectric contrast between the NCs and the environment \cite{Brus1983,Rodina2016JETP}. We assume $V_{\rm C}({\bm r})=V_{\rm C}({r})$ to be isotropic and depending only on the distance $r=|{\bm r}_e-{\bm r}_h|$ between the carriers.

As first emphasized by Dresselhaus \cite{Dresselhaus1956}, the complex structure of the valence band prevents the possibility of a complete decoupling of the relative and translation motions in the exciton Hamiltonian \eqref{ExcitonHamilt}. The problem of the exciton translational motion in bulk within the four-band model was addressed in Refs. \cite{Kane1974,Altarelli1977,Gelmont1984}. The six-band Luttinger Hamiltonian for the translational motion of the excitons in CuCl nanocrystals was considered in Ref. \cite{Ekimov_jtp}, however, the relations between the valence band hole and the exciton Luttinger parameters were not specified. 

Our aim here is to find the zero-order approximation for the relative and center of mass exciton motion within the six-band model. We note that for the kinetic energies being much smaller than $\Delta_{\rm SO}$, the ground state exciton formed with the $\Gamma_7$ hole can be considered separately. For such exciton in the case of weak confinement with the confinement energies being much smaller than the exciton binding energy $E_{\rm b}$, there is an exact separation of the relative motion with ${\bm r}={\bm r_e} - {\bm r_h}$ and the center of mass motion with the center of mass coordinate ${\bm R}$ defined as
\be \label{Rc} {\bm R} = \frac{m_e{\bm r}_e +m_s{\bm r}_h}{m_e+m_s}
\ee
and $m_{s} = m_0/\gamma_1$. To keep this limiting case valid we use the same center of mass definition of Eq. \eqref{Rc} for all states within the six-band model.

Then the Hamiltonian \eqref{ExcitonHamilt} can be rewritten as:
\begin{widetext}
   \begin{equation}
\label{ExcitonHamiltSep}
\widehat{H}_{\rm ex}= E_g-\frac{1}{3}\Delta_{\rm SO}\left[(\hat{\bm I}\hat{\bm s}_h)-1\right] + 
\widehat{H}_{\rm kin}({\bm r})
+ V_{\rm C}({\bm r})  + \widehat{H}_{\rm kin}({\bm R}) + V_{{\bm k}{\bm K}}({\bm r},{\bm R}) + V_{\rm loc}({\bm r},{\bm R}) \, ,
\end{equation} 
\end{widetext}
where $V_{\rm loc}({\bm r},{\bm R})=V_{\rm ext}({\bm r}_e)+V_{\rm ext}({\bm r}_h)$. 

The kinetic energy of the relative motion is described by $\widehat{H}_{\rm kin}({\bm r})$ from Eq.~\eqref{Hkin} with ${\bm k}_h$ replaced by ${\bm k}= - {\bm \nabla}_{\bm r}$ and $\gamma_1$ replaced by $\gamma_1^{\rm r}$: $$ \gamma_1^{\rm r} = \gamma_1 + \frac{m_0}{m_e}.$$

The kinetic energy of the center of mass motion is described by $\widehat{H}_{\rm kin}({\bm R})$ from Eq.~\eqref{Hkin} with  ${\bm k}_h$ replaced by ${\bm K}= - {\bm \nabla}_{\bm R} = {\bm k}_e+{\bm k}_h$ and $\gamma_1$ and $\gamma$ replaced by $\gamma_1^{\rm ex}$ and $\gamma^{\rm ex}$:
\begin{eqnarray} \label{ggCM}
&&\gamma_1^{\rm ex} =\gamma_1 \frac{m_s}{m_e +m_s} = \gamma_1 \frac{m_0}{m_e \gamma_1+m_0}, \\ && \gamma^{\rm ex} = \gamma \left( \frac{m_s}{m_e+m_s} \right)^2=\gamma \left( \frac{m_0}{m_e \gamma_1+m_0} \right)^2 \, . \nonumber
\end{eqnarray}
These parameters are related to the heavy, $M_{h}$, the light, $M_l$, and the spin-orbit split, $M_s$, exciton translational masses by \cite{Ekimov_jtp} 
\begin{eqnarray} \label{translationmass}
\frac{m_0}{M_{h}} = \gamma_1^{\rm ex}-2\gamma^{\rm ex}, \, \, \frac{m_0}{M_{l}} = \gamma_1^{\rm ex}+2\gamma^{\rm ex}, \, \, \frac{m_0}{M_{s}}=\gamma_1^{\rm ex}\,. \nonumber \\
    \end{eqnarray}
Note, that $M_s = m_e + m_s$ as expected for an exciton formed from two two-fold degenerate parabolic bands. However, this simple relation  does not work for excitons formed from four-fold degenerate $\Gamma_8$ holes because of the spin-orbit-induced mixing  of the light and heavy holes, and excitons in the kinetic energy Hamiltonians $\widehat{H}_{\rm kin}({\bm r_h})$ and $\widehat{H}_{\rm kin}({\bm R})$ are described by different parameters $\gamma$ and $\gamma^{\rm ex}$, respectively. 

Both, the localizing NC potential $V_{\rm loc}({\bm r},{\bm R})=V_{\rm ext}({\bm r}_e)+V_{\rm ext}({\bm r}_h)$ and  the perturbation 
\begin{widetext}
\begin{eqnarray} \label{VKk}
V_{{\bm k}{\bm K}} =- \frac{2\hbar^2\gamma}{ m_0}\frac{ m_h}{m_e+m_h}\left[2{\bm k}{\bm K} -3 \left( \sum_{\alpha=x,y,z}k_{\alpha}K_{\alpha}\hat{I}_{\alpha}^2+2\sum_{\alpha\neq\beta}\{k_{\alpha} K_{\beta}\}\{\hat{I}_\alpha \hat{I}_\beta\}\right)\right] ,
\end{eqnarray}
\end{widetext}
mix the relative and the center of mass exciton motion. 

We will consider here only the case $a_{\rm ex} \ll a$ and introduce the replacement $V_{\rm loc}({\bm r},{\bm R})=V_{\rm ext}({\bm R})=V_{\rm ext}(R)$. Then, without accounting for the $V_{{\bm k}{\bm K}}$ term, the relative motion is described by $\widehat{H}_{\rm kin}({\bm r})+ V_{\rm C}({\bm r})$ and is not affected by the localizing NC potential due to the spherical symmetry of the $V_{\rm C}({\bm r})$ potential. The relative motion states of electron and hole in the exciton are characterized by the orbital angular momentum   ${\bm l}={\bm r} \times {\bm k}$. The difference between the binding energies of the excitons with the contribution from the $\Gamma_7$ and $\Gamma_8$ holes, $E_{\rm b}^{\Gamma_7}$ and $E_{\rm b}^{\Gamma_8}$, respectively, can be neglected if $\gamma/\gamma_1^r \ll 1$ for the relative motion and the mixing of the $l=0$ ($s$-symmetry) and $l=2$ ($d$-symmetry) harmonics is small. We estimate the ratio as $\gamma/\gamma_1^r \approx 0.1$ for CuCl (see the Section \ref{sec:param}) and maintain the simplifying assumption with $E_{\rm b}^{\Gamma_7}=E_{\rm b}^{\Gamma_8}=E_{\rm b}$ in our consideration. 

The motion of the exciton center of mass in the spherical potential $V_{\rm ext}(R)$ is characterized by the orbital angular momentum ${\bm{{\cal L}}}= {\bm R} \times {\bm K}$. The $V_{{\bm k}{\bm K}}$ perturbation mixes the states with $l$ and $l+1$ and also with ${\cal L}$ and ${\cal L}+1$. We will neglect the mixing of $s$ and $p$ symmetry excitons, characterized by the internal orbital angular momentum $l=0$ and $l=1$, respectively, as they are separated by an energy of about 150 meV (3/4 of the exciton binding energy $E_{\rm b} \sim 200$ meV). Thus, we will consider only the lowest states with $l=0$ for the relative motion. 

As a result, the considered exciton states in spherical NCs are characterized by the total exciton angular momentum ${\bm{{\cal F}}}$ including the orbital angular momentum ${\bm{{\cal L}}}$ in addition to all the angular momenta described in the Introduction for the bulk band-edge excitons: ${\bm{{\cal F}}} = {\bm s}_e  + {\bm s}_h + {\bm I} + {\bm{{\cal L}}} = {\bm s}_e  + {\bm j_h} + {\bm{{\cal L}}}={\bm s}_e  + {\bm{{\cal J}}}$.  The spin-orbit-induced term $\propto \gamma^{\rm ex}$ in the six-band kinetic energy exciton Hamiltonian  $\widehat{H}_{\rm kin}({\bm R})$ is mixing the ${\cal L}$ and ${\cal L}+2$ states, as well as the $\Gamma_7$ and $\Gamma_8$ excitons. Again, when the mixing of the relative and center of mass exciton motions is neglected, the parity of the exciton wave function is conserved and it is constructed with either even or odd ${\cal L}$.  The optically active states are the ${\cal S}$-type exciton states with ${\cal J} = 1/2$ and ${\cal J} = 3/2$, which contain the spherical harmonics with ${\cal L}=0$. Without account of the electron-hole exchange interaction, these states are $(2{\cal J}+1)$-fold degenerate with respect to the projection ${\cal J}_{z} = -{\cal J}, ...,+{\cal J}$ and doubly degenerate with respect to the electron spin projection. In the following, we denote the confined exciton states as $n{\cal S}_{3/2}$, $n{\cal S}_{1/2}$, and $1{\cal P}_{1/2}$ (for the lowest exciton state with ${\cal L}=1$)in a manner similar as we denoted the confined hole states. 

Then, the two-particle exciton function $\Phi^{\rm ex}({\bm r}_e,{\bm r}_h)=\Phi^{\rm ex}_{s_z,{\cal J},{\cal J}_{z}}({\bm r},{\bm R})$in the considered approximation can be written as 
\be
\Phi^{{\rm ex},n}_{s_z,{\cal J},{\cal J}_{z}}({\bm r},{\bm R})= u^c_{1/2,s_z} \phi({\bm r})\Psi^{{\rm ex},n}_{{\cal J},{\cal J}_{z}}({\bm R}) \, , 
\ee  where $s_z=\pm 1/2$, the function $\phi({\bm r})$ is the $s$-type ($l=0$) wave function describing the ground state relative  motion in the exciton, $n=1,2..$ numerates the exciton states of a given symmetry, and the function $\Psi^{{\rm ex},n}_{{\cal J},{\cal J}_{z}}({\bm R})$ describing the center of mass exciton motion which is constructed in a similar way as the hole wave function of Eq. \eqref{PsiJJ}: 
\begin{multline}
\Psi^{{\rm ex},n}_{{\cal J},{\cal J}_{z}}({\bm R})= \sqrt{2{\cal J}+1}\sum_{{\cal L}} (-1)^{{\cal L}-j_h+{\cal J}_{z}} (i)^{{\cal L}} R^{{\cal J},j_h}_{{\cal L},n}(R)\times \\ \times\sum_{{\cal L}_z+\mu = {\cal J}_{z}, j_h=3/2,1/2}
\left(
\begin{array}{ccc}
{\cal L} & j_h&{\cal J} \\ {\cal L}_z&\mu&-{\cal J}_{z}
\end{array}
\right)Y_{{\cal L},{\cal L}_z} u_{j_h,\mu}^v \, .
\label{PsiJJex}
\end{multline}
The energy levels $E_{n{\cal S}_{1/2}}$, $E_{n{\cal S}_{3/2}}$, and $E_{n{\cal P}_{1/2}}$ for the quantum confined states of the center of mass exciton can be obtained from the boundary conditions $\Psi^{{\rm ex},n}_{{\cal J},{\cal J}_{z}}({\bm R})|_{R=a}=0$. They are the same as given in the Appendix B for the hole states with $\gamma_1$ and $\gamma$ substituted by $\gamma_1^{\rm ex}$ and $\gamma^{\rm ex}$, respectively. 
 
The oscillator strength of the ${\cal S}$-type exciton optical transition can be calculated from
\be \label{os}
f_{n{\cal S}_{\cal J}} \propto N_{\cal J}|\phi(0)|^2 \int {\bm d}^2{\bm R} |\Psi^{{\rm ex},n}_{{\cal J},{\cal J}_{z}}({\bm R}) |^2 \propto \frac{a^3}{a_{\rm ex}^3} {\tilde f}_{n{\cal S}_{\cal J}} , 
\ee
where 
\be \label{osr}
{\tilde f}_{{\cal J},n} = N_{\cal J}\delta_{J,j_h}  \left|\int_0^1 {\tilde R}^2 d {\tilde R} R_{0}^{{\cal J},j_h}({\tilde R}) \right|^2 ,  \quad  {\tilde R}=R/a ,
\ee 
is the relative oscillator strength in comparison to the oscillator strength ${\tilde f}_{1{\cal S}_{1/2}}^0=1$ of the ground state $1{\cal S}_{1/2}$ exciton in large NCs considered without admixture of the $\Gamma_8$ states ($E_{1{\cal S}_{1/2}}^0 \ll |\Delta_{\rm SO}|$). The factors $N_{1/2}=1$ and $N_{3/2}=2$ reflect the relative intensities of the optical transitions between the Bloch states of conduction and valence band for $j_h=1/2$ and  $j_h=3/2$, respectively. The factor $a^3/a_{\rm ex}^3$ in Eq. \eqref{os} describes the effect of giant oscillator strength caused by the localization of the exciton as a whole. Note that the relative oscillator strength $ {\tilde f}_{n{\cal S}_{\cal J}}$ might weakly depend on $a$ because of the exciton state mixing.

With account of the electron-hole exchange interaction, a fine structure of the exciton states with total angular momentum $\bm{{\cal F}}=\bm{{\cal J}}+{\bm s}_e$ will emerge. For the considered states with ${\cal J}=3/2$ and ${\cal J}=1/2$ the fine structures are the same as shown in the Introduction in Fig. \ref{Fig:ebands}(e). The bright exciton states are the states with total angular momentum ${\cal F}=1$. The value of the electron-hole exchange interaction for the exciton in the considered model is determined by the internal motion and expected not to be size-dependent, but coincides with the bulk value of about 6 meV \cite{Certier1969}. For this reason, the energy shift of the bright ${\cal F}=1$ excitons can be considered to be identical for all exciton states and is included into the value of $E_{\rm b}$ for the following comparison with the experimental data. 

\subsubsection{Evaluation of the valence band parameters from the experimental exciton absorption spectra} \label{sec:param}

In the considered approximation, the exciton optical transition energies are given by 
\be
E_{X}({n{\cal S}_{\cal J}}) = E_g - E_{\rm b} + E_{n{\cal S}_{\cal J}} \, .
\ee
The calculated results of $E_{n{\cal S}_{\cal J}}$ in dependence on $1/a^2$ are shown in Fig.~\ref{Fig:AbsModel} with the green lines for the exciton states with ${\cal J}=1/2$ and the blue lines for ${\cal J}=3/2$. The orange line shows the energy of the lowest $1{\cal P}_{1/2}$ exciton. For $\beta^{\rm ex} = (\gamma_1^{\rm ex}- 2\gamma^{\rm ex})/(\gamma_1^{\rm ex}+ 2\gamma^{\rm ex}) > 0.18$, $E_{1{\cal S}_{1/2}} < E_{1{\cal P}_{1/2}}$ for all NC radii $a \ge 1.8$ nm. The size dependences of the energies of the $1{\cal S}_{1/2}$ and $1{\cal P}_{1/2}$ excitons are also shown in Figs. \ref{ehsp}(c) and  \ref{ehsp}(d). One can see that for $a>2$ nm, the $E_{1{\cal S}_{1/2}}$ exciton energy dependence is well approximated by the $E_{1{\cal S}_{1/2}}^0$ energy dependence, calculated without account of the admixture of the $\Gamma_8$ states.  

The green and blue squares show the relative oscillator strengths ${\tilde f}_{n{\cal S}_{\cal J}}$ for the ${n{\cal S}_{1/2}}$ and ${n{\cal S}_{3/2}}$ exciton states. Similar to the hole states, one can see in Fig.~\ref{Fig:AbsModel} the anti-crossing of the ${\cal S}_{3/2}$ arising from the $\Gamma_7$ (mostly $d$-like) and $\Gamma_8$ (mostly $s$-like) valence subbands.  However, the states with ${\cal J}=3/2$ have nonzero oscillator strength only, when they contain a significant ${\cal S}$-type contribution with the $j_h=3/2$ holes from the $\Gamma_8$ valence band. Their energies (the upper branches after anti-crossing) and the relative oscillator strengths  are well reproduced by the energies and oscillator strengths of the $n{\cal S}_{3/2}^{\rm 4b}$ states, calculated within the four-band model. Due to the light hole to heavy hole state mixing, not only the lowest exciton transitions with ${\cal J} = j_h=3/2$ arising from the $\Gamma_8$ excitons contribute to the absorption, but the excited states as well. Moreover, we obtain ${\tilde f}_{2{\cal S}_{3/2}}^{\rm 4b}>{\tilde f}_{1{\cal S}_{3/2}}^{\rm 4b}$ for $\beta^{\rm ex} < 0.2$. This explains the larger slope of the exciton energy shift observed experimentally for the $\Gamma_8$ exciton ($Z_{1,2}$ line) compared to the $\Gamma_7$ exciton ($Z_{3}$ line): $C(Z_{1,2})>C(Z_3)$. Note that due to the admixture of the $\Gamma_8$ states, also the excited ${\cal J} = j_h=1/2$ states (the $\Gamma_7$ excitons) become optically active. The effect is weaker for the $\Gamma_7$ excitons than for the $\Gamma_8$ excitons, however, the excited state transition can be observed in absorption and is responsible for the doublet structure of the $Z_3$ line (see the red circles in Fig.~\ref{Fig:AbsModel}).

\begin{figure}[h!]
\begin{center}
\includegraphics[width=8cm]{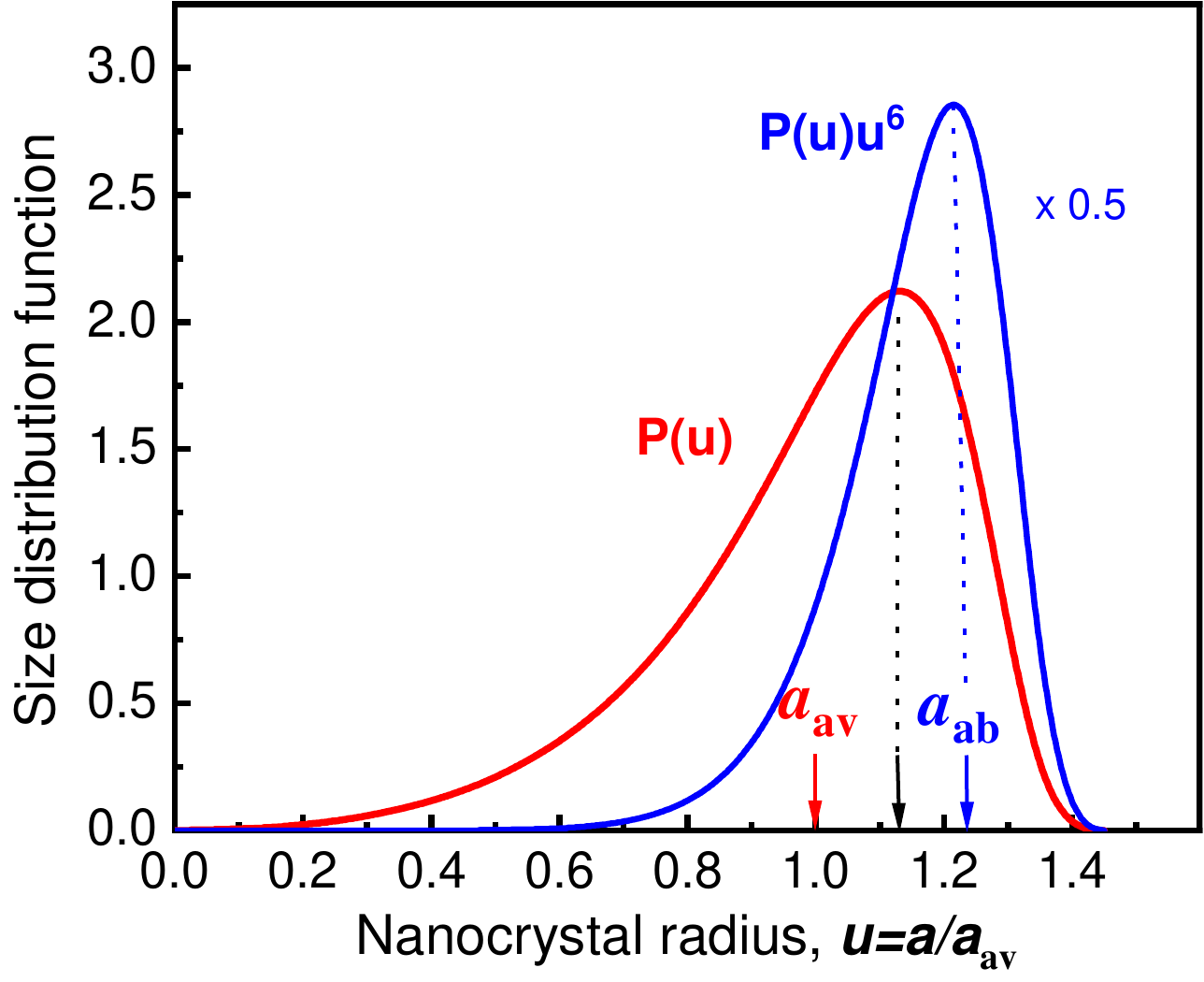}
\caption{Lifshitz-Slesov size distribution function $P(u)$ \cite{LS1959} (red line). Red, black and blue arrows correspond to the average radius $a_{\rm av}$ of the ensemble, the maximum of $P(u)$, and the radius $a_{\rm ab}$. $a_{\rm ab}\approx 1.22a_{\rm av} $ gives the maximum of the $P(u)u^6$ function (blue line) and corresponds to the NCs giving maximum absorption in the NC ensemble.} \label{LS}
\end{center}
\end{figure}

For comparison of the calculated exciton spectrum with the experiment, we show in Fig.~\ref{Fig:AbsModel} the experimental data from Refs.~\onlinecite{Ekimov_jtp,Ekimov1985} and from our study, which are given already in Fig.~\ref{Fig:Abs}(b). The energies of the experimental exciton transitions in Fig.~\ref{Fig:AbsModel} are calculated according to $E_g-E_{\rm b}=3.2$~eV. The experimental data are plotted as function of $1/a^2$ with the radius $a=a_{\rm av}$, where $a_{\rm av}$ is the average radius for the NC size distribution described by the Lifshitz-Slezov function $P(a/a_{\rm av})$, obtained in Ref.~\onlinecite{LS1959} and plotted in Fig.~\ref{LS}.

Note that this size distribution is asymmetric with respect to  $a_{\rm av}$ which is shown by the red arrow in  Fig.~\ref{LS}.  This radius is smaller than the radius at which  $P(u)$ reaches its maximum, also shown by an arrow. As discussed in Ref. \cite{Ekimov_jtp}, the dependence of the lowest $\Gamma_7$ exciton transition energy on $1/a^2$ should be described by $C(Z_3)=0.67/M_s$ with $M_s=1.9m_0$. The factor $0.67$ arises from averaging over the Lifshitz-Slezov size distribution with the size dependence of the transition energies $\propto 1/a^2$  multiplied by the oscillator transition strength $\propto a^3$.  Note that the introduction of the factor 0.67 is equivalent to the replacement of $a=a_{\rm av}$ with $a=a_{\rm ab}\approx 1.22a_{\rm av}$, corresponding to the maximum of the function $P(u) u^6$ that describes the absorption efficiency.  The function's maximum (shown by the blue arrow) in  Fig.~\ref{LS}  corresponds to the NCs that give maximum contribution to the absorption. 
Therefore, we use $a=a_{\rm ab}$ for plotting the calculated exciton energy spectra in Fig.~\ref{Fig:AbsModel}. 
 
To model the experimental data, we use  $\gamma_1^{\rm ex}$  corresponding to $M_s=1.9 m_0$, previously determined in Ref.~\onlinecite{Ekimov1985}, and vary the exciton center of mass parameter $\gamma^{\rm ex}$  to optimize the maximum slope of the excited state $E_{2S_{3/2}}^{\rm 4b}$ energy with ${\tilde f}_{2{\cal S}_{3/2}}^{\rm 4b}> {\tilde f}_{1{\cal S}_{3/2}}^{\rm 4b}$. The dependence of the energy shift slope for the $1{\cal S}_{3/2}^{\rm 4b}$ and $2{\cal S}_{3/2}^{\rm 4b}$ states and their relative oscillator strength with respect to the $1S_{1/2}^0$ state on $\beta^{\rm ex}$ are shown in Figs. \ref{ebeta}(a) and \ref{ebeta}(b) of Appendix B. The optimum value is chosen as $\beta^{\rm ex} \approx 0.19$. 

Thus, the modeling of the experimental data allows us to get the exciton center of mass parameters: $\gamma_1^{\rm ex}=0.53$ and $\gamma^{\rm ex}=0.18$.  The evaluated parameters correspond to the following exciton translational masses: $M_s=1.9 m_0$, $M_{h}=5.9 m_0$, and $M_l=1.1 m_0$. Information about all previously reported exciton translational masses in CuCl is collected in Table~\ref{x_masses}. 

Next, using the electron mass $m_e=0.4m_0$, we obtain the hole Luttinger parameters: $\gamma_1=0.67$ and $\gamma=0.29$. The corresponding hole masses are: $m_{s}=m_h=1.5 m_0$, $m_{hh}=10.8 m_0$, $m_{lh}=0.8 m_0$. Note, that the masses of the $\Gamma_8$ holes are sensitive even to small changes in the Luttinger parameters as $\beta=m_{lh}/m_{hh} \approx 0.07$ is small. In Ref.~\onlinecite{khan1970g} the calculated $\Gamma_8$ holes masses are in the range of $(3-13.7) m_0$. Information about all previously reported hole masses is collected in Table~\ref{h_masses}.  Importantly, for such a small value of $\beta$ the mixing of the hole states from different subbands becomes extremely important. 

For the relative motion of the hole in the exciton, $\gamma_1^r = 3.17$. The exciton binding energies for the resulting $\beta^r=(\gamma_1^r - 2\gamma)/(\gamma_1^r + 2\gamma)=0.66$, calculated within the six-band model with the Coulomb interaction potential, are found to be very close. This justifies neglecting their difference in our modeling. 
\newline \newline

\subsection{Effect of external magnetic field on the electron and hole states in CuCl nanocrystals}
\label{sec:gfactors}

An external magnetic field breaks the time-reversal symmetry and splits the electron and hole states with opposite angular momentum projections on the magnetic field direction. In the low field regime, this Zeeman splitting depends linearly on the magnetic field strength. It is given by the electron (hole) effective $g$-factor or Land\'e factor $g_e$ $(g_h)$. The corresponding Zeeman Hamiltonian for the conduction band electrons is 
\bea \label{g_e}
\widehat{H}_{Z}^{e}&=& \mu_B g_e ({\bm B} {\bm s}_e)= \mu_B g_e B s_{ez}\,,~{\rm where} \nonumber\\
 g_e&=&\frac{E_{1/2}-E_{-1/2}}{\mu_{B} B}~.	
\eea
Here $\mu_{B}$  is the Bohr magneton (positive), and $E_{\pm 1/2}$ are the energies of the states with spin projection $s_{ez}=\pm 1/2$, where the $z$ axis is chosen to be parallel to the direction of the magnetic field. We use the same definition for the size-dependent $g$-factor of the lowest $1S_e$ electron state confined in a spherical NC, $g_e(a)$.  

For the bulk $\Gamma_7$ holes from the topmost valence subband with $j_h=1/2$, we use the following definition of the effective $g$-factor and effective Zeeman Hamiltonian:
\bea\label{g_h}
\widehat{H}_{Z}^{h}&=& -\mu_B g_h ({\bm B} {\bm j}_h)= -\mu_B g_h B j_{hz} \,,~{\rm where} \nonumber\\
g_h&=&\frac{E_{-j_{hz}}-E_{+j_{hz}}}{2j_{h}\mu_B B}=\frac{E_{-1/2}-E_{+1/2}}{\mu_B B}.
\eea
Here $j_{hz}$ is the hole spin projection on the magnetic field direction. For the lowest $1S^h_{1/2}$ state of the hole confined in a spherical NC $j_h$ and $j_{hz}$ in Eq.~\eqref{g_h} should be replaced with $J_h=1/2$ and $J_{hz}=\pm 1/2$. Note, that the sign of the hole Zeeman term in Eq. \eqref{g_h} is opposite to that in Eq. \eqref{g_e}, so that a positive $g_h$ corresponds to a hole ground state with a positive spin projection $j_{hz}$ or $J_{hz}$. Here, we follow the approach suggested in Refs. \cite{Efros1996,Efros_book} and widely used for colloidal NCs, e.g. in Ref. \cite{Semina2021} for the $\Gamma_8$ holes.  

It is worthwhile to note that in CuCl and the lead halide perovskites the states in the vicinity of the band gap have spin $1/2$. Therefore, our theory is analogous to the theory for perovskite bulk crystals \cite{Yu2016,Huynh2022,Kirstein2022} and nanocrystals \cite{Nestoklon2023} when accounting that the formalism for electrons (holes) in CuCl is appropriate for holes (electrons) in perovskites. The main difference between them is the importance of the mixing of the $\Gamma_7$ and $\Gamma_8$ states, provided by the relatively small value of the spin-orbit splitting in the CuCl valence band.

In the following we calculate the size dependence of the electron and hole $g$-factors in  CuCl NCs in order to analyze which of them contributes to the spin dynamics observed in the experiment. \newline \newline

\subsubsection{Electron $g$-factor in bulk CuCl}

In bulk semiconductors, the isotropic electron $g$-factor at the bottom of the $\Gamma_6$ conduction band  can be calculated using  second-order perturbation theory as suggested by Roth~\cite{Roth1959} (see the general expression in \ref{AC}).  Within the eight-band model, the interaction of the two electron bands with the six valence bands is considered explicitly, while the interaction with  other (remote) bands are still included as second-order perturbation correction.  As a result,  the electron $g$-factor at the bottom of the conduction band is sensitive to the spin-orbit interaction in the valence band and is given by
\begin{eqnarray}\label{R8}
    g_e=g_0+g_{\rm rb}-\frac{2E_p}{3} \left( \frac{1}{E_g^{\Gamma_8}} - \frac{1}{E_g^{\Gamma_7}}\right) \,.
    \end{eqnarray}
Here $g_0=2.0023$ is the $g$-factor of the free electron, $g_{\rm rb}$ is the contribution from all other energy bands (so called remote bands), $E_g^{\Gamma_8} = E_{\Gamma_6}-E_{\Gamma_8}$ and $E_g^{\Gamma_7}=E_{\Gamma_6}-E_{\Gamma_7}$ are the energy gaps between the bottom of the conduction band and the top of the $\Gamma_8$ and $\Gamma_7$ valence bands, $E_p=2P^2/m_0$ is the Kane energy, and $P=-i\langle S| \hat p_x| X \rangle$ is the Kane matrix element. In CuCl, $E_g=E_g^{\Gamma_7}=3.4$ eV. The negative spin-orbit splitting $\Delta_{\rm SO}=E_g^{\Gamma_7}-E_g^{\Gamma_8}<0$ ($\Delta_{\rm SO}=-0.07$ eV) results in the electron $g$-factor at the bottom of the conduction band of bulk CuCl to be $g_e=2.03$ \cite{khan1970g}, slightly larger than $g_0$. 
  
We have not found any information in the literature for the values of $E_p$ and $g_{\rm rb}$  in CuCl. The Kane energy  $E_p$ can be estimated from the known value of the electron effective mass $m_e=0.4 m_0$ at the bottom of the conduction band given within the same model by 
\begin{eqnarray}\label{R8}
    \frac{m_0}{m_e}= 1+\alpha_{\rm rb}+\frac{E_p}{3} \left( \frac{2}{E_g^{\Gamma_8}} + \frac{1}{E_g^{\Gamma_7}}\right) \, ,
    \end{eqnarray}
where $\alpha_{\rm rb}$ is the contribution from the remote bands. For $\alpha_{\rm rb}=0$ we estimate $E_p=5$~eV. However, as it is often the case in semiconductors with wide band gaps, in CuCl the energy separation $E_g'\approx7$~eV  \cite{Doran1979}  from $E_{\Gamma_6}$ to the nearest above lying $p$-type conduction bands ${\Gamma_8^c}$ and  ${\Gamma_7^c}$ is comparable to $E_g=3.4$ eV. Therefore, the interaction with these bands can give significant contributions $\alpha_{\rm rb}<0$ and $g_{\rm rb}$ to the electron effective mass and the $g$ factor within the fourteen-band model (see Eq. \eqref{rb} in \ref{AC}) as was shown, for example, for GaN \cite{Rodina2001a}. The sign of $g_{\rm rb}$ depends on the sign of the spin-orbit splitting of the remote conduction band $\Delta_{\rm SO}'=E_{\Gamma_8^c}-E_{\Gamma_7^c}$. 

To estimate the parameters of the eight-band and fourteen-band models for the following calculations of the electron $g$ factor size dependence in CuCl NCs we keep the values $m_e=0.4 m_0$, $g_e=2.03$ and consider four values of $\alpha_{\rm rb}<0$. The  calculated values of $E_p$, $g_{\rm rb}$, $E_p'$ (Kane interaction energy between the $E_{\Gamma_6}$ and remote conduction bands) and $\Delta_{\rm SO}'$ are also summarized in Table \ref{tab:Ep} of \ref{AC}.

\subsubsection{Size dependence of electron $g$-factor in CuCl NCs}

The quantum size effect leads to an increase of the effective band gap energy with decreasing NC radius. As a result, the contribution from the valence band to the electron $g$-factor is expected to become smaller than in bulk. In CdSe NCs this effect is well known and was observed in the size dependence of the electron $g$-factor, namely as an increase of $g_e(a)$ with decreasing NC radius $a$~ \cite{Gupta2002,Rodina2003,Semina2021,Gang2022}. In CuCl, however, because of $\Delta_{\rm SO}<0$, one may expect the opposite tendency -- a decrease of the absolute value of $g_e(a)$ with decreasing NC radius $a$.  Indeed, the $g$-factor of the ground state of an electron confined in a spherical NC can be written as \cite{Rodina2003}:
\begin{equation} \label{ga}
g_e(a)=g_0+ {\cal A}(E_e)\Delta g_e(E_e)+g_{\rm surf}(a),
\end{equation}
where $\Delta g_e(E_e) = {\tilde g}_e(E_e) - g_0$ describes the difference between the electron $g$-factor in bulk at the energy $E_e$ calculated from $E_{\Gamma_6}$ and $g_0$. The factor ${\cal A}(E_e)=\int_0^a dr \, r^2 |R_0^c(r)|^2  \le  1$  (see the analytical expression in Ref. \cite{Semina2021}) accounts for the contribution from the valence band states to the electron wave function at the energy $E_e$ in the eight-band Kane model. As the band gap energy in CuCl is quite large, we consider in Section \ref{sec:ehex} the electron confinement in the simple band model where ${\cal A}(E_e) = {\cal A}(0)=1$. Within the eight-band Kane model,  ${\cal A}(E_e) \ge 0.95$ for the ground state electron in NCs with $a \ge 1.8$~nm and ${\cal A}(E_e) \ge 0.99$ in NCs with $a \ge 4$~nm.

The last term $g_{\rm surf}(a)= \Delta g_e(E_e)|R_0^c(a)|^2/3$ in Eq.~\eqref{ga} describes the surface contribution to the electron magnetic moment which comes from the surface-induced spin-orbit interaction \cite{Rodina2003}. We have neglected this contribution  so far, assuming the standard boundary conditions  $R_0^c(a)=0$, where the electron wave function vanishes at the NC surface. With the two approximations ${\cal A}(E_e) = {\cal A}(0)=1$ and $g_{\rm surf}(a)=0$, the size dependence of the electron $g$-factor is determined totally by its energy dependence: $g_e(a) \approx {\tilde g}_e(E)$ with $E=E_0^e(a)=E_{{1S_e}}^0(a)$ for the ground state of the electron confined in a NC.

The energy dependence of the electron $g$-factor in bulk CuCl within the eight-band Kane model can be described by:
\bea
 {\tilde g}_e(E)&=&g_0+g_{\rm rb}-\frac{2E_p}{3}\left(  \frac{1}{E_g+|\Delta_{\rm SO}|+E} - \frac{1}{E_g+E}\right) \nonumber \\
    &= &g_e + E \alpha_{\rm so}(E), ~{\rm where} \nonumber\\ 
 \alpha_{\rm so}(E)& = &\frac{2E_p}{3}\left( \frac{1}{(E_g+|\Delta_{\rm SO}|)(\tilde{E_g}+|\Delta_{\rm SO}|)} -  \frac{1}{E_g\tilde{E}_g}\right). \nonumber\\ \label{geE}
\eea
Here $E_g+|\Delta_{\rm SO}|=E_g^{\Gamma_8}$ and $\tilde{E}_g=E_g+E$, and for electron energies $E \ll E_g$ the deviation from the bulk $g$-factor $g_e$  given by $E \alpha_{\rm so}(0)$ can be obtained through a Taylor expansion.   

\begin{figure}[h!]
\begin{center}
\includegraphics[width=8cm]{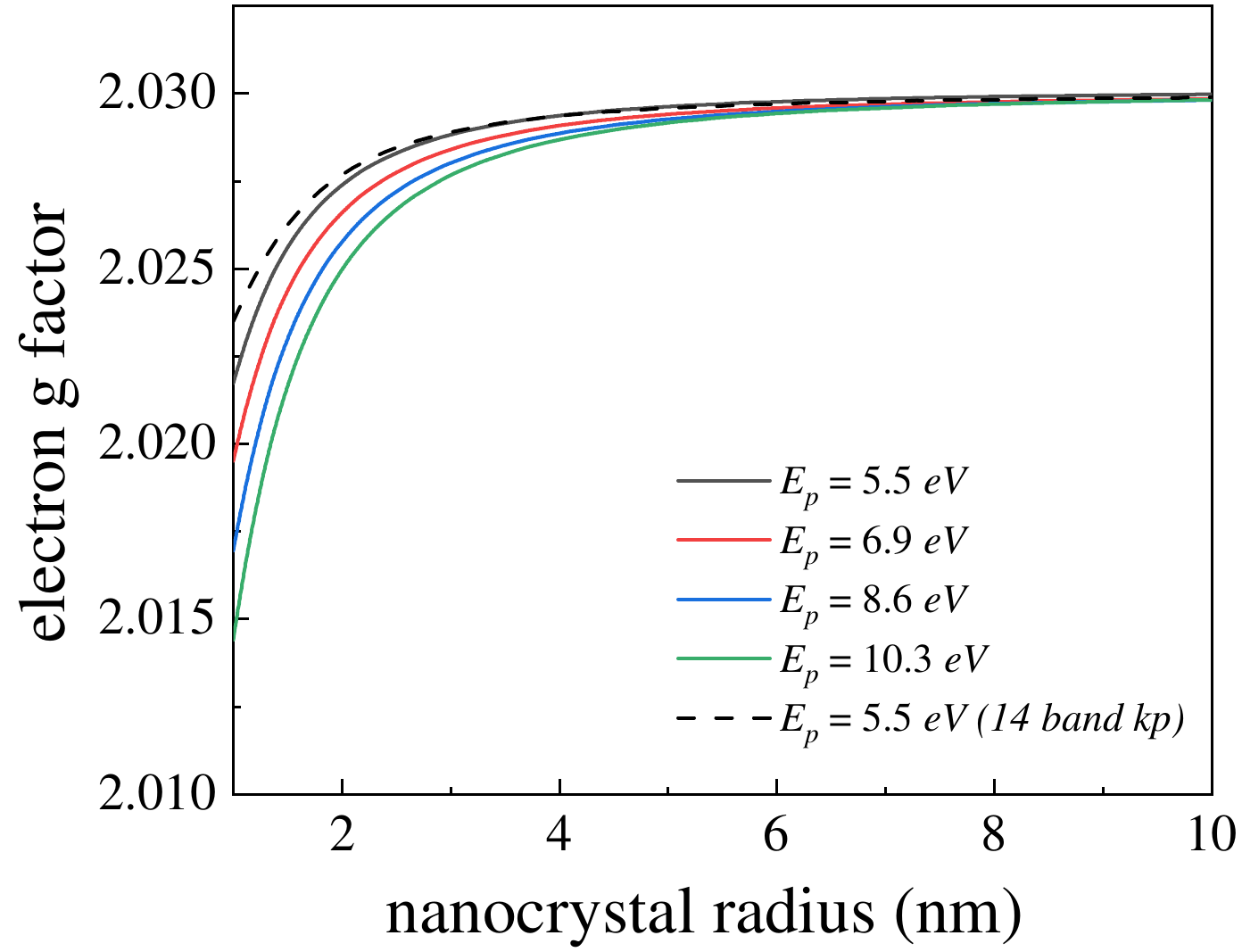}
\caption{Dependence of the electron $g$-factor on CuCl NC radius, $g_e(a)$, calculated within the eight-band ${\bm k}{\bm p}$-model (the solid lines for different Kane energies. Dashed line shows the result of calculations within the fourteen-band ${\bm k}{\bm p}$-model. The parameters in the calculations are given in the text and Table~\ref{tab:Ep}. }
\label{ge8band}
\end{center}
\end{figure}

Importantly, due to $\Delta_{\rm SO} <0$ (and thus $\alpha_{\rm so} < 0$) in CuCl, the function  ${\tilde g}_e(E_e)>0$ is decreasing  with increasing electron energy $E_e$. This results in a decrease of $g_e(a)$ with decreasing NC radius $a$, regardless of the bulk parameters chosen, as shown in Fig.  \ref{ge8band}. This size-dependence of the electron $g$-factor $g_e(a)$ is opposite to what we have found experimentally in this work, see Tables~\ref{Tab_g-factors}, ~\ref{Table 1} and  Fig.~\ref{fig gfactor}. 
Accounting for the nonzero surface-induced contribution $g_{\rm surf} \ne 0$ with consequently assuming that $R_0^c(a)\neq 0$, cannot reverse the tendency for an electron localized inside the NC (not at the surface) as $|R_0^c(a)|^2/3 \ll 1$. Thus, the eight-band Kane model cannot explain the $g_e(a)$ dependence and the large electron $g$-factors observed experimentally by SFRS in small-sized CuCl NCs.
 
Within the fourteen-band model,  an additional energy dependence of the electron $g$-factor appears, arising from the interaction with the nearest $p$-type conduction band, as was shown in Ref.~\cite{Rodinaemrs} for ZnO. In this case, the $g_{\rm rb}$ of Eq. \eqref{rb} should be replaced with the $g_{\rm rb}(E_e)$ given by:
  \begin{eqnarray} \label{rb}
		g_{\rm rb}(E_e)=- \frac{2E_p'}{3}\left(\frac{1}{E_{\Gamma_8^c}-E_e} - \frac{1}{E_{\Gamma_7^c}-E_e}\right) \,  .
		\end{eqnarray}

In Figure~\ref{ge8band} we show by the dashed line the results of the calculation within the fourteen-band model with the parameters from Table~\ref{tab:Ep} (first row). Clearly, only an increasing positive $g_{\rm rb}(a)$ (negative $\Delta_{\rm SO}'$, first two rows in Table ~\ref{tab:Ep}) can help to reverse the $g_e(a)$ dependence. However, as one can see in  Fig.~\ref{ge8band},  $g_e(a)$ is still decreasing with decreasing NC radius $a$.  To reverse this tendency and obtain the large values of the electron $g$-factor observed in SFRS, one has to assume quite unrealistic values of $E_{p}'$ and $\Delta_{\rm SO}'$, namely $E_{p}'=37$~eV and $\Delta_{\rm SO}'=-2.8$~eV. 

Within the fourteen-band model additional spin-orbit terms that are cubic in the wave vector ${\bm k}$, arise for the conduction band electrons due to the lack of the bulk inversion symmetry in the $T_d$ point group (so-called BIA or Dresselhaus terms) \cite{Dresselhaus1955}. In magnetic field, they could also contribute to the splitting of the electron states.  However, it is easy to see that the contributions from the BIA terms both in zero and weak external magnetic fields vanish for the electrons in the ground state of spherical NCs due to the isotropy of its wave function. 

We consider also another possibility to reverse the $g_e(a)$ tendency for the resident electron $g$-factor in NCs. For that we revisit the ${\bm k}{\bm p}$ model for the electron $g$-factor in spherical NCs and take into account the interaction with the quantum confined levels in the valence band, calculated in the previous section within the six-band Luttinger Hamiltonian. The details of this consideration are given in the Appendix C. In effect, accounting for the quantized states in the valence band is equivalent to  replacing $E$ with $E=E_0^e(a)+E_0^h(a)$ in ${\tilde g}_e(E)$ of Eq. \eqref{geE}, where the hole energy is 
$E_0^h(a)=E_{1{\cal S}_{1/2}}^0(a)=E_{1{\cal S}_{3/2}}^0(a)$. However, this does not allow one to reverse the $g_e(a)={\tilde g}_e(E(a))$ dependence as well. 

\subsubsection{Hole $g$-factor size dependence in CuCl nanocrystals}

The holes in an external magnetic field are described by the following Hamiltonian, which we rewrite in the hole representation as the sum of three contributions: 
\begin{equation}
\label{HamiltB}
\widehat{H}_h^{\rm B}({\bm r}_h)=\widehat{H}_h({\bm r}_h)+\widehat{H}^{(h)}_{\text{Z}}+
\widehat{H}^{\text{B}}_{\rm kin}({\bm r}_h),
\end{equation}
where  $\widehat{H}_h({\bm r}_h)$ is the hole Hamiltonian in zero magnetic field, described by Eqs. \eqref{Hamilt}, \eqref{lutt6} and \eqref{Hkin}. The Zeeman part of the hole Hamiltonian in bulk, $\widehat{H}_{\rm Z}^{(h)}$,  in the hole representation has the form \cite{Luttinger1956}:
\begin{equation}\label{HZ}
\widehat{H}_{Z}^{(h)}=-\mu_B(1+3\varkappa)(\hat{\bm {I}}_h \bm B)+\mu_B g_0(\hat{\bm s}_h \bm B),
\end{equation}
where  $\varkappa$ is the magnetic Luttinger parameter \cite{Luttinger1956}.

The matrix form of the Hamiltonian \eqref{HZ} can be found in the Appendix B of Ref. \cite{Semina2021}.  
Note that in the hole and in the electron representation the relative sign of the hole Zeeman Hamiltonian \eqref{HZ} and the kinetic energy \eqref{lutt6} is the same.  In bulk semiconductors with a large value of the spin-orbit splitting of the valence band $|\Delta_{\rm SO}|$, the Hamiltonian \eqref{HZ} describes the Zeeman splitting of the four-fold  degenerate hole states at the top of the $\Gamma_8$ valence band ($j_h=3/2$) with the bulk effective $g$-factor $g_h  = 2\varkappa$ \cite{Semina2021} and of the two-fold  degenerate hole states at the top of the $\Gamma_7$ valence band ($j_h=1/2$) according to Eq. \eqref{g_h} with the bulk effective $g$-factor  $g_h = g_h^{(1/2)} = 2+4\varkappa$ \cite{Kiselev2001}. 

The next contribution to the hole Hamiltonian \eqref{Hamilt} is the orbital (kinetic energy) contribution in the magnetic field, $\widehat{H}^{\text{B}}_{\rm kin}$. It can be obtained by replacing the hole wave vector  ${\bm k}_h$ with ${\bm k}_h -\frac{e}{c\hbar}\bm A$ in the Hamiltonian $\widehat{H}_{\rm kin}$ of Eq. \eqref{Hkin}, where $e=|e|$ is the absolute value of the free electron charge and $\bm A$ is the vector potential of the magnetic field.  In a weak magnetic field one can keep in $\widehat{H}^{\text{B}}_{\rm kin}$only the terms linear in $\bm A$ and thus linear in $B$. The explicit matrix form of the six-band orbital magnetic field contribution  $\widehat{H}^{\text{B}}_{\rm kin}$ for the field directed along the $z$ axis and using the Landau gauge ${\bm A} =(0,Bx,0)$ can be  found in the Appendix B of Ref. \cite{Semina2021}. Here we present its general operator form using the symmetric gauge ${\bm A}=(1/2)[{\bm B}\times{\bm 
 r}]$ :
\begin{widetext}
\begin{equation}\label{HB}
 \widehat{H}_{\rm kin}^{\rm B}({\bm r}_h)= -\mu_{\rm B}\left[\left(\gamma_1+4\gamma\right)({\bm L}_h{\bm B})+3\gamma ({\bm I}_h{\bm B}) - 6\gamma \{({\bm k}_h{\bm I}_h)([{\bm r}_h \times {\bm I}_h]{\bm B}) \}\right] \, .
\end{equation}
\end{widetext}
Note that Eq.~\eqref{HB} is similar to the perturbation Hamiltonian obtained in Ref. \cite{Malyshev1998} in the momentum representation for the hole localized at an acceptor center. For its derivation the commutation relation $[k_x,k_y]= ie/(c\hbar) B_z$ \cite{Luttinger1956} for the hole wave vector in the magnetic field is used.   

For the hole localized in some potential, for example in a spherical NC, the mixing of the  states from different subbands  in zero magnetic field leads to renormalization of the hole $g$-factor. As a result, the Zeeman splitting is determined by the total hole angular momentum ${\bm J}_h$ and its projection on the magnetic field. 
The renormalization of the ground hole $g$-factor by the NC confinement was considered by us within the six-band model for semiconductors with  $\Delta_{\rm SO}>0$ \cite{Semina2021,Semina2023}.   Here we focus on the  hole states with $J_h=1/2$ arising from the two-fold degenerate $\Gamma_7$ valence subband. Accounting for the valence band warping, which is not included in the Hamiltonians $\widehat{H}_{\rm kin}$ and $\widehat{H}^{\text{B}}_{\rm kin}$, would not lead to an anisotropy of the hole $g$-factor \cite{Semina2023}. We consider both the even $1S_{1/2}$ and odd $1P_{1/2}$ hole state $g$-factors, as the energy levels of these states are close to each other and interchange with decreasing NC radius. 

After substituting the wave functions \eqref{PsiJJ} into Eqs. \eqref{HB} and \eqref{HZ} and averaging over the angle variables, we obtain the following expression for the effective hole $g$-factor in the weak magnetic field regime:
\begin{multline}\label{gfactor}
g_{h}({1S_{1/2}})=2+4\varkappa-6\varkappa I_2+12\sqrt{2}\gamma I_1-\\-(2-2\gamma_1-2\gamma)I_2+4\sqrt{2}\gamma I_3,
\end{multline}
where
\bea \label{eq:ints}  I_1&=&\int_0^\infty R_s(r_h)R_2(r_h) r_h^2 dr_h,~ I_2=\int_0^\infty R^2_2(r_h) r_h^2 dr_h,~ \nonumber\\
I_3&=&\int_0^\infty R_s(r_h)\frac{d R_2(r_h)}{dr_h} r_h^3 dr_h.
\eea

The expression given in Eq. \eqref{gfactor} is valid for a hole that confined by an arbitrary spherically symmetric potential. In the limit $|\Delta_{\rm SO}|\rightarrow \infty$ or large NCs, resulting in small size quantization energies for holes, all integrals in Eq.~\eqref{eq:ints} tend to zero and $g_{h}(a) \rightarrow g_{h}^{(1/2)} = 2+4\varkappa$. Note, that this tendency differs from the one for the $\Gamma_8$ holes, where even in large NCs the hole $g$-factor is renormalized compared to the bulk value $2\varkappa$ due to heavy and light hole mixing in the spherically-symmetric potential  \cite{Gelmont1973,Semina2021}.

The resulting expression for the $1P_{1/2}$ hole states, the effective $g$-factor in a weak magnetic field regime is:
\begin{multline}
\label{gfactorP}
g_h(1P_{1/2})=-\frac{2}{3} \bigg((1+2\varkappa-2\gamma_1)-I_4(1+7\varkappa-5\gamma-3\gamma_1) \\+I_5 2\sqrt{2}(2+2\varkappa-7\gamma)-I_6\sqrt{2}\gamma \bigg) \,,
\end{multline}
where
\begin{align} \label{I456}
    &I_4=\int_0^{\infty}R_1(r_h)^2r_h^2dr_h, I_5=\int_0^{\infty}R_1(r_h)R_{s1}(r_h)r_h^2dr_h, \nonumber \\
    &I_6=\int R_{s1}(r_h)\frac{dR_1(r_h)}{dr_h}r_h^3dr_h \,.
\end{align}
Note that in NCs with large radius, the integrals \eqref{I456} tend to zero. However, the $g$-factor of the odd $1P_{1/2}$ hole state is different from the bulk $g$-factor $2+4\varkappa$ due to the orbital $L_h=1$ contribution.  

\begin{figure}[h!]
\includegraphics[width=7cm]{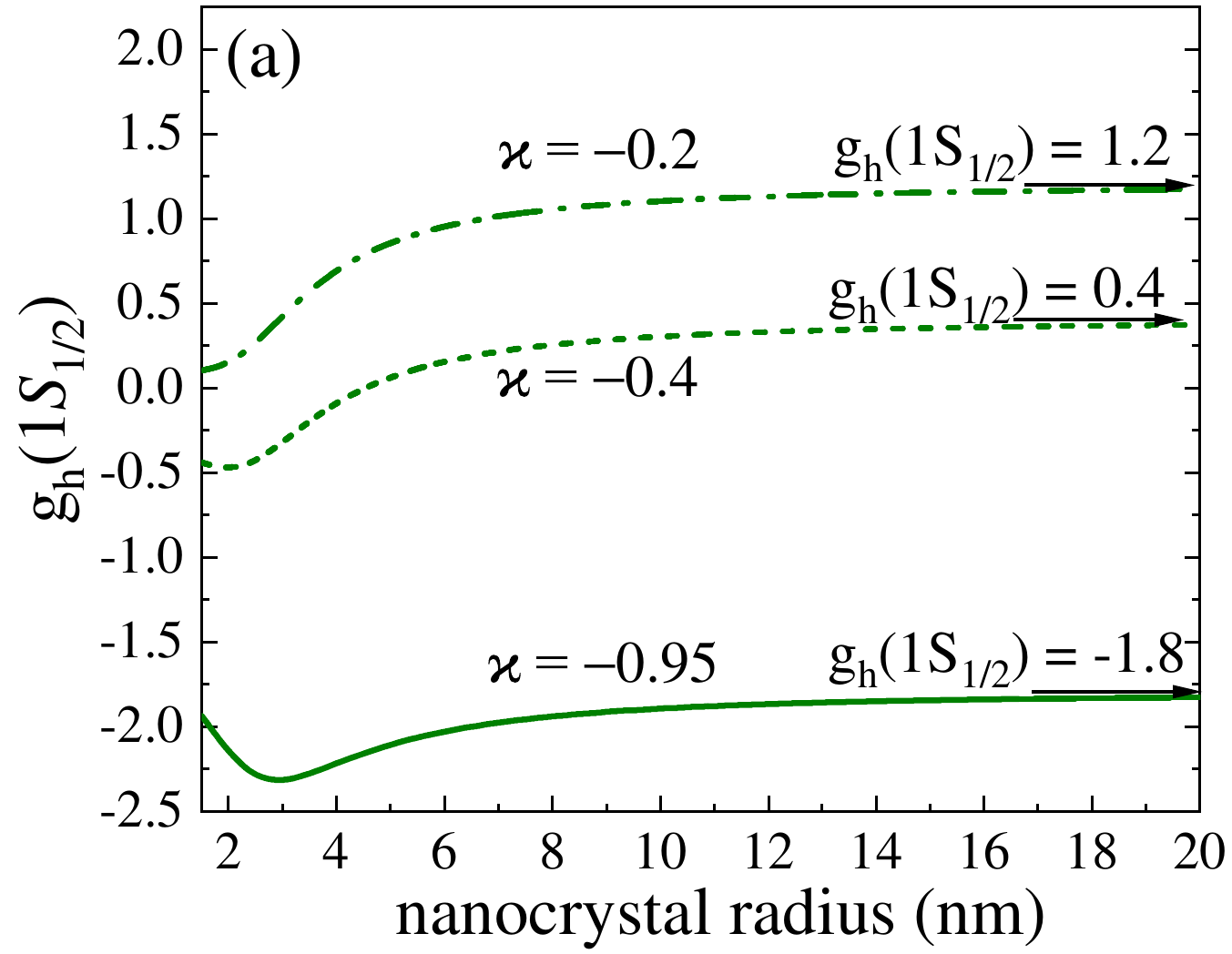}
\includegraphics[width=7cm]{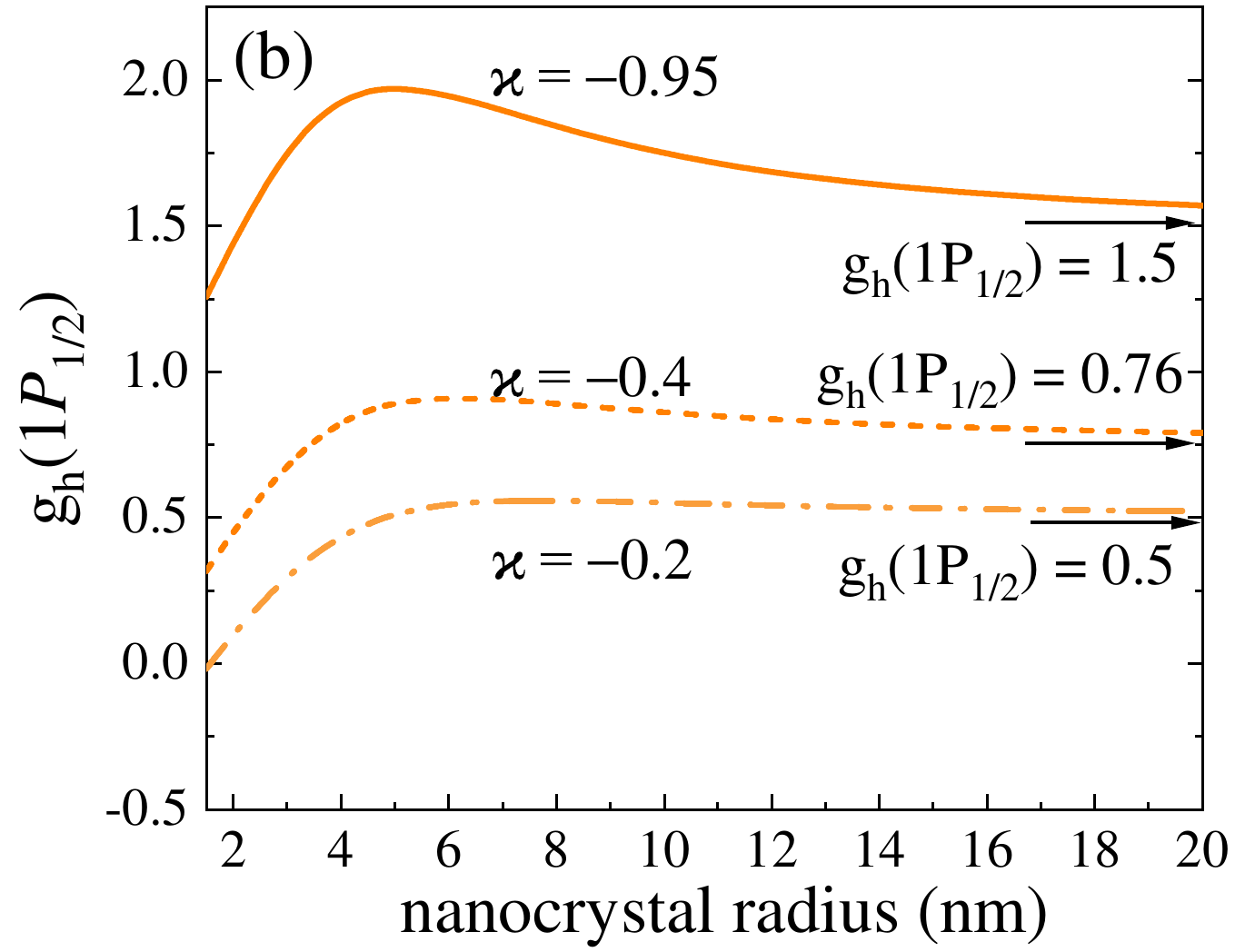}
\caption{\label{ghR} Size dependence $g_h(a)$ of the $g$-factor of the hole states $1S_{1/2}$ (a) and $1P_{1/2}$ (b) in spherical CuCl NCs for different values of $\varkappa$. In large NCs, the dependencies tend to the bulk values $g_h(1S_{1/2})=g_h^{1/2}=(2+4\varkappa)$ and $g_h(1P_{1/2})=-(2+4\varkappa-4\gamma_1)/3$.}   
\end{figure}

The magnetic Luttinger parameter $\varkappa$ is not known for CuCl. Using the relationship obtained by perturbation theory in Refs. \cite{Roth1959,Dresselhaus1955} in some approximation:
\begin{equation}
 \varkappa\approx-\frac{2}{3}-\frac{1}{3}\gamma_1+\frac{5}{3}\gamma   
\label{eq:45}
\end{equation}
we can calculate $\varkappa = -0.4$ and $g_{h}^{(1/2)} = 2+4\varkappa = 0.4 $ for CuCl. The assumptions used in the derivation  of  Eq.~\eqref{eq:45} have never been justified and it is desirable to obtain an experimentally measured value of $\varkappa$. Therefore, we peform calculations for three values: $\varkappa=-0.4$  with the corresponding $g_{h}^{(1/2)} = 0.4 $,  $\varkappa=-0.2$ ($g_{h}^{(1/2)} =  1.2$), and   $\varkappa=-0.95$ ($g_{h}^{(1/2)} = -1.8 $). The calculated size dependences $g_h(a)$ for $1S_{1/2}$ are shown in  Fig.~\ref{ghR}(a). 

One can see, that the size dependences $g_h(a)$ for the $1S_{1/2}$ hole  are similar for all values of $\varkappa$ and larger ($a\gtrsim 3$~nm) NCs. Namely, the value of $g_h(a)$ is decreasing with decreasing $a$. For the value $\varkappa=-0.95$ and the negative bulk value $g_{h}^{(1/2)} = -1.8 $, this leads to an increase of the absolute value of $|g_h(a)|$ and the size dependence resembles the experimentally measured one. However, the bulk value $g_{h}^{(1/2)} = -1.8 $  does not correspond to the values known from literature. For example, the value $g_{h} = +1.1 $ was reported in Ref. \cite{Certier1969} from fitting of magnetic field dependence of the dark exciton splitting, and in Ref. \cite{khan1970g} the value $g_{h}=+1.44$ was calculated. 

In Fig.~\ref{ghR} (b) we show the size dependences of the $g$-factor of the $1P_{1/2}$ hole state, calculated for the same three values of the magnetic Luttinger parameter $\varkappa$ as used for the $1S_{1/2}$ hole state calculation.

Note that the size dependence of the hole $g$-factor originates from the admixture of the $j_h=3/2$  ($\Gamma_8$) hole states to the $j_h=1/2$ ($\Gamma_7$) hole ground state, while the admixture of the conduction band states is not included. A similar size dependence of the $j_h=3/2$ ground state hole $g$-factor due to the admixture of the spin-orbit-split $j_h=1/2$ states in the case of $\Delta_{\rm SO}>0$ was found for CdSe and InP NCs in Ref.~\cite{Semina2021}.  In both cases, the effect is controlled by the ratio $E_h(a)/|\Delta_{\rm SO}|$, where $E_h(a)$ is the size-dependent hole ground state energy calculated from the top of the respective valence band. The size dependences of $E_h(a)/|\Delta_{\rm SO}|$ for $E_h=E(1S_{1/2}^h)$ and $E_h=E(1P_{1/2}^h)$ are shown in  Fig.~\ref{ehsp}. 

\subsubsection{Size dependence of the exciton effective $g$-factor}

For the sake of completeness, we consider here also the effect of the magnetic field on the lowest $\Gamma_7$ exciton state.  The exciton Zeeman term is the sum of the electron and hole terms: $$\widehat{H}_{Z}^{e}+\widehat{H}_{Z}^{h}=\mu_{B}B (g_e s_{ez} - g_h j_{hz}).$$

The exciton $g$-factor for the bright exciton with ${\cal F}=1$ comprising the hole $j_h=1/2$ in bulk (or $J_h=1/2$ in nanocrystals) is given by $g_{\rm X}=g_e-g_h$  for our definition of the hole $g$-factor. Importantly, here $g_h=g_h^{\rm ex}$ is the $g$-factor of the hole bound in the exciton. In the considered approximation of the exciton in the weak confinement regime, the hole state in the exciton is not affected by the external potential, so it is the same as in the bulk exciton at rest and in the exciton confined in spherical NC as a whole. 

For the relative motion of the hole in the exciton, we have neglected the difference of the $\Gamma_7$ and $\Gamma_8$ exciton binding energies. As we mentioned above, this can be justified by the smallness of $\gamma/\gamma_1^r \approx 0.09$ ($\beta^r = 0.66$) for the parameters determined. In this case, the admixture of the $d$ state with $L_h=2$ to hole wave function describing the relative motion of the $j_h=1/2$ exciton is small. However, even a small $d$-symmetry contribution results into same renormalization of the hole $g$-factor $g_h^{\rm ex}$ exciton with respect to the bulk value $g_h^{(1/2)}=2+4\varkappa$. The calculations using Eq.~\eqref{gfactor} for $g_h^{\rm ex}$ with the radial functions of the hole bound in the Coulomb potential give  $g_h^{\rm ex} \approx 0.16$ for $\varkappa=-0.4$, $g_h^{\rm ex} \approx -2$ for $\varkappa=-0.95$, and $g_h^{\rm ex} \approx 1.1$ for $\varkappa= -0.2$. As  mentioned above, the value $g_h^{\rm ex} = 1.1$ was obtained in Ref. \cite{Certier1969} from the lowest Zeeman components of the exciton with zero projection on the magnetic field, which is in fact the dark exciton state ${\cal F}=0$ mixed with the ${\cal F}_z=0$ component of the bright exciton by the magnetic field. This mixing is determined by the dark exciton $g$-factor $g_{\rm dark}=g_e+g_h^{\rm ex}$ which was estimated as $3.1$ in Ref.~\cite{Certier1969}. 

Note that there is no orbital contribution to the exciton $g$-factor related to the admixture of the orbital state ${\cal L}=2$ to the wave function of the exciton center of mass. Indeed, as ${\bm K}={\bm k}_e +\frac{e}{c\hbar}{\bm A}+{\bm k}_h-\frac{e}{c\hbar}{\bm A} = {\bm k}_e +{\bm k}_h$, it is not modified in external magnetic field because of the opposite charges of the electron and hole.

Thus, the size dependence of the exciton effective $g$-factor could arise only from the size dependence of the electron effective $g$-factor $g_e(a)$. The value of the bright exciton $g$-factor $g_{\rm X}=g_e-g_h^{\rm ex} \approx 4$ in the case $\varkappa=-0.95$. In the SFRS experiment one observed signals with $g_{\rm X}$ and $g_{\rm X}/2$ \cite{Rodina2024}. Oscillations of the ellipticity signal due to the exciton spin polarization could be observed with $g_{\rm X}/2$ \cite{Gang2022}. However, the size dependence  of $g_{\rm X}$ would not agree with the observed experimental dependence both in the SFRS and pump-probe FE. 

\section{Discussion}
\label{sec:discussion}

Let us discuss the origin of the spin signals that we have measured experimentally, at this point on the basis of both the experimental results and the comprehensive model considerations. For that, we also provide in Table~\ref{gfactreview} an overview of the literature data for the $g$-factors in CuCl bulk and NCs. Our experimental data measured on CuCl NCs with radii in the range $1.8-28$~nm are collected in Fig.~\ref{fig gfactor}. The $g$-factor values increase with reducing NC size (i.e., increasing confinement) from $1.92$ in 28~nm NCs up to $2.23$ in 1.8~nm NCs. Generally, there are several possible origins to consider: bright excitons, electrons and holes within the exciton, resident electrons and holes.

The bright excitons can be safely excluded for several reasons. As one can see in Table~\ref{gfactreview}, its $g$-factor values $g_{Z_{3}}=0.3-0.9$ are too small. Also, coherent spin dynamics are detected for delay times as large as 1~ns, which considerably exceeds the exciton lifetime.

The hole $g$-factor at the top of the valence band in bulk CuCl is in the range $g_h~(\Gamma_7)=1.1-1.44$, see Table~\ref{gfactreview}.  Our calculations for NCs presented in Fig.~\ref{ghR}(a) show that $g_{h}({1S_{1/2}})$ shows a weak dependence on NC size down to 5~nm. The bulk values are matched with $\varkappa=-0.2$, but they considerably differ from our measured data. In order to get the absolute value of $|g_{h}({1S_{1/2}})|$ closer to 2, a large $\varkappa=-0.95$ is required, which in turn would result in an unrealistically large $g_{\rm Z_3}=g_{\rm X}\approx 4$. Not to say, that $\varkappa=-0.95$ results in a negative sign of $g_{h}({1S_{1/2}})$, which contradicts the bulk data. In principle, a large, positive hole $g$-factor can be obtained for the ${1P_{1/2}}$ hole state (see Fig.~\ref{ghR}(b)), but with $\varkappa=-0.95$, which corresponds to a too large value of $g_{\rm Z_3}\approx 4$. Therefore, we conclude that the resident holes cannot be the origin of our experimental observations.  

With that finding we come to the conclusion that resident electrons are the most realistic origin of the spin signals measured in our CuCl NCs. Indeed, the $g$ factor values are very close to what is expected for $g_e$ in CuCl bulk and NCs, see Table~\ref{gfactreview} and Fig.~\ref{ge8band}. In NCs, the theoretical values are in the range $2.015-2.030$ with a small decrease with decreasing NC size, which becomes pronounced only for NCs smaller than 5~nm. By closer comparison of the experiment with theory, one meets two problems. The first one is the value of $g_e=1.92$ measured in the largest NCs at 3.194~eV, as the theoretical values for an electron on a quantum confined level cannot fall below $g_0=2.0023$. We suggest that in this case the electron may be bound on a neutral donor, for  which $g$ factor values smaller than $g_0$ were experimentally reported, see $g_{D^{0}}=1.925-1.997$ in Table~\ref{gfactreview}. Note that at the close spectral energy of 3.195~eV, $g_e=2.08$, and there is also a finite error bar for the experimentally measured values. The second problem is the small increase of the experimentally observed $g$ factor with decreasing NC size (i.e. with increasing confinement energy), despite the decreasing trend suggested by theory. We tested theoretically various mechanisms which could change this trend but could not find the one that would match experimental dependence. We suggest that there is some factor that we have not identified yet.  

\section{Conclusions} 

We have studied experimentally and theoretically the spin properties of charge carriers confined in semiconductor CuCl nanocrystals of different sizes (radius from 1.8~nm up to 28~nm) embedded in a glass matrix. Using experimental techniques involving cryogenic temperatures, strong magnetic fields, high spectral or picosecond time resolution, and resonant optical excitation, we have collected comprehensive information on the carrier Land\'e $g$-factors and the spin and population dynamics of these carriers. Coherent spin dynamics have been detected in the wide temperature range of $1.6-120$~K. Theoretically, we reexamine the level structure of exciton transitions involving two valence subbands, and have calculated its modification in NCs of various sizes. The developed theory consistently explains the increase of the exciton transition energies with decreasing NC size observed that scales experimentally stronger for the excited than for the ground states. The valence band parameters describing the quantum confined energy levels of the resident holes have been determined. A theory of electron and hole $g$-factors in CuCl NCs has been developed and model calculations with various band parameters have been made. These results have been compared with experimental data, which allows us to conclude, that the spin signals in the studied CuCl NCs are most probably provided by resident electrons.

\renewcommand{\thesection}{Appendix A}

\section{Parameters for CuCl found in literature}

\setcounter{equation}{0}
\setcounter{figure}{0}
\setcounter{table}{0}
\renewcommand{\thefigure}{A\arabic{figure}}
\renewcommand{\theequation}{A\arabic{equation}}
\renewcommand{\thetable}{A\arabic{table}}

Here, we summarize the information about the electron, hole, and exciton effective masses as well as $g$-factors for CuCl reported in literature. All values in the tables are given with the same accuracy as in their source.

\begin{table}[hbt]
\caption{Electron effective mass in $m_e/m_0$ units. }
\begin{tabular}{| l | l |l |l |}
\hline
 calculation & experiment & reference \\
\hline
$0.417$     &  &    \cite{Kleinman1979}  \\ \hline
     & $0.5$ &    \cite{Honerlage}  \\ \hline
     $0.3$&  & \cite{Ferhat} \\ \hline
     $0.35$ & & \cite{Nair2020accurate} \\ \hline 
    & $0.415$& \cite{Nikitine1967} \\ \hline

   $0.48$ & & \cite{Calabrese1973} \\ \hline 

$0.43$ &  & \cite{Ferhat1996}\\ \hline
 & $0.43$& \cite{Kato1974} \\ \hline
 & $0.43$ & \cite{Goto1973} \\ \hline
   
\end{tabular}
\label{e_masses}
\end{table}

\begin{table}[hbt]
\caption{Hole effective mass.}
\begin{tabular}{| l | l |l |l |}
\hline
 \specialcell{$m_{s}/m_0$ \\ calculation} &\specialcell{$m_{s}/m_0$ \\ experiment}  & $\Gamma_8$ holes, calculation & reference\\  \hline
 $1.5$& &\specialcell{$m_{hh}=10.8m_0$\\ $m_{lh}=0.8m_0$}& this work \\ \hline
 $1.48$  &   &\specialcell{$m_{hh}=3.1m_0$\\ $m_{lh}=0.96m_0$} &      \cite{Kleinman1979} \\ \hline
 & $2.0$ &  &  \cite{Honerlage}   \\ \hline
 $1.6$&  & &\cite{Ferhat}  \\ \hline
   $1.95$ & & &\cite{Nair2020accurate} \\ \hline 
& $3.6$ &    &\cite{Kato1974}   \\ \hline
& $4.2$& & \cite{Goto1973} \\ \hline
  & $20.4$& &\cite{Nikitine1967} \\ \hline
  13.5 & & $(3-13.7) m_0$  &\cite{khan1970g}  \\ \hline
     
\end{tabular}
\label{h_masses}
\end{table}

\begin{table}[hbt]
\caption{Exciton translation mass. }
\begin{tabular}{| l | l |l |l |}
\hline
 $M_s/m_0$ & $M_{h}/m_0$ & $M_l/m_0$ & reference  \\  \hline

 $1.9$ & $5.9$ & $1.1$ & this work   \\  \hline
 $1.9$ & $2.6$ & $1.5$ & \cite{Ekimov1985,Ekimov_jtp}   \\  \hline
 $2.3$ & $1.9$ & $0.8$ & \cite{Itoh1988}  \\  \hline
 $1.9$ & $3.7$ & $1.8$ & \cite{Chinese}  \\  \hline

\end{tabular}
\label{x_masses}
\end{table}

\begin{table*}[h!]
\begin{tabular}{|c|>{\centering\arraybackslash} m{0.07\textwidth}|>{\centering\arraybackslash} m{0.07\textwidth}|>{\centering\arraybackslash} m{0.07\textwidth}|>{\centering\arraybackslash} m{0.08\textwidth}|>{\centering\arraybackslash} 
m{0.08\textwidth}|>{\centering\arraybackslash} 
m{0.08\textwidth}|>{\centering\arraybackslash} m{0.1\textwidth}|>{\centering\arraybackslash} m{0.14\textwidth}|>{\centering\arraybackslash} m{0.15\textwidth}|}
\hline
$g_{Z_{3}}$& $g_{Z_{1,2}}$&$g_e~(\Gamma_6)$ & $g_h~(\Gamma_7)$ & $g_{A^{0}X}$&$g_{A^-}$ &$g_{D^{0}}$   & $g_{A^{0}}$ & Method &Ref. \\ \hline
&  &$2.03$     &$1.44$  &  & & & &Theory &\cite{khan1970g}\\ \hline
0.56& $-0.3$ &     &  &  & & &  & Theory & \cite{Suga1971}\\ \hline
&  &     && & & & &&\\ \hline
$0.9$&  &  $2.05$     &$1.1$& &  & & & Absorption& \cite{Certier1969}\\ \hline
0.61 &  &$2.02$    &$1.38$ & & &  & & Faraday rotation&\cite{Koda1970}\\ \hline
0.61&  $-0.4$ &     &  &  & & & & Faraday rotation & \cite{Suga1971}\\ \hline
0.447& $-0.18$  &    & & &  &   & & MCD & \cite{Nomura1993large} NCs \\ \hline
$0.28-0.34$&  &      & & &  &  & & MCD & \cite{Yasui1996} \\ \hline
&  &     & & $2.05$ & &&$-1.1~$ & Absorption&\cite{Certier1969}\\ \hline
&  &      & & $2.3$ & &  & $-1.2~$ & Absorption&\cite{Certier1969}\\ \hline
&  &    &&$2.3$ & & & & Absorption & \cite{Sauder}\\ \hline

&  &     & &  &$2.036$ && & EPR & \cite{Goltzene1972}\\ \hline
&  &      & & &$ ~2.07$ && & EPR & \cite{Goltzene1972}\\ \hline
&  &      & &   &$2.11$ &&& EPR & \cite{Goltzene1972}\\ \hline
&  &$1.998$    &&  & & & & EPR & \cite{Goltzene1975}\\ \hline
&  &    &&  & &$1.925$ & & EPR & \cite{Goltzene1975}\\ \hline
&  &    && & &$1.997$  & & EPR & \cite{Goltzene1975}\\ \hline
\end{tabular}
\caption{Collection of reported $g$-factor values. All results are for bulk CuCl, except of the Nomura 1993 data~\cite{Nomura1993large}, which are for CuCl NCs with $a=4.6$~nm size. Note that the sign of the hole $g$-factor ($g_h~(\Gamma_7)$ and also $g_{A^{0}}$) are given according to the definition used in the present paper, see Eq.~\eqref{g_h}. In the original papers \cite{khan1970g,Certier1969,Koda1970}, $g_h~(\Gamma_7)$ is negative (and $g_{A^{0}}$ is positive), as it is given in the electron representation, where the negative $g$-factor corresponds to the ground state of the electron $-1/2$ (the ground state of the hole $+1/2$). In the used electron representation, the $g$-factor of the $Z_3$ exciton is $g_{Z_3}=g_e-g_h$. In Ref.~\onlinecite{Goltzene1975}, the $g$-factor value of 1.998 was only tentatively assigned to either a conduction band electron or to an electron localized at a donor.}
\label{gfactreview}
\end{table*}

Table~\ref{gfactreview} summarizes the literature data on $g$-factors in bulk CuCl and CuCl NCs, measured by various experimental techniques. Also theory data are included. The values for the bright exciton $Z_3$ are in the range $g_{Z_{3}}=0.3-0.9$. The electron and hole $g$-factors ($g_e~(\Gamma_6)$ and $g_h~(\Gamma_7)$) were not measured directly, but rather evaluated from the exciton Zeeman splitting using $g_{Z_3}=g_e-g_h$. For that, the calculated values of $g_e=2.02-2.05$ were used. The hole $g$-factor for the top of the valence band is in the range $g_h~(\Gamma_7)=1.1-1.44$. Interestingly, the positive sign of the free hole $g$-factor changes to negative for a hole localized at a neutral acceptor $g_{A^{0}}=-(1.1-1.2)$. This is provided by the strong admixture of the higher lying valence bands. Such strong changes do not occur for electrons bound to a neutral donor, where the value $g_{D^{0}}=1.925-1.997$ is close to that of the band electron $g_e~(\Gamma_6)$. The EPR data for the negatively charged acceptors $g_{A^-}=2.036-2.11$ are close to those of those for the $g_e~(\Gamma_6)$ electron, so most probably the signals are provided by electrons here. 

\renewcommand{\thesection}{Appendix B}
\section{Equations for the hole energy levels in CuCl NCs}

\label{Ap:eqs}

\setcounter{figure}{0}
\setcounter{equation}{0}
\setcounter{table}{0}
\renewcommand{\thefigure}{B\arabic{figure}}
\renewcommand{\theequation}{B\arabic{equation}}
\renewcommand{\thetable}{B\arabic{table}}

Here we provide the equations for determining the size dependent hole energy levels which we use for the calculations in our paper. The equations are adopted from Ref. \cite{Ekimov1993} using the notations of the present work. 

To find the hole energy levels in the six-band model, we introduce the positive hole energy $\epsilon$ calculated from the top of the $\Gamma_7$ valence band and increasing with the increasing confinement (decreasing NC radius $a$). The top of the $\Gamma_8$ valence band corresponds to $\epsilon=|\Delta_{\rm SO}|$. The heavy, light, and spin-orbit split hole wave vectors can be found at the given  $\epsilon$ from \cite{Ekimov1993}:

\begin{eqnarray}
  && k_h^2=\frac{2m_0(\epsilon-|\Delta_{\text{SO}}|)}{\hbar^2(\gamma_1-2\gamma)}=\frac{2m_0\tilde{\epsilon}}{\hbar^2(\gamma_1-2\gamma)} , \\
&&k_{l,s}^2=\frac{m_0 \left(2\epsilon \left(\gamma_1 +\gamma
   \right) -\gamma_1 |\Delta_{\text{SO}}| \right)}{\hbar^2\left(\gamma_1 -2\gamma \right) \left( \gamma_1 +4\gamma
   _1\right) }\mp\\
   &&\mp\frac{m_0 }{\hbar^2\left(\gamma _1-2 \gamma \right) \left( \gamma _1+4 \gamma\right)
  }\{(2 \epsilon (\gamma_1 +\gamma)
   -\gamma_1 |\Delta_{\text{SO}}|)^2-  \nonumber \\
   &&-4 \epsilon(\epsilon-|\Delta_{\text{SO}}|)\left(\gamma _1-2 \gamma
\right) \left(\gamma_1 +4\gamma \right)\}^{1/2}. \nonumber
\end{eqnarray}

Note that for the energies between the tops of the $\Gamma_7$ and $\Gamma_8$ subbands, $\epsilon<|\Delta_{\rm SO}|$ (${\tilde{\epsilon}=\epsilon-|\Delta_{\rm SO}|<0}$), the wave vectors of the light and heavy holes are imaginary ($k_h^2<0$ and $k_l^2<0$). 

To obtain the hole energy levels for the $S_{1/2}$ states, one has to solve the following equations:
\begin{multline} \label{eq:s12}
   F_{l}(\epsilon) \frac{j_2\left(a k_l\right) j_0\left(a k_s\right)}{k_l^2}-F_{s}(\epsilon)\frac{j_0\left(a k_l\right)
   j_2\left(a k_s\right) }{k_s^2 }=0 \, ,
\end{multline}
where $$F_{l,s}(\epsilon)=\frac{\hbar^2\gamma_1}{2m_0}  k_{l,s}^2 -\epsilon.$$

The approximate energy of the $1S_{1/2}$ state, starting from the top of the $\Gamma_7$ valence band, can be found from \cite{Ekimov_jtp}:
\be\label{Sapr}
E_{1S_{1/2}}(a) \approx \frac{\hbar^2 \pi^2}{2m_sa^2} -\frac{2}{|\Delta_{\rm SO}| }\frac{\gamma^2}{\gamma_1^2} \left(\frac{\hbar^2 \pi^2}{m_sa^2} \right)^2\left(1- \frac{6}{\pi^2}\right) \, .
   \ee
	
For the $P_{1/2}$ states the dispersion equation reads:
\begin{equation}
    j_1(a k_l)j_1(a k_s)=0,
\end{equation}
and the exact expressions for the energy levels can be obtained analytically:\begin{multline}
E_{nP_{1/2}}=  \frac{|\Delta_{\rm SO}|}{2}+\frac{\hbar^2 \phi_{1,n}^2}{2m_0a^2}(\gamma_1+\gamma)\pm\\ \pm\frac{\hbar^2|\Delta_{\rm SO}|}{2m_0 a^2}\sqrt{\left(\frac{m_0a^2}{\hbar^2}\right)^2+\frac{2m_0a^2\gamma\phi_{1,n}^2}{\hbar^2|\Delta_{\rm SO}|}+\frac{9\gamma^2\phi_{1,n}^4}{\Delta_{\rm SO}^2}}, \label{Penergy}
\end{multline}
with the "-" and "+" for levels symbolizing the $\Gamma_7$ and $\Gamma_8$ subbands, respectively, where $\phi_{1,n}\approx 4.49$ is the $n$-th root of the spherical Bessel function $j_1(r)$.
In the limit case $\hbar^2/(2m_0a^2) \ll |\Delta_{\rm SO}|$, the energy of the $1P_{1/2}$ state arising from the $\Gamma_7$ subband can be found approximately as
\be\label{Papr}
E_{1P_{1/2}} \approx \frac{\hbar^2 \phi_{1,1}^2}{2m_sa^2} -\frac{2}{|\Delta_{\rm SO}| }\frac{\gamma^2}{\gamma_1^2} \left(\frac{\hbar^2 \phi_{1,1}^2}{m_sa^2} \right)^2 \, ,
\ee
where $\phi_{1,1}\approx 4.49$. 

In the limit of a large spin-orbit splitting of the valence band, $|\Delta_{\text{SO}}|\rightarrow \infty $ or $\gamma/\gamma_1\rightarrow  0$ ($\beta \rightarrow 1$), corresponding to noninteracting $\Gamma_7$ and $\Gamma_8$ valence subbands, we have 
\bea
\label{e0sp}
&&E_{1S_{1/2}} \approx E_{1S_{1/2}}^0=\frac{\hbar^2 \pi^2}{2m_sa^2} ,  \,  {\rm and } \\
&&E_{1P_{1/2}} \approx=E_{1P_{1/2}}^0= \frac{\hbar^2 \phi_{1,1}^2}{2m_sa^2} . \nonumber
\eea

\begin{figure*}[h!]
\begin{center}
\includegraphics[width=7.2cm]{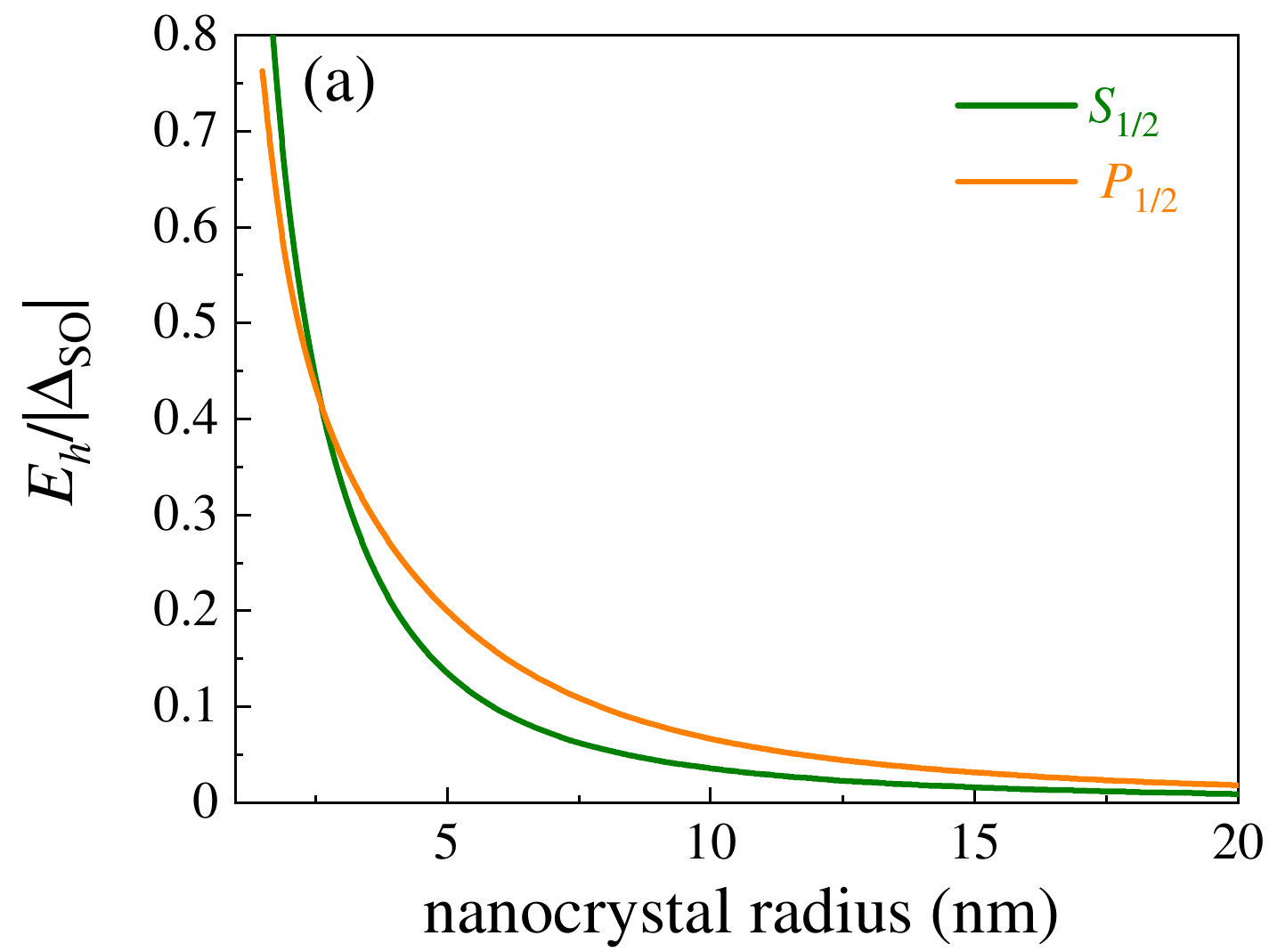}\includegraphics[width=7cm]{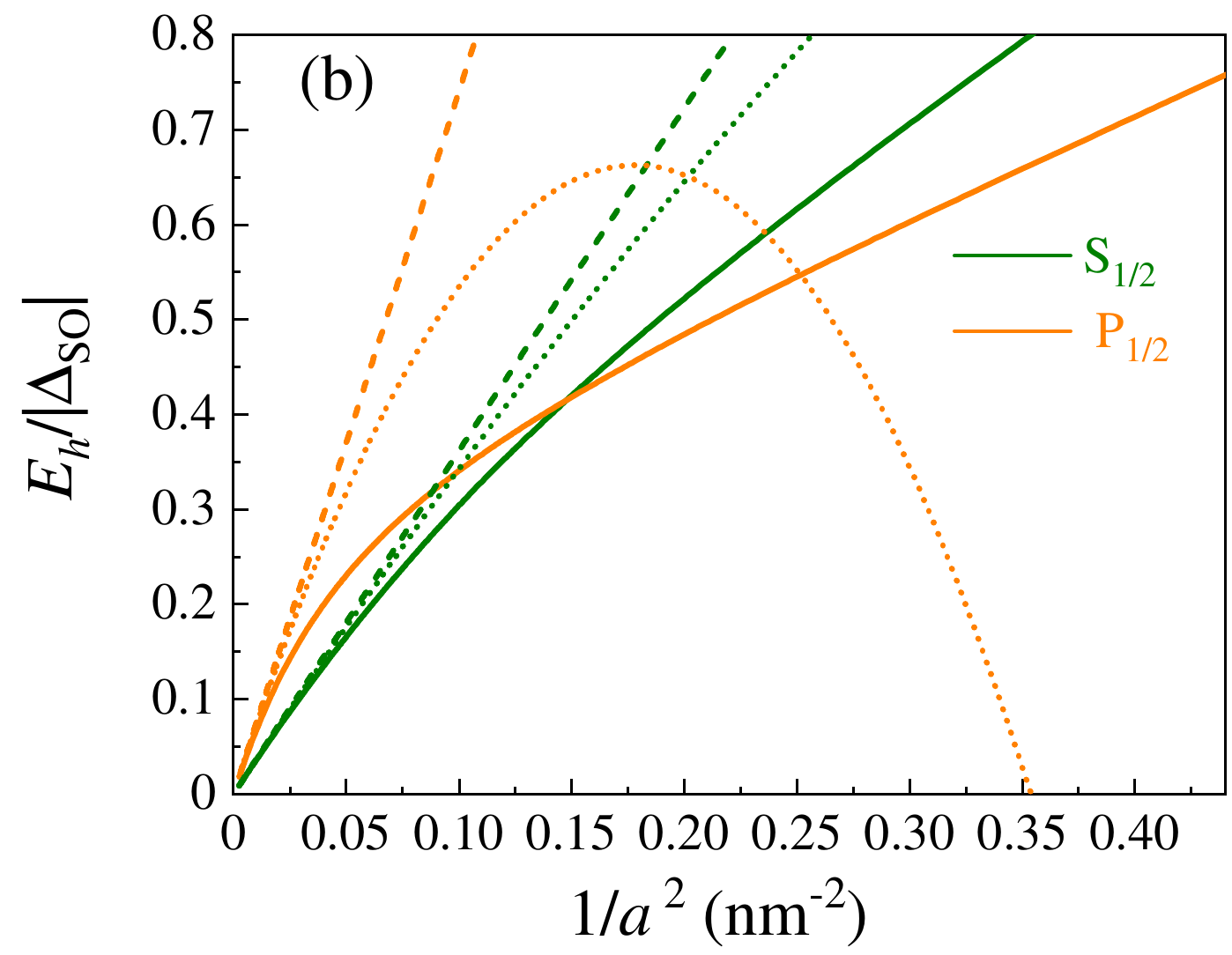}
\includegraphics[width=7.2cm]{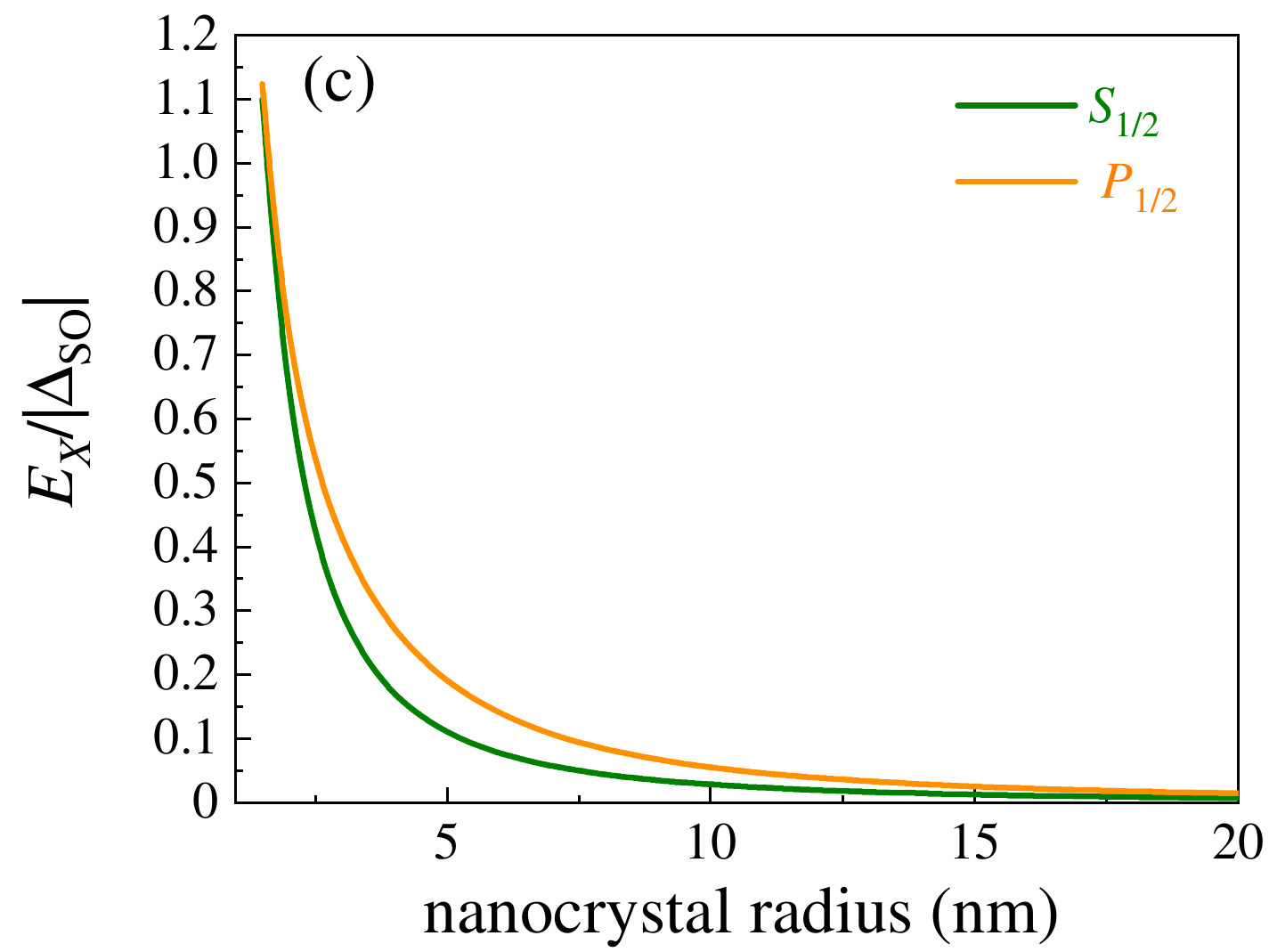}\includegraphics[width=7cm]{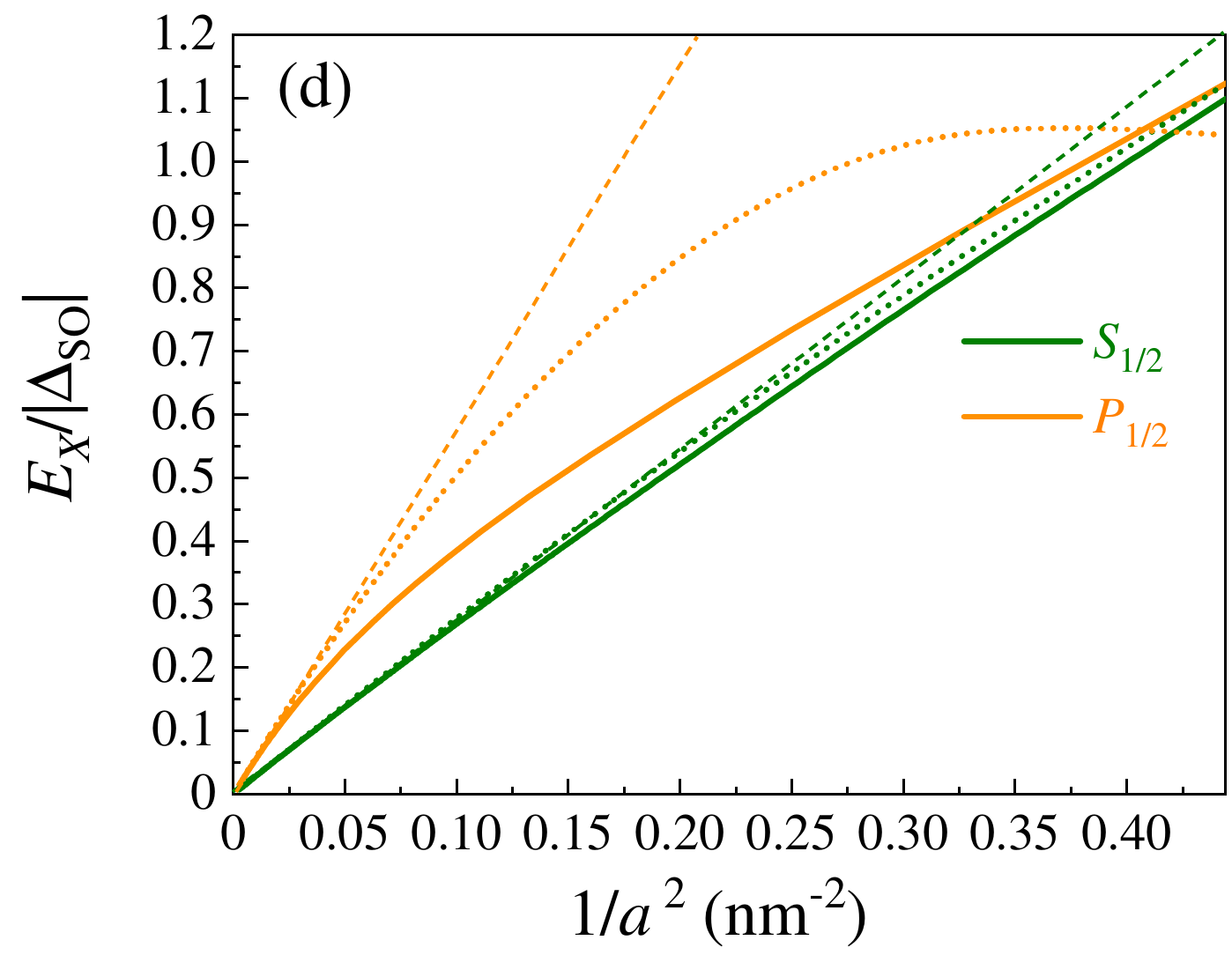}
\caption{\label{ehsp} Quantum confinement energies $E_h$ of the first two hole states $1S_{1/2}$ (green) and $1P_{1/2}$ (orange) in spherical CuCl NCs in units of $|\Delta_{\rm SO}|$ as function of the NC radius $a$ (a) and of $1/a^2$ (b). Quantum confinement energies $E_X$ of the first two exciton states $1{\cal S}_{1/2}$ (green) and $1{\cal P}_{1/2}$ (orange) in spherical CuCl NC in units of $|\Delta_{\rm SO}|$ as function of the NC radius $a$ (c) and of $1/a^2$ (d). The Luttinger parameters for the hole are $\gamma_1=0.67$ and $\gamma=0.29$. The Luttinger parameters for the exciton are $\gamma_1^{\rm ex}=0.53$ and $\gamma^{\rm ex}=0.18$. Solid lines correspond to numerically (for $1S_{1/2}$ and $1{\cal S}_{1/2}$) and analytically by Eq. \eqref{Penergy} (for $1P_{1/2}$ and $1{\cal P}_{1/2}$) calculated energies. Dashed lines in (b) and (d) correspond to Eq.~\eqref{e0sp}, dotted lines  in (b) and (d) correspond to  Eqs.~\eqref{Sapr} and \eqref{Papr}.} 
\end{center}
\end{figure*}

\begin{figure*}[h!]
\begin{center}
\includegraphics[width=12cm]{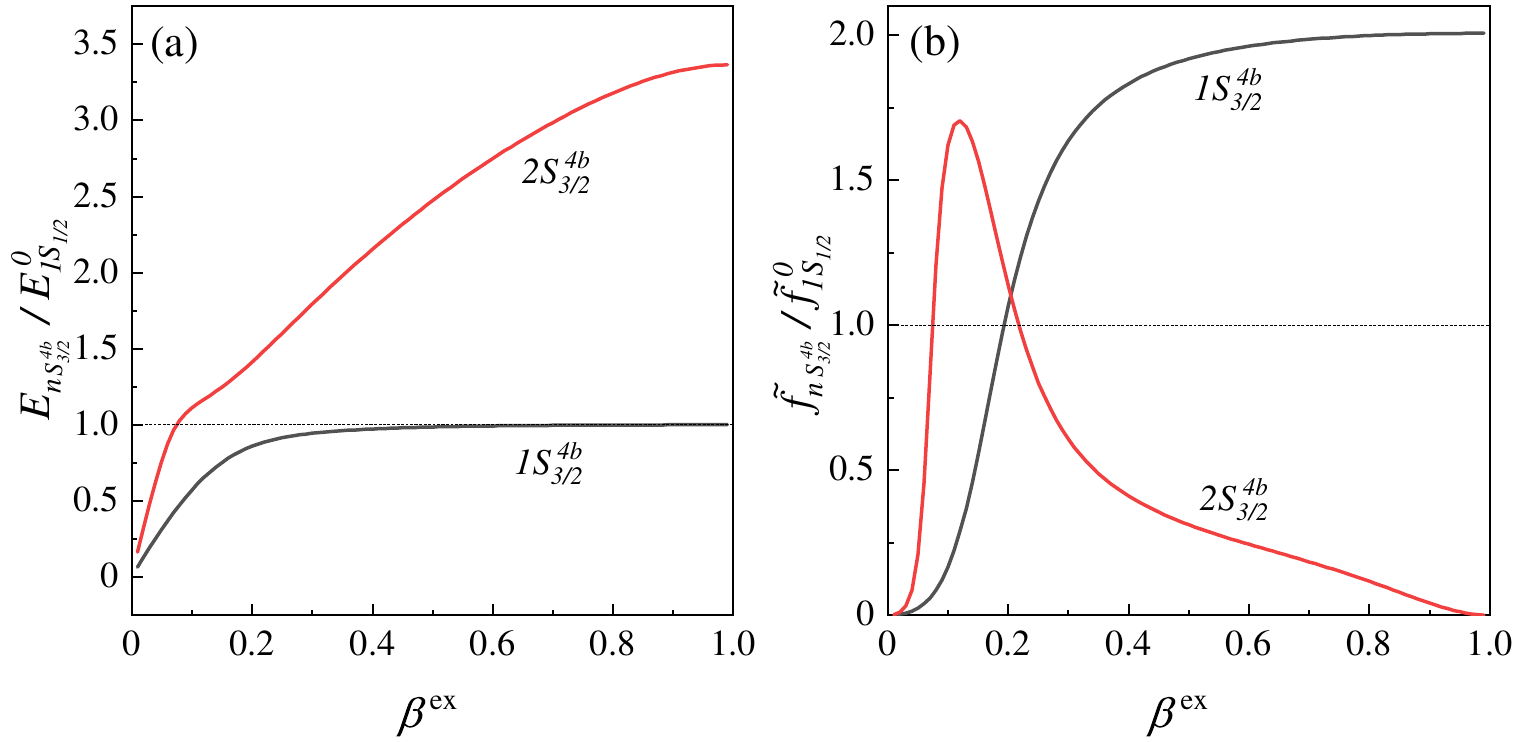}
\caption{\label{ebeta} (a) The ratio of the quantum confinement exciton energies $E_{n{\cal S}_{3/2}^{\rm 4b}}/E_{1{\cal S}_{1/2}}^0$ (a) and of the relative oscillator strengths ${\tilde f}_{n{\cal S}_{3/2}^{\rm 4b}}/{\tilde f}_{1{\cal S}_{1/2}}^0$ (b) as function of the exciton light-hole to heavy hole mass ratio $\beta^{\rm ex}=(\gamma_1^{\rm ex} - 2\gamma^{\rm ex})/(\gamma_1^{\rm ex} + 2\gamma^{\rm ex})=M_l/M_h$ for $n=1$ (black lines) and $n=2$ (red lines).} 
\end{center}
\end{figure*}

The dispersion equation for the $S_{3/2}$ state reads:
\bea
&&2m_0\tilde{\epsilon}/\hbar^2(k_s^{-2}-k_l^{-2})j_2(k_la)j_0(k_ha)j_2(k_sa)  \\&&-\frac{(\gamma_1+2\gamma)k_s^2-2m_0\tilde{ \epsilon}/\hbar^2}{k_s^2}j_2(k_s a)j_0(k_la)j_2(k_h a) \nonumber \\ &&+\frac{(\gamma_1+2\gamma)k_l^2-2m_0\tilde{\epsilon}/\hbar^2}{k_l^2}j_0(k_sa)j_2(k_l a)j_2(k_h a)=0. \nonumber
\eea
If we neglect the mixing between the $\Gamma_7$ and $\Gamma_8$ bands, the dispersion equation for the $S_{3/2}^{\rm 4b}$ states  within the four-band model is given by:
\begin{align}
j_0(k_la)j_2(k_ha)+j_2(k_la)j_0(k_ha)=0 \,.
\end{align}
If, additionally, $\beta=1$, we obtain $E_{S_{3/2}}^{\rm 4b}=E_{S_{1/2}}^0$. 

The dispersion equation for the $P_{3/2}$ states reads:
\bea
&&2m_0\tilde{\epsilon}/\hbar^2(k_s^{-2}-k_l^{-2})j_1(k_la)j_3(k_ha)j_1(k_sa)  \\&&-9\frac{(\gamma_1+2\gamma)k_s^2-2m_0\tilde{ \epsilon}/\hbar^2}{k_s^2}j_3(k_s a)j_1(k_la)j_1(k_h a) \nonumber\\ &&+9\frac{(\gamma_1+2\gamma)k_l^2-2m_0\tilde{\epsilon}/\hbar^2}{k_l^2}j_3(k_sa)j_1(k_l a)j_1(k_h a)=0. \nonumber
\eea

All the expressions above are valid also for calculation of the center of mass exciton confined states after the replacements $\gamma_1 \rightarrow \gamma_1^{\rm ex}$ and $\gamma \rightarrow \gamma^{\rm ex}$. 

The size depedences of the  energy levels $E_h$ for the hole $1S_{1/2}$, $1P_{1/2}$ states and $E_{X}$ for the $1{\cal S}_{1/2}$, $1{\cal P}_{1/2}$  exciton states in $\Delta_{\rm SO}$ units, calculated for the determined in the text parameters of CuCl, are shown in Figure \ref{ehsp}. For comparison, the results of the calculations using approximate analytical expressions are also shown. One can see that the analytical expressions can be used only for large NCs where $E_h,E_X \ll \Delta_{\rm SO}$. The approximate expression describes the exciton states (Fig. \ref{ehsp}(d))  better than the hole states (Fig. \ref{ehsp}(b)) because of the larger value of the effective mass ratio $\beta^{\rm ex} > \beta$.  

We present here also  explicit expressions for the  wave functions of the $nS_{1/2}$ and $nP_{1/2}$ states arising from the $\Gamma_7$ valence band. The even wave function $\Psi^{h,S}_{J_h=1/2,J_{hz}=\pm 1/2}$ (see Eq. \eqref{PsiJJ})  can be written as follows \cite{Ekimov1985,Grigoryan1990}: 
\begin{widetext}
 \begin{equation}\label{Psi_plus}
\Psi_{1/2,+1/2}^{h,S}=\left(
\begin{array}{c}
 \frac{1}{\sqrt{10}}R_2(r)Y_{2-1}(\theta,\varphi)  \\
-\frac{1}{\sqrt{5}}R_2(r)Y_{20}(\theta,\varphi) \\
 \sqrt{\frac{3}{10}}R_2(r)Y_{21}(\theta,\varphi)\\
  -\sqrt{\frac{2}{5}}R_2(r)Y_{22}(\theta,\varphi) \\
R_s(r)Y_{00}(\theta,\varphi)\\
0\\
\end{array}
\right), \quad 
\Psi_{1/2,-1/2}^{h,S}=\left(
\begin{array}{c}
 \sqrt{\frac{2}{5}}R_2(r)Y_{2-2}(\theta,\varphi)  \\
-\sqrt{\frac{3}{10}}R_2(r)Y_{2-1}(\theta,\varphi) \\
 \frac{1}{\sqrt{5}}R_2(r)Y_{20}(\theta,\varphi)\\
  -\frac{1}{\sqrt{10}}R_2(r)Y_{21}(\theta,\varphi) \\
0\\
R_s(r)Y_{00}(\theta,\varphi)\\
\end{array}
\right).
\end{equation}
\newline \newline
Here, the radial wave functions $R_s(r)\equiv R_0^{J_h=1/2,j_h=1/2}(r_h)$ and $R_2(r)\equiv R_2^{J_h=1/2,j_h=3/2}(r_h)$ are given by:
\bea\label{Rs}
  &&R_s(r)= N_s^{1/2} \left[j_0\left(r
  k_s(\epsilon )\right) -  \frac{j_0\left(a k_s(\epsilon )\right)}{j_0\left(a
   k_l(\epsilon )\right)} j_0\left(r k_l(\epsilon )\right)\right] , \, \,  {\rm and} \\ 
  &&R_2(r)= N_s^{1/2} \left[\frac{j_2\left(r k_s(\epsilon )\right) \left(F_{l}(\epsilon) -|\Delta_{\rm SO}|
  \right)}{ \frac{\sqrt{2}\hbar^2 \gamma}{m_0}  k_s^2(\epsilon
   )}-\frac{j_0\left(a k_s(\epsilon )\right)}{j_0\left(a
   k_l(\epsilon )\right)} \frac{j_2\left(r
  k_l(\epsilon )\right) \left(F_{s}(\epsilon)-|\Delta_{\rm SO}|
   \right)}{ \frac{\sqrt{2}\hbar^2 \gamma}{m_0}  k_l^2(\epsilon
   )}\right]~, 
\eea
where  $N_s$ is determined from the the normalization condition 
$\int_0^\infty \left[R_s^2(r)+R_2^2(r)\right] r^2 \, dr=1$. 

The odd wave functions   $\Psi^h_{J_h=1/2,J_{hz}=\pm 1/2}$, see Eq.~\eqref{PsiJJ},  can be written explicitly as: 
 \begin{equation}\label{Psip_plus}
\Psi_{1/2,+1/2}^{h,P}=\left(
\begin{array}{c}
 i\frac{1}{\sqrt{2}}R_1(r)Y_{1-1}(\theta,\varphi)  \\
-i\frac{1}{\sqrt{3}}R_1(r)Y_{10}(\theta,\varphi) \\
 i\sqrt{\frac{1}{6}}R_1(r)Y_{11}(\theta,\varphi)\\
  0 \\
-i\sqrt{\frac{1}{3}}R_{s1}(r)Y_{10}(\theta,\varphi)\\
i\sqrt{\frac{2}{3}}R_{s1}(r)Y_{11}(\theta,\varphi)\\
\end{array}
\right), \quad 
\Psi_{1/2,-1/2}^{h,P}=\left(
\begin{array}{c}
0 \\
i\frac{1}{\sqrt{6}}R_1(r)Y_{1-1}(\theta,\varphi) \\
 -i\sqrt{\frac{1}{3}}R_1(r)Y_{10}(\theta,\varphi)\\
  i\sqrt{\frac{1}{2}}R_1(r)Y_{11}(\theta,\varphi)\\
-i\sqrt{\frac{2}{3}}R_{s1}(r)Y_{1-1}(\theta,\varphi)\\
i\sqrt{\frac{1}{3}}R_{s1}(r)Y_{10}(\theta,\varphi)\\
\end{array}
\right).
\end{equation}
Here, the radial wave  functions $R_{s1}(r)\equiv R_1^{J_h=1/2,j_h=1/2}(r_h)$ and $R_1(r)\equiv R_1^{J_h=1/2,j_h=3/2}(r_h)$   are given by \cite{Grigoryan1990}
\begin{equation}\label{Rs}
  R_{s1}(r)= N_p^{1/2}\sqrt{\frac{1}{8}}\left( \frac{2 m_0\tilde{\epsilon}/\hbar^2-(\gamma_1+2\gamma)k_s^2}{\gamma k_s^2}j_1(r k_s) - \frac{j_1\left(a k_s\right)}{j_1\left(a
   k_l\right)} \frac{2 m_0\tilde{\epsilon}/\hbar^2-(\gamma_1+2\gamma)k_l^2}{\gamma k_l^2}j_1(r k_l)\right),  \, \, {\rm and}
\end{equation}
\begin{equation}\label{R2}
  R_1(r)= N_p^{1/2}\left(j_1(r k_s) - \frac{j_1\left(a k_s\right)}{j_1\left(a
   k_l\right)}j_1(r k_l)\right)  .
\end{equation}
and satisfy the normalization condition 
$\int_0^\infty \left[R_{s1}^2(r)+R_1^2(r)\right] r^2 \, dr=1$.
\end{widetext}

\renewcommand{\thesection}{Appendix C}

\section{The ${k}{p}$ model for the electron $g$-factor}\label{AC}

\setcounter{figure}{0}
\setcounter{equation}{0}
\setcounter{table}{0}
\renewcommand{\thefigure}{C\arabic{figure}}
\renewcommand{\theequation}{C\arabic{equation}}
\renewcommand{\thetable}{C\arabic{table}}

The general expression for the electron $g$ factor within second-order ${\bm k}{\bm p}$ perturbation theory reads:
\begin{widetext}
\be \label{Roth}
g_e = g_0 \left(1 - \frac{i}{m_0} \sum_{n}^\infty \frac{\langle u^c_{1/2,\pm 1/2}| \hat p_x | u_{n}\rangle \langle u_{n}|\hat p_y| u^c_{1/2,\pm 1/2}\rangle  - \langle u^c_{1/2,\pm 1/2}| \hat p_y | u_{n}\rangle \langle u_{n}|\hat p_x| u^c_{1/2,\pm 1/2}\rangle }{E_{\Gamma_6} - E_{n}}\right) \, .
\ee
\end{widetext}
Here the $u_n$ and $E_n$ are the Bloch functions and extremal energies of all energy bands interacting with the conduction band (having nonzero matrix elements of the components $\hat p_\alpha$ ($\alpha=x,y,z$) of the momentum operator $\hat {\bm p} = - i\hbar {\bm \nabla}$ acting on the Bloch functions).

The contribution of the $p$-type bands nearest to the $s$-type conduction band to the electron effective mass and $g$-factor are given by:
  \bea \label{rb}
		g_{\rm rb}&=&g_{\rm rb}'- \frac{2E_p'}{3}\left(\frac{1}{E_{\Gamma_8^c}} - \frac{1}{E_{\Gamma_7^c}}\right) ,~\\
\alpha_{\rm rb}&=&\alpha_{\rm rb}'- \frac{E_p'}{3}\left(\frac{2}{E_{\Gamma_8^c}} + \frac{1}{E_{\Gamma_7^c}}\right) \,.
\eea
Here $\Gamma_8^c$ and $\Gamma_7^c$ are the energies of the bottom of the $p$ symmetry conduction bands split by the spin-orbit interaction with energy $\Delta_{\rm SO}'=E_{\Gamma_7^c}-E_{\Gamma_8^c}$. $E_{g}'=E_{\Gamma_8^c}-E_{\Gamma_6}$ for $\Delta_{\rm SO}'>0$ and $E_{g}'=E_{\Gamma_7^c}-E_{\Gamma_6}$ for $\Delta_{\rm SO}'<0$,  where $E_p'$ is the Kane matrix element describing the interaction between the $s$-type and $p$-type conduction bands.  

To estimate the unknown parameters of the eight-band and fourteen-band models we take the parameters $m_e=0.4 m_0$, $g_e=2.03$, $E_g=3.4$eV, $\Delta_{\rm SO}=-0.07$ eV, $E_g'=7$ eV, and assume that there is no contribution from other remote bands $g_{\rm rb}'=\alpha_{\rm rb}'=0$. We chose four values of $\alpha_{\rm rb}<0$ and list all estimated parameters in Table~\ref{tab:Ep}.  
	\begin{table}[h!] 
		\caption{Parameters of the eight-band and fourteen-band ${\bm k}{\bm p}$ models for CuCl, corresponding to $m_e=0.4 m_e$ and $g_e=2.03$ with the fixed values $E_g=3.4$  eV, $\Delta_{\rm SO}=-0.07$ eV, $E_g'=7$ eV, and $\alpha_{\rm rb}'=g_{\rm rb}'=0$.  }
		\begin{tabular}{| l | l |l |l |l |}
			\hline
			 $\alpha_{\rm rb}$ &$E_p$ (eV)  & $g_{\rm rb}$ & $E_p'$ &  $\Delta_{\rm SO}'$  \\  \hline
						 $-0.1$ & $5.5$ & $0.006$ &  $0.65$&   $-0.61$ \\  \hline
			 $-0.5$ & $6.9$ & $0.0004$ &  $3.5$ &   $-0.008$ \\  \hline
			$-1.0$ &$8.6$ &  $-0.006$ & $7.0$ & $0.07$   \\  \hline
  		$-1.5$ & 	$10.3$ &  $-0.013$ & $10.5$  & $0.09$ \\ \hline
			
		\end{tabular}
		\label{tab:Ep}
	\end{table}

To take into account the quantum-confined energy levels of the electron in the valence band in spherical NCs we reformulate the eight-band ${\bm k}{\bm p}$ model for the energy dependent electron $g$-factor and translate it into the size-dependence as follows:
\begin{widetext}
\begin{eqnarray}\label{gkp}
g_e(a) = &&g_0+g_{\rm rb}- \frac{2i}{m_0} \sum_{{n;J_h=3/2,1/2;J_{hz}=\pm 3/2,\pm1/2 }} \\
&&\frac{\langle \Psi^c_{1/2,\pm 1/2}| \hat p_x | \Psi^{v,n}_{J_h,J_{hz}}\rangle \langle \Psi^{v,n}_{J_h,J_{hz}}|\hat p_y| \Psi^c_{1/2,\pm 1/2}\rangle  - \langle \Psi^c_{1/2,\pm 1/2}| \hat p_y | \Psi^{v,n}_{J_h,J_{hz}}\rangle \langle \Psi^{v,n}_{J_h,J_{hz}}|\hat p_y| \Psi^c_{1/2,\pm 1/2}\rangle  }{E_e(a) - E_{nJ_h}^{v}(a)} \, . \nonumber
\end{eqnarray}
\end{widetext}
Here $E_e(a) = E_0^{e}=E_{1S_e}$ is the energy of the $1S_e$ electron state, the energies $E_{nJ_h}^{v}(a)=-E_g-E_{nJ_h}^{h}(a)$ and the wave functions of the valence band $\Psi^{v,n}_{J_h,J_{hz}}$ should be taken in the electron representation. 

The matrix elements of the momentum operator are given by:
\bea
&&\langle \Psi^c_{1/2,\pm 1/2}| \hat p_\alpha | \Psi^{v,n}_{J_h,J_{hz}}\rangle = \\
&&\frac{(-1)^{j_h-\mu}\delta_{J_h,j_h}\delta_{J_{hz},\mu}}{\sqrt{2j_h+1}}
  \langle u^c_{1/2,\pm 1/2}| \hat p_\alpha | u_{j_h,\mu}^v \rangle {\cal I}_{nJ_h}
   \,  , \nonumber 
\eea
where $\langle u^c_{1/2,\pm 1/2}| \hat p_\alpha | u_{j_h,\mu}^v \rangle $ are the band-edge matrix elements and 
\be
 {\cal I}_{nJ_h} = \int_0^a r^2 dr R_0^c(r) R_{0,n}^{J_h,J_{h}}(r)
\ee
is the electron-hole overlap integral between the radial wave function of the $1S_e$ electron state $R_0^c(r)$ and the radial wave function with $L_h=0$ of the $nS_{J_h}$ hole state, see Eq.~\eqref{PsiJJ}, $R_{0,n}^{J_h,J_{h}}(r)$. Note that in case of the simple band model, or $\gamma=0$, we obtain ${\cal I}_{nJ_h}=\delta_{n,1}$. The interaction of the $n=1$ electron with excited hole states becomes allowed when the complex structure of the valence band is considered \cite{Ekimov1993}.  Then the condition 
$
\sum_{n=1}^\infty |{\cal I}_{nJ_h}|^2 = 1$
holds. 

The size dependence of ${\tilde g}_e(a)$ arises not only from the electron energy $E_0^e(a)$, but also from the size dependence of the valence band levels $E_{nJ_h}^{h}(a)$ and can be written as
\begin{widetext}
\begin{eqnarray} \label{gcah}
   g_e(a)=g_0+g_{\rm rb} -\frac{2E_p}{3}\sum_{n=1}^\infty \left(\frac{|{\cal I}_{n3/2}|^2}{E_g+|\Delta_{\rm SO}|+E_0^e(a)+E_{n3/2}(a)} - \frac{|{\cal I}_{n1/2}|^2}{E_g+E_0^e(a)+E_{n1/2}(a)} \right) \, .
\end{eqnarray}
\end{widetext}
Note that the energies $E_{n3/2} = E_{n3/2}^{h} - |\Delta_{\rm SO}|$ and  $E_{n1/2}= E_{n1/2}^{h}$ in Eq. \eqref{gcah} are calculated from the top of the $\Gamma_8$ and $\Gamma_7$ valence bands, respectively, as only the states with $J_h=j_h$ may have $I_{nJ_h} \ne 0$. Assuming that only levels with $E_{nJ_h}^h \ll E_g$ are giving a noticeable contribution to the sum, the energies $E_{n3/2}^{h}$ and overlap integrals $I_{n3/2}$ in Eq. \eqref{gcah} can be replaced by $E_{n3/2}^{\rm 4b}$ and the overlap integrals $I_{n3/2}^{\rm 4b}$, calculated within the four-band model so that Eq. \eqref{gcah} can be simplified to 
\begin{widetext}
\begin{eqnarray} \label{gcahav}
  g_e(a)\approx &&g_e + E_0^e(a) \alpha_{\rm so}(E_0^e(a))+ \frac{2E_p}{3}\left(  \frac{E_{3/2}(a)}{(E_g+E_0^e(a)+|\Delta_{\rm SO}|)^2} -\frac{E_{1/2}(a) }{(E_g+E_0^e(a))^2} \right) \nonumber  \\
  \approx && g_e + (E_0^e(a)+E_0^h(a)) \alpha_{\rm so}(E_0^e(a)+E_0^h(a)) \, . 
\end{eqnarray}
\end{widetext}
Here $E_{1/2}(a) = \sum_{n=1}^\infty E_{n1/2}^h |{\cal I}_{n1/2}|^2 $ and 
$E_{3/2}(a) = \sum_{n=1}^\infty E_{n3/2}^h |{\cal I}_{n3/2}|^2 = \sum_{n=1}^\infty E_{n3/2}^{\rm 4b} |{\cal I}_{n3/2}^{\rm 4b}|^2 $
are the averaged energies in the valence band. Direct calculations show that $E_{1/2}(a)=E_{3/2}(a)=E_{1S_{1/2}}^0(a)=E_{1S_{3/2}}^0(a)=E_0^h(a)$  are the same for the $\Gamma_7$ and $\Gamma_8$ valence bands. For this reason, accounting for the quantum confined levels in the valence band does not help to reverse the size dependence of the electron $g$-factor. \newline \newline 

\textbf{AUTHOR INFORMATION}

{\bf Corresponding Authors} \\

Dmitri R. Yakovlev,  Email: dmitri.yakovlev@tu-dortmund.de\\
Anna V.~Rodina,  Email: anna.rodina@mail.ioffe.ru   \\

\textbf{ORCID}\\
Dennis~Kudlacik:          0000-0001-5473-8383 \\
Evgeny~A.~Zhukov:         0000-0003-0695-0093\\
Dmitri R. Yakovlev:       0000-0001-7349-2745 \\
Gang~Qiang:               0000-0003-2241-0409  \\
Marina A.~Semina          0000-0003-3796-2329 \\
Aleksandr A.~Golovatenko  0000-0003-2248-3157 \\
Anna V.~Rodina           0000-0002-8581-8340  \\ 
Alexander L.~Efros:       0000-0003-1938-553X \\
Manfred~Bayer:            0000-0002-0893-5949 \\

{\bf Notes}\\
The authors declare no competing financial interests.

\textbf{Acknowledgments.}
The theoretical work of M.~A.~S., A.~A.~G., and A.~V.~R. on the the size-dependent exciton optical transitions and carrier $g$-factors was supported by the Russian Science Foundation  (Grant No. 23-12-00300). Al.~L.~E. acknowledges  support  of  the US Office of Naval Research (ONR) through the Naval Research Laboratory’s Basic Research Program.

\end{document}


\newcount\timehh  \newcount\timemm
\timehh=\time \divide\timehh by 60
\timemm=\time
\count255=\timehh\multiply\count255 by -60 \advance\timemm by \count255
\title{Supplementary Information: \\"Land\'e $g$-factors and spin dynamics of charge carriers \\in CuCl nanocrystals in a glass matrix"}

\author{Dennis~Kudlacik$^{1}$,  Evgeny~A.~Zhukov$^{1,2}$, Dmitri~R.~Yakovlev$^{1,2}$,  Gang~Qiang$^{1}$,  \\ Marina~A.~Semina$^{2}$, Aleksandr A.~Golovatenko$^{2}$, Anna~V.~Rodina$^{2}$, \\ Alexander L. Efros$^{3}$,  Alexey~I. Ekimov$^{4}$,   and Manfred~Bayer$^{1}$}

\affiliation{$^{1}$Experimentelle Physik 2, Technische Universit\"at Dortmund, 44221 Dortmund, Germany}
\affiliation{$^{2}$Ioffe Institute, Russian Academy of Sciences, 194021 St. Petersburg, Russia}
\affiliation{$^{3}$Naval Research Laboratory, Washington, DC 20375, USA}
\affiliation{$^{4}$Nanocrystals Technology Inc., New York, NY, USA }

\maketitle
\crefalias{sectiom}{supp}

\onecolumngrid

\setcounter{equation}{0}
\setcounter{figure}{0}
\setcounter{table}{0}
\setcounter{page}{1}
\setcounter{section}{0}
\makeatletter
\renewcommand{\thepage}{S\arabic{page}}
\renewcommand{\theequation}{S\arabic{equation}}
\renewcommand{\thefigure}{S\arabic{figure}}
\renewcommand{\thetable}{S\arabic{table}}
\renewcommand{\thesection}{S\arabic{section}}
\renewcommand{\bibnumfmt}[1]{[S#1]}
\renewcommand{\citenumfont}[1]{S#1}

\section{Photoluminescence spectra}
\label{sec:SI_PL}

Photoluminescence (PL) spectra of the studied CuCl NCs are measured under continuous wave and pulsed laser excitation in order to provide an optical characterization of the samples. The two excitation regimes allow us to obtain spectra for varying excitation density. In both cases, the measured PL is time-integrated, as it is detected after dispersions with an 0.5 meter spectrometer using a silicon charge-coupled-device camera cooled by liquid nitrogen. The samples are placed in a helium cryostat and measured at temperatures of 1.9~K or 4.2~K.  

\begin{figure*}[hbt]
\begin{center}
\includegraphics[width=16cm]{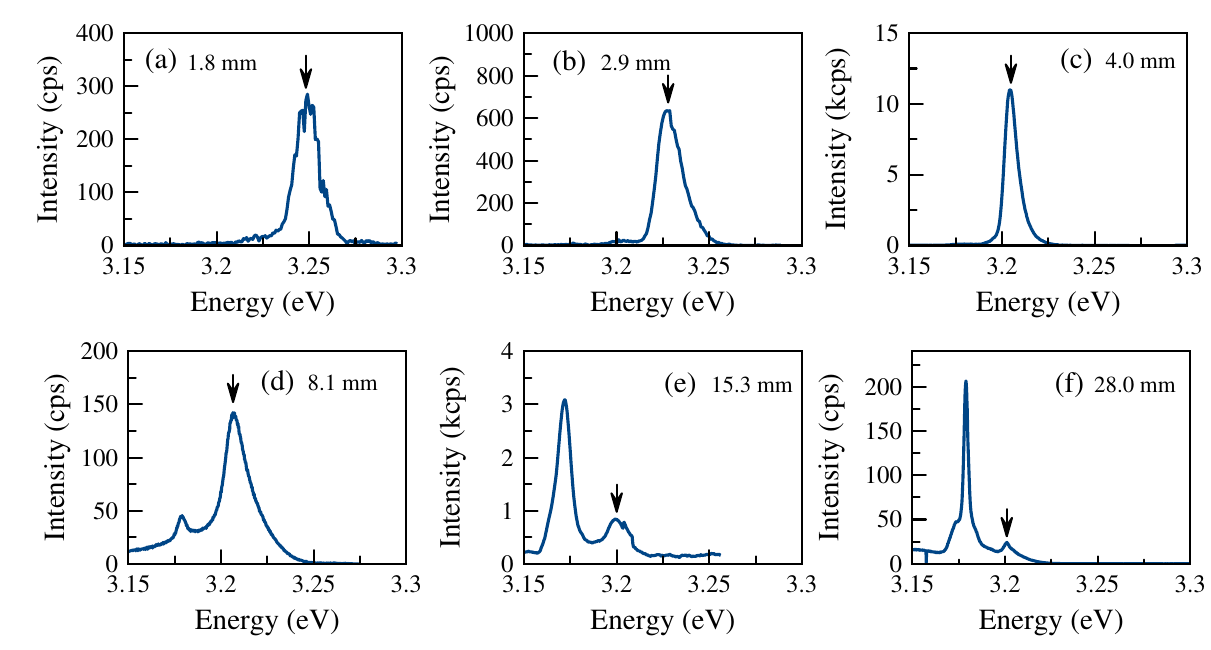}
\caption{Photoluminescence spectra of CuCl NCs of various sizes measured under continuous wave excitation at $E_{exc} = 3.306$~eV excitation for $T = 1.9$~K. The arrows mark the exciton emission line.
}
\label{fig PL}
\end{center}
\end{figure*}

For continuous wave excitation, a laser with photon energy of $E_{exc} = 3.306$~eV is used. The laser spot diameter on the sample is about 360~$\mu$m, resulting in an excitation density of $3$~W/cm$^2$ for $0.7$~mW excitation power. The PL spectra are shown in Fig.~\ref{fig PL}. With the 'X' label we mark the line assigned to the Z$_3$ exciton emission.  One can see that it exhibits a significant high energy shift with decreasing NC radius, which is provided by the quantum confinement effect. The energies of the X lines in PL are in good agreement with the lowest lying peaks of the Z$_3$ exciton in absorption, see Table I.

\begin{figure*}[hbt]
\begin{center}
\includegraphics[width=14cm]{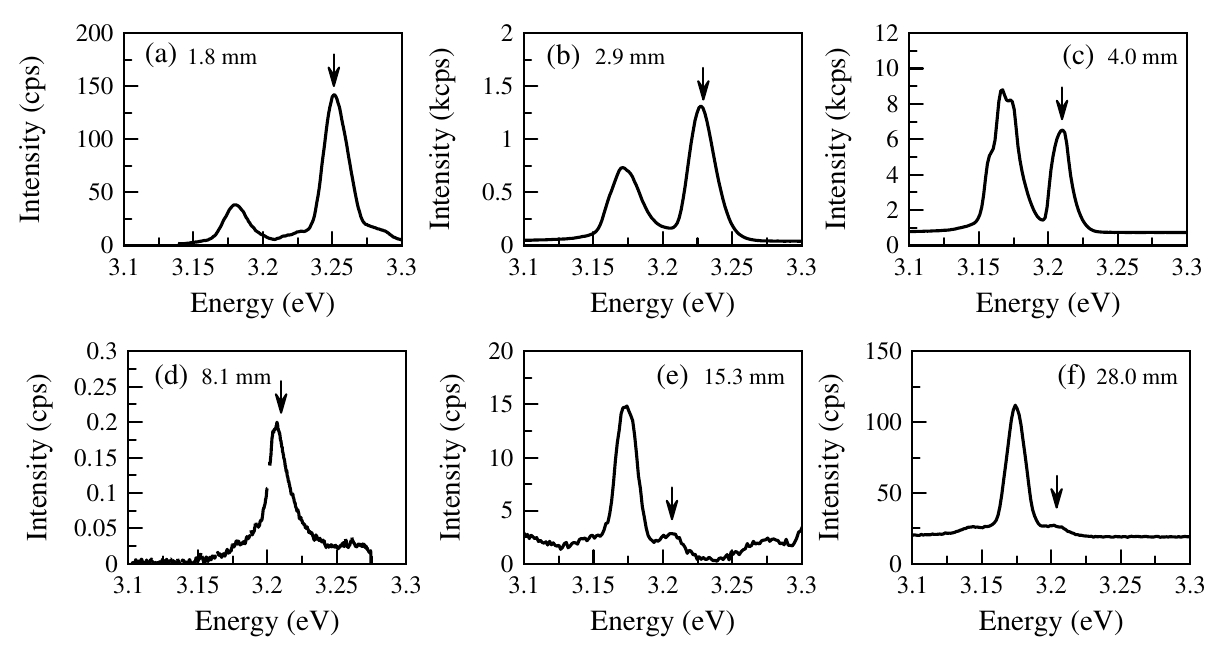}
\caption[]{\label{fig_PLall_4K} Time-integrated photoluminescence spectra of CuCl NCs of various sizes measured for pulsed excitation at $E_{exc} = 3.493$~eV photon energy for $T = 4.2$~K. The arrows mark the exciton emission line-}
\end{center}
\end{figure*}

For pulsed excitation, a laser with photon energy of $E_{exc} = 3.492$~eV (355 nm) is used. The pulse duration is 2~ns and the repetition frequency is 1 kHz. The laser spot diameter on the sample is about 300~$\mu$m and the excitation power is 170~$\mu$W at the sample surface. The excitation density for pulsed excitation is $4-5$ orders of magnitude larger than for continuous wave excitation. Therefore, in the PL spectra in Fig.~\ref{fig_PLall_4K}, additionally to the exciton X line one can see lines shifted to lower energies, which can be assigned to biexcitons and bound exciton complexes. Their details are beyond the scope of our study. 

\clearpage

\section{Spin-flip Raman scattering on $\bf 2.9~nm$ NCs}
\label{sec:SI_SFRS}

\begin{figure*}[hbt]
\begin{center}
\includegraphics[width=16cm]{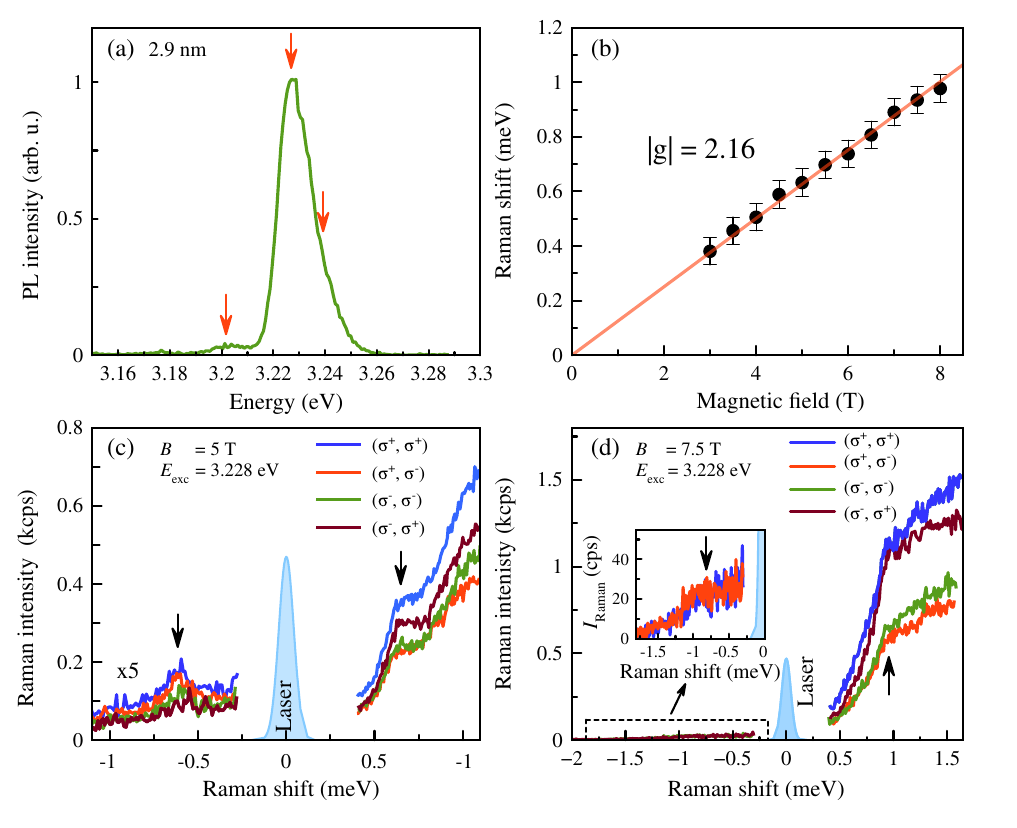}
\caption{\label{fig4}(a) Photoluminescence spectrum of 2.9~nm CuCl NCs measured at $T = 1.9~$K for $B=0~$T. The laser energies used for recording the spin-flip Raman spectra are indicated by the red arrows. (b) Magnetic field dependence of the Raman shift for the Stokes shifted line measured in the co-polarized configuration ($\sigma^+$, $\sigma^+$). $T = 1.9~$K, $E_{exc} = 3.228~$eV, and $P_{exc} = 2~$mW. Red line is a linear fit to the data with $|g|=2.16$. (c,d) Spin-flip Raman spectra in various cross- and co-polarized configurations at 5~T and 7.5~T. For better visibility in (c) the anti-Stokes spectra are multiplied by a factor of five.}
\end{center}
\end{figure*}

Results of spin-flip Raman scattering on 2.9~nm NCs are collected in Fig.~\ref{fig4}. Tuning the laser excitation into resonance with the exciton PL at 3.228~eV, one can monitor Raman signals in both the Stokes (positive energy shifts) and anti-Stokes (negative energy shifts) spectral ranges, see Fig.~\ref{fig4}(c). The spectra are shown for $B=5$~T in various cross- and co-polarization configurations. One can see that the polarization selection rules are not very pronounced here, which is a common feature for the measured CuCl NCs. The spin-flip lines, which are marked by the arrows, appear on top of the background signal of resonantly excited photoluminescence. Both the Stokes and anti-Stokes lines are shifted form the laser line by about 0.66~meV. The anti-Stokes signal is almost 5 times weaker than the Stokes one. With increasing magnetic field to 7.5~T the spin-flip line increases its energy separation, but their intensities become weaker and they become less pronounced, especially in the anti-Stokes range, Fig.~\ref{fig4}(d). We can follow the Stokes shifted line in the magnetic field range of $3-8$~T. It shifts linearly with magnetic field and can be fitted with  $|g|=2.16\pm 0.05$ without demonstrating an offset at zero field, see Fig.~\ref{fig4}(b). 

\begin{figure*}[hbt]
\begin{center}
\includegraphics[width=17.2cm]{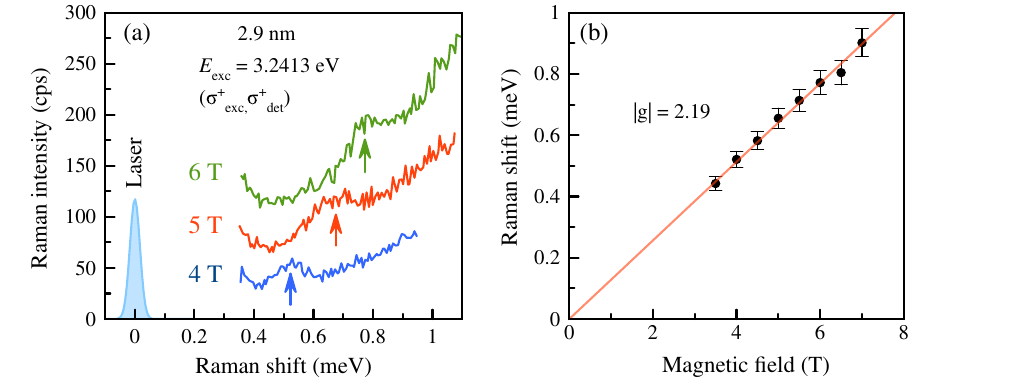}
\caption{\label{fig6} (a) Spin-flip Raman spectra of the 2.9~nm CuCl NCs for excitation on the high energy flank of the exciton emission line ($E_{exc} = 3.241~$eV), measured at $B = 4$, $5$, and $6$~T in the co-polarized configuration ($\sigma_{exc}^+$, $\sigma_{det}^+$). $T = 1.9~$K, $P_{exc} = 1.1~$mW. The spectra are shifted vertically for clarity. The maxima of the spin-flip lines are marked by the arrows. The laser at zero Raman shift is shown by the cyan filled profile. (b) Magnetic field dependence of the Raman shift. Red line is a linear fit to the data with the function $\Delta E = |g| \mu_B B $, giving $|g|=2.19$. }
\end{center}
\end{figure*}

When the excitation laser is tuned to the high-energy side of the PL line at 3.241~eV, minor differences are observed in the Raman shift and line width, Fig.~\ref{fig6}(a). However, the Raman intensity decreases significantly in comparison to excitation at the PL maxima. The $g$-factor evaluated from the Raman shift with magnetic field (Fig.~\ref{fig6}(b)) increases slightly at this spectral energy to $|g|=2.19\pm 0.05$ 

\begin{figure*}[hbt]
\begin{center}
\includegraphics[width=15cm]{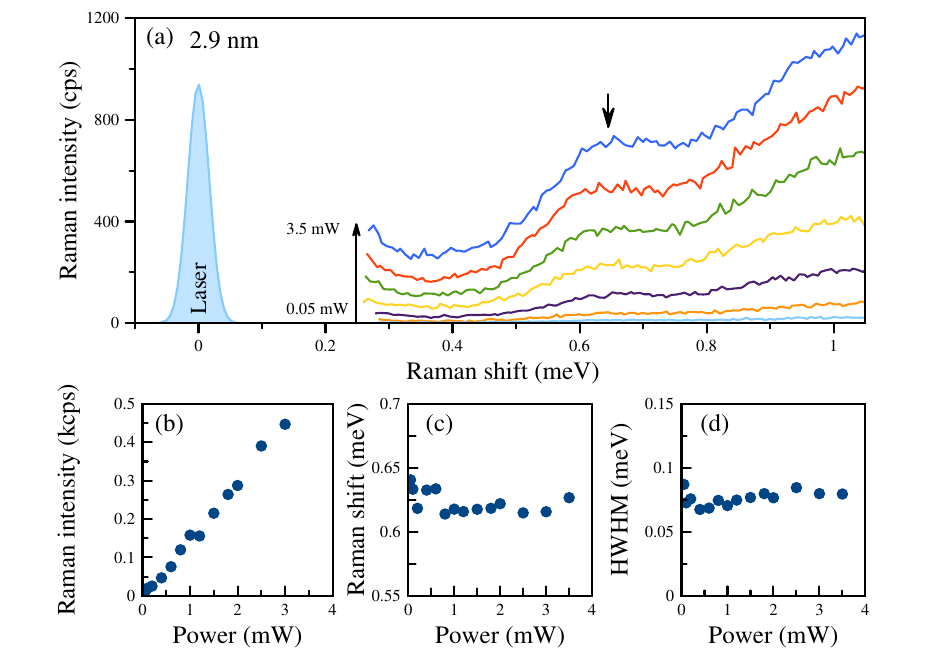}
\caption{\label{fig5} Effect of excitation power on SFRS spectra of 2.9 nm CuCl NCs. $T = 1.9$~K, $B = 5$~T, $E_{exc} = 3.228$~eV. (a) Raman spectra for varying laser power measured in the co-polarized configuration ($\sigma^+$, $\sigma^+$). (b)-(d) Excitation power dependences of the Raman intensity, Raman shift, and half width at half maximum (HWHM). }
\end{center}
\end{figure*}

Photochaging effects in NCs typically depend on the excitation density and demonstrate a nonlinear response to changing this density, while the respective nonlinear regimes depend strongly on the charging mechanisms and the dynamics of the charging and neutralization processes. In order to search for corresponding indications, we measure SFRS spectra on the 2.9~nm NCs for excitation powers varied in range of $0.05-3.5$~mW, see Fig.~\ref{fig5}~(a). The spectra are measured in the co-polarized configuration ($\sigma^+$, $\sigma^+$) at $T = 1.9~$K, for $B = 5$~T using excitation at the PL maximum at $E_{exc} = 3.228~$eV. One can see, that in this power range the spectral shape of the Raman lines is maintained. The Raman shift and linewidth of the SFRS line remain constant, see Figs.~\ref{fig5}~(c,d). The signal intensity increases linearly with power, see Fig.~\ref{fig5}~(b). From that we conclude that the NC photocharging is not modified for these excitation powers.

\section{Spin-flip Raman scattering in $\bf 28~nm$ NCs}

\begin{figure*}[t!]
\begin{center}
\includegraphics[width=16cm]{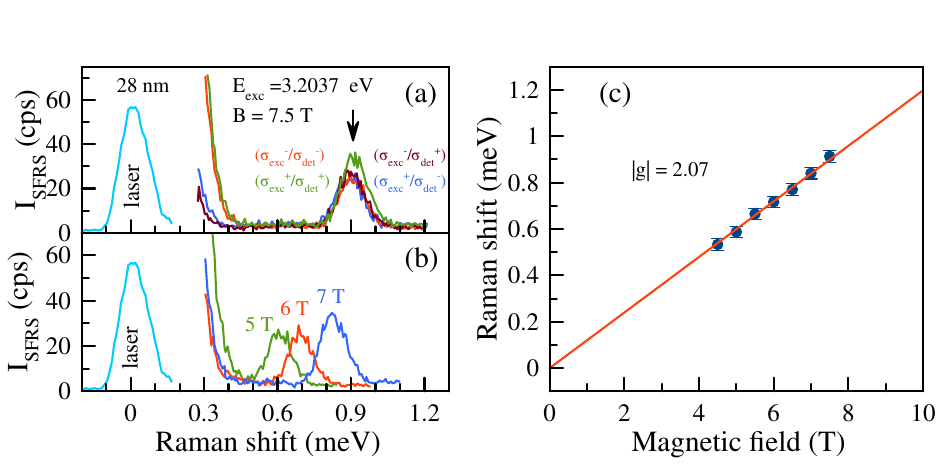}
\caption{SRFS on 28~nm CuCl NCs measured at $T = 1.9~$K, $E_{exc} = 3.204~$eV, $P_{exc} = 2~$mW. (a) Raman spectra in different cross- and co-polarized configurations at $B = 7.5~$T. (b) Raman spectra in the co-polarized configuration at various magnetic fields. (c) Magnetic field dependence of the Raman shift in the Stokes range. Red line shows a linear fit with $|g|=2.16$.}
\label{fig7}
\end{center}
\end{figure*}

Spin-flip Raman scattering spectra taken on the 28~nm CuCl NCs are shown in Fig.~\ref{fig7}. Here, pronounced spin-flip lines are observed in the anti-Stokes spectral range, see Figs.~\ref{fig7}(a,b). For the Stoles range they are less pronounced on the strong PL background (not shown here). Similar to the 2.9~nm NCs shown in Fig.~\ref{fig4}(c), the spin-flip signals show no sensitivity to the polarization selection rules, see Fig.~\ref{fig7}(a).  The shift of the spin-flip line has a linear dependence on magnetic field strength which can be fitted with $|g|=2.07$, see Fig.~\ref{fig7}(c). 

\clearpage

\section{Spin dynamics in $\bf CuCl$ NCs}
\label{sec:SI_TR}

In Figure~\ref{Fig:4-2,9 nm} the dynamics of FE and $\Delta T/T$ are shown for the 4.0~nm and 2.9~nm NCs. Similar to the 1.8~nm NCs, amplitude oscillations related to spin precession in magnetic field are not visible in the FE dynamics. Also the shape of the dynamics sdoes not depend on magnitude and direction of the magnetic field. In the 4.0~nm NCs, the spin relaxation dynamics have two components with $21$~ps and $60$~ps decay time. The decay of the differential transmission signal (population relaxation) shows also two decay times of $21$~ps and $73$~ps  [Fig.~\ref{Fig:4-2,9 nm}(c)].

The dynamics in the 2.9~nm NCs [Figs.~\ref{Fig:4-2,9 nm}(d,e,f)] differ somewhat from the results obtained for the other samples. In this sample, there is an additional long-lived component in the FE and $\Delta T/T$ signals. In the first case its decay time is 430~ps, and in the second case it is 6~ns.

\begin{figure*}[hbt!]
\begin{center}
\includegraphics[width=16cm]{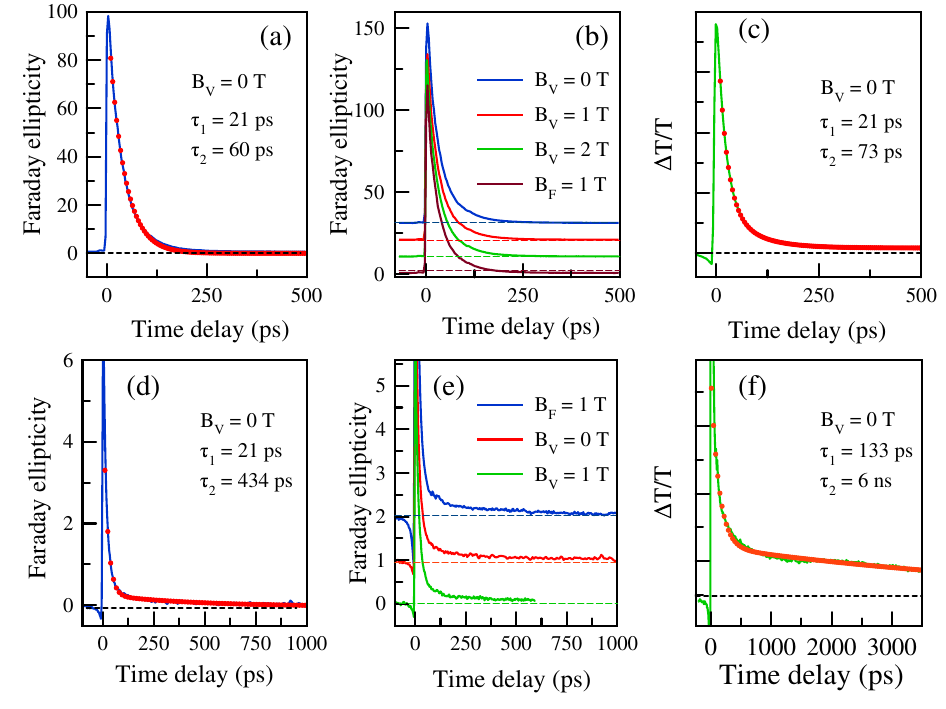}
\caption{Relaxation dynamics of carriers in the 4.0~nm and 2.9~nm NCs measured at  $T=1.7$~K. (a) FE dynamics at $B_V = 0$~T (blue line) and its fit with Eq.~(4) (red circles). (b) FE dynamics at various magnetic fields $B_V$ and $B_F$ in Voigt and Faraday configuration, respectively, shifted vertically relative to each other. (c) Dynamics of differential transmission at zero magnetic field (green line) and fit using Eq.~(4) (red circles). Panels (a, b, c) are for the 4.0~nm NCs measured at $E_{pump}=3.202$~eV. Panels (d, e, f) are for the 2.9~nm NCs measured at $E_{pump}=3.228$~eV. }
\label{Fig:4-2,9 nm}
\end{center}
\end{figure*}

\begin{figure*}[hbt!]
\begin{center}
\includegraphics[width=16cm]{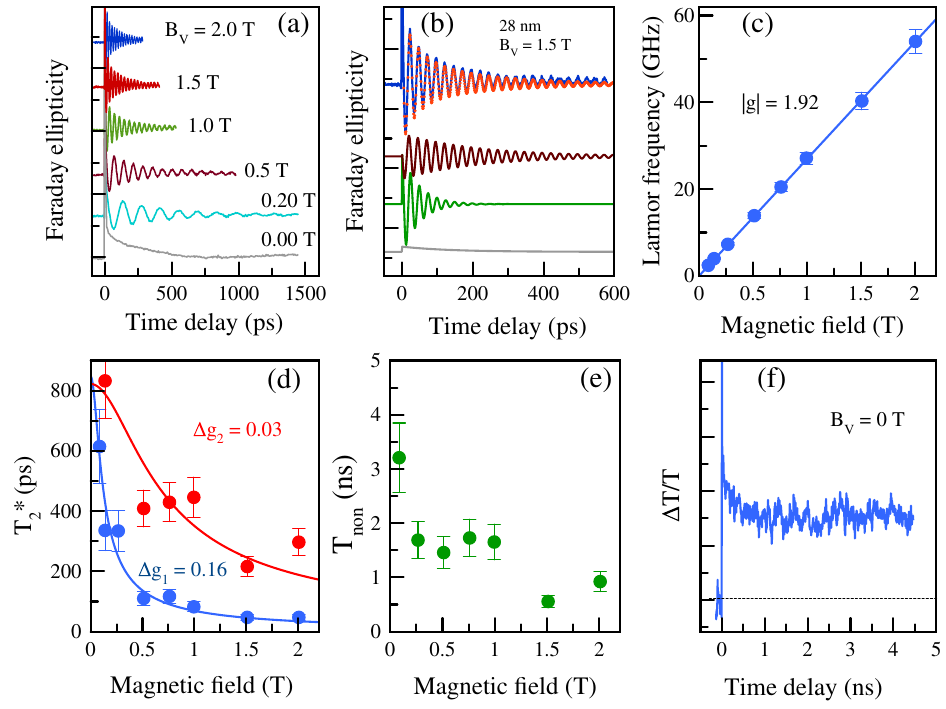}
\caption{Spin dynamics of carriers in the 28~nm NCs measured at $T=1.7$~K for $E_{pump} =3.194$~eV. (a) FE dynamics at various magnetic fields $B_{\rm V}$. (b) FE dynamics at $B_{\rm V} = 0.5$~T (blue line) and fit of this curve using Eq.~(5) (red circles). The components contributing to the fit are shown in brown, green and grey color. They are shifted vertically for better visibility. (c) Dependence of Larmor precession frequency  on magnetic field and linear fit of these data with $|g|=1.92$. (d) Dependence of spin dephasing time on magnetic field and fit using Eq.~(5) for the short-lived (blue) and long-lived (red) oscillating components. (e) Dependence of decay time of the nonoscillating component on magnetic field. (f) Dynamics of differential transmission at zero magnetic field. }
\label{Fig:28 nm}
\end{center}
\end{figure*}

The dynamics of FE and $\Delta T/T$ measured for various magnetic fields $B_{\rm V}$ on the 28~nm NCs are shown in Fig.~\ref{Fig:28 nm} and for the 8.1~nm NCs in Fig.~\ref{Fig:8,1 nm}.

\begin{figure*}[hbt!]
\begin{center}
\includegraphics[width=16cm]{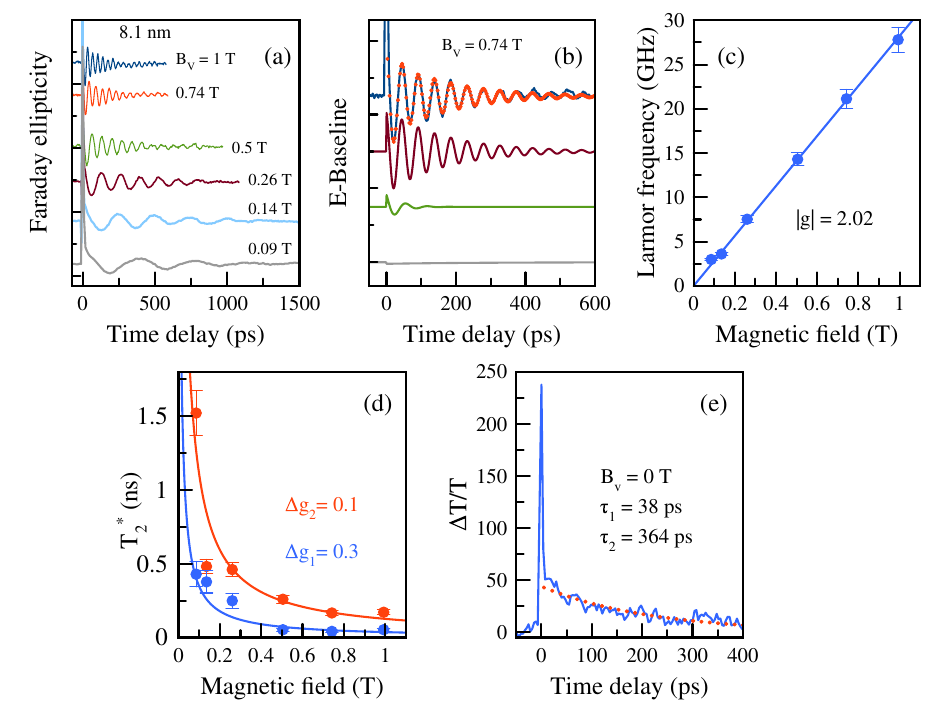}
\caption{Spin dynamics of carriers in the 8.1~nm NCs measured at $T=1.7$~K fpor $E_{pump} =3.200$~eV. (a) FE dynamics at various magnetic fields $B_{\rm V}$. (b) FE dynamics at $B_{\rm V} = 0.74$~T (blue line) and fit of this curve using Eq.~(5) (red circles). The components contributing to the fit are shown in brown, green and grey color.  They are shifted vertically for better visibility. (c) Dependence of Larmor precession frequency on magnetic field and linear fit of these data with $|g|=2.02$. (d) Dependence of spin dephasing time on magnetic field and fit using Eq.~(5) for the short-lived (blue) and long-lived (red) oscillating components. (e) Dynamics of differential transmission at zero magnetic field.  Red dotted line is fit with the decay times of 38~ps and 364~ps.}
\label{Fig:8,1 nm}
\end{center}
\end{figure*}

\clearpage
\section{Notations of the physical parameters used in the text}
 
\begin{table*}[hbt]
	\small
	\caption{Notations used in the manuscript}
	\begin{tabular*}{1\textwidth}{rl}
		Notation & Definition\\ \hline
		$a$ -& nanocrystal radius \\	
  	$a_{\rm ex}$ -& exciton Bohr radius \\	
     	$E_g$ -& band gap energy \\
      $\Delta_{\rm SO}$ -& spin-orbit energy splitting in the valence band  \\
      $E_{\rm b}$ -& exciton binding energy\\
        $E_{\rm X}(a)$ -& size-dependent exciton optical transition energy\\
     	     	$g_{e}$ -& electron $g$-factor\\
     	$g_{h}$ -& hole $g$-factor\\
      $g_{\rm X}$ -& exciton $g$-factor\\
      $E_p$ -& Kane energy for the valence band -- conduction band interaction \\
       $\alpha_{\rm rb}$   ($g_{\rm rb}$) -& remote band contribution to the electron effective mass ($g$-factor) \\
 \hline
	\end{tabular*}
	\label{tab:notation}
\end{table*}